\documentclass[a4paper,fleqn,usenatbib]{mnras}
\usepackage[T1]{fontenc}
\usepackage{ae,aecompl}
\usepackage[utf8]{inputenc}

\usepackage{graphicx}	
\usepackage{amsmath}	
\usepackage{amssymb}	

\title[]{Ionization and feedback in Ly$\alpha$ halos around two radio galaxies at z$\sim$2.5 \thanks{Based on observations carried out with the GTC as part of the observing proposal 2-GTC2/11BIACMEX.}}

\author[S. G. Morais et al.]{S. G. Morais$^{1}$, A. Humphrey$^{2}$, M. Villar-Mart\'{i}n $^{3,4}$, P. Lagos$^{2}$, M. Moyano$^{5,6}$, 
\newauthor R. Overzier$^{7}$, S. di Serego Alighieri$^{8}$, J. Vernet$^{9}$, C. A. C. Fernandes$^{7}$\\
$^{1}$ Faculdade de Ci\^{e}ncias da Universidade do Porto, Rua do Campo Alegre, 4150-007 Porto, Portugal\\
$^{2}$ Instituto de Astrof\'{i}sica e Ci\^{e}ncias do Espa\c{c}o, CAUP, Rua das Estrelas, 4150-762 Porto, Portugal\\
$^{3}$ Centro de Astrobiolog\'{i}a (INTA-CSIC), Ctra de Torrej\'{o}n a Ajalvir, km 4, E-28850 Torrej\'{o}n de Ardoz, Madrid, Spain\\
$^{4}$ Astro-UAM, UAM, Unidad Asociada CSIC, Facultad de Ciencias, Campus de Cantoblanco, E-28049 Madrid, Spain\\
$^{5}$ Observatorio Astron\'{o}mico de Cordoba, Universidad Nacional de C\'{o}rdoba, Argentina\\
$^{6}$ Facultad de Matem\'{a}tica, Astronomia y F\'{i}sica, Universidad Nacional de C\'{o}rdoba, Argentina \\
$^{7}$ Observat\'{o}rio Nacional, Rua Jos\'{e} Cristino, 77. CEP 20921-400, S\~{a}o Crist\~{o}v\'{a}o, Rio de Janeiro-RJ, Brazil\\
$^{8}$ INAF - Osservatorio Astrofisico di Arcetri, Largo E. Fermi 5, 50125 Firenze, Italy\\
$^{9}$ European Southern Observatory, Karl Schwarzschild Strasse 2, 85748 Garching bei M\"{u}nchen, Germany
}

\pubyear{2016}

\begin{document}
\label{firstpage}
\pagerange{\pageref{firstpage}--\pageref{lastpage}}
\maketitle

\begin{abstract}
{We present new spectroscopic observations of two high redshift radio galaxies, TXS 0211-122 (z=2.34) and TXS 0828+193 (z=2.57), known to be associated with large Ly$\alpha$ halos. The observations were taken with the slits placed perpendicularly to the radio axis. With access to pre-existing Keck II observations taken with the slit placed along the radio axis we are able to compare the properties of the gas in different regions of the galaxies. 

In both objects we detect spatially extended Ly$\alpha$ emission perpendicularly to the radio axis. In TXS 0211-122, the flux and velocity profiles of Ly$\alpha$ are strongly affected by HI absorption/scattering. In line with previous studies, we find evidence for outflowing gas along the radio axis which may be the result of jet-gas interactions. 
In the slit oriented perpendicularly to the radio axis we find less perturbed gas kinematics, suggesting outflows of ionized gas in this object are focused along the radio jet axis. Additionally, we find evidence for a giant, UV-emitting arc or shell-like structure surrounding the radio galaxy Ly$\alpha$ halo, possibly resulting from feedback activity. 

In TXS 0828+193 a large Ly$\alpha$ halo ($\sim$56 kpc) is detected perpendicularly to the radio axis. Along both slit position angles we find evidence for outflowing gas, which we argue is part of an approximately spherical, expanding shell or bubble of gas powered by feedback activity in the central regions of the galaxy. Our results suggest a diversity in the spatial distribution of ionized outflows in powerful radio galaxies at z$\sim$2.5.}
\end{abstract}



\section{Introduction}\label{sec:Introduction}

High redshift radio galaxies (z\textgreater2, hereinafter HzRGs)  are useful for understanding the evolution of massive galaxies.
They are amongst the largest, most massive and most luminous objects at any epoch \citep[e.g.][]{Jarvis2001a,Jarvis2001b, Breuck2002, Rocca2004}. 
Near-IR and HST imaging of radio galaxies between redshifts 2 and 3 reveal clumpy morphologies, as expected if galaxies are forming through mergers and in agreement with hierarchical models of galaxy evolution \citep[e.g.][]{Dubinski1998}. 
They are often found in dense cluster-type regions \citep[e.g.][]{McCarthy1993, Pentericci2000, Venemans2004,Galametz2012,Mayo2012,Wylezalek2013,Dannerbauer2014}.
They are believed to be the progenitors of massive elliptical and cD galaxies \citep[e.g.][]{Pentericci1999, Miley2008, Adams2009, Hatch2009} since it appears that most massive galaxies undergo a phase of radio-loudness during their formation \citep[e.g.][]{Willott2001}.

Many powerful HzRGs are embedded in large ($\sim$ several 10s and up to 200 kpc) luminous 
Ly$\alpha$\footnote{In the interest of simplicity, we refer to the lines Ly$\alpha\lambda$1216, HeII$\lambda$1640 and the doublets NV$\lambda\lambda$1240, CIV$\lambda\lambda$1549 and CIII]$\lambda\lambda$1909 as Ly$\alpha$, HeII, NV, CIV and CIII].}
nebulae \citep{Ojik1997,VM2003,Reuland2003} which have clumpy, irregular morphologies \citep[e.g.][]{Reuland2003}, and are often aligned with the radio source axis \citep{McCarthy1987, McCarthy1995}. Alignment between the radio jets and CO(1-0) has also been found \citep[e.g.][]{Klamer2004, Nesvadba2009, Emonts2014}. \\ 
The main ionizing mechanism in the nebulae seems to be photoionization from Active Galactic Nuclei \citep[hereinafter AGN, e.g.][]{Fosbury1982}, however other mechanisms may also make some contribution: photoionization from stars \citep[e.g.][]{VM2007a}, photoionization by X-rays emitted by shocked warm gas and collisional excitation from shocks \citep[e.g.][]{Bicknell2000}.

Extended HI absorbers are often found associated with HzRGs. \citet{Ojik1997} discovered that 90\% of HzRGs with small radio sources (\textless 50 kpc) show associated absorption systems while only 25\% of HzRGs with larger radio sources had associated HI absorbers. In a few cases, absorption features are also observed in metal lines, such as CIV \citep{Binette2000,Jarvis2003}.
These absorption lines are thought to be formed in cold/warm gas between the emission source and the observer. The observed properties of this absorbing gas suggests that it is part of a super shell of gas with a radius larger than 50 kpc, that is expanding out of the host galaxy \citep{Binette2000, Humphrey2008b, Swinbank2015}. As such, they hold important clues about the mass assembly, feedback, and dispersion of metals through galaxies and the intergalactic medium. 

The origin, gas distribution, chemical composition and the source of ionization of the gas in the extended Ly$\alpha$ halos are still unclear. 
Ly$\alpha$ halos may provide important information about the early stages of the formation of massive elliptical galaxies therefore the study of their properties is essential to determine how hosts of powerful radio galaxies form and evolve. Numerous spectroscopic studies of HzRGs have been performed using the long-slit technique. In most cases the slit was aligned with the radio structures \citep[e.g.][]{Legrand1997,Vernet2001,VM2003}. This is the region where the impact of jet-gas interactions and the ionizing radiation of the AGN are likely to be strongest. 
In this paper we use long-slit spectra to study the properties of the regions far from the radio jets, outside the ionizing beams of the AGN, with the goal of better understanding the nature and origins of the Ly$\alpha$ halos. This gas traces the properties of the reservoir within which the radio galaxies are embedded, without the distortion produced by the radio structures and/or the excitation by the hard ionizing radiation field of the quasar.

This paper is organized as follows. Section \ref{sec:PreviousResults} summarizes the results obtained for TXS 0211-122 and TXS 0828+193 in previous works. The observations and data reduction are described in Section \ref{sec:Observations}. In Section \ref{sec:Models} the photoionization models used in this work are presented. In Section \ref{sec:Results} the observational results for TXS 0211-122 and TXS 0828+193 are presented, and analyzed. In Section \ref{sec:Discussion} the results are discussed. Finally, we summarise our main results and final conclusions in Section \ref{sec:Conclusions}.

A flat universe with $\Omega_{\Lambda}$=0.73, $\Omega_{m}$=0.27 and H$_{0}$=71 km s$^{-1}$ Mpc$^{-1}$ is adopted in this paper. Using this cosmology 1\arcsec \, corresponds to 8.295 kpc at z=2.34 (TXS 0211-122), and to 8.144 kpc at z=2.57 (TXS 0828+193).

\section{Overview of our sample of galaxies}\label{sec:PreviousResults}

TXS 0211-122 and TXS 0828+193 belong to the Texas Sky Survey catalogue \citep{Douglas1980}, a catalog containing low-frequency radio sources. They were first selected as high redshift candidates on the basis of their ultra-steep radio spectrum (USS). Optical identification and determination of the redshift of these objects was performed by \citet{Rottgering1993} confirming that they are high redshift radio galaxies.

\subsection{TXS 0211-122}

TXS 0211-122 is a powerful radio source at z=2.34. It has a radio flux at 8.2 GHz of 24 mJy \citep{Rottgering1993}, a very large radio structure \citep[$\sim$135 kpc,][]{Carilli1997} and an optical R-band magnitude of 22.7 \citep{Carilli1997}. 
\citet{Breuck2010} derived a maximum stellar mass of the host of this radio galaxy of \textless1.45 $\times$ 10$^{11}$ M$_{\sun}$.

The radio structure consists of a bright core and a smaller clump. A jet extends from the core towards the south, bends and reaches the eastern lobe \citep{Pentericci1999}. 
\citet{Pentericci1999} found that the UV continuum emission of this object is aligned with the axis of the radio source. 
This radio galaxy is associated with an extended Ly$\alpha$ ($\sim$110 kpc), CIV and HeII nebula \citep{Ojik1994, VM2003}, and with a $\geq$ 100 kpc scale neutral hydrogen (HI) absorbing structure \citep{Ojik1994}. 

\citet{Ojik1994} found that the Ly$\alpha$ emission is weak when compared to higher ionization lines and that the line-emitting gas is overabundant in nitrogen. They concluded that the galaxy is undergoing a massive starburst which would produce the dust necessary to absorb the Ly$\alpha$ emission and that the enhancement of nitrogen could be due to photoionization or shocks. 

\citet{VM2003} and \citet{Humphrey2006,Humphrey2007a} found a clear difference in kinematics between the gas inside (FWHM$\sim$700 km s$^{-1}$) and outside (FWHM$\leq$400 km s$^{-1}$) the radio structures. The authors propose that the inner gas is in outflow triggered by the interaction between the radio structures and the ambient gas, while the outer, more quiescent gas is in the process of infall.

This source has a high rest-frame UV continuum polarization of around 19\% \citep{Vernet2001}, and the Ly$\alpha$ nebula associated with this object has a polarization of 16.4$\pm$4.6\% in its outer, eastern region \citep{Humphrey2013TXS0211}, indicating that the nebula is partly powered by the scattering of Ly$\alpha$ photons by HI. 

\citet{Humphrey2013TXS0211} detected two continuum sources at a distance of $\sim$7\arcsec\, from the centre of the radio source. The sources were detected at the edges of the Ly$\alpha$ halo, one of which was found to be polarized and was also detected in H$\alpha$ confirming its association with this radio galaxy. They proposed that the two continuum sources were part of a shell of gas and dust around the Ly$\alpha$ halo, visible because it was illuminated by the active nucleus.

\subsection{TXS 0828+193}
			
TXS 0828+193 has a large radio source \citep[98 kpc,][]{Carilli1997}, with a total radio flux at 4.7 GHz of 22 mJy. It is found at z=2.57 \citep{Ojik1995} and has an R-band magnitude of 20.7 \citep{Carilli1997}. 

The radio structure consists of several clumps \citep{Pentericci1999}, and has a double morphology \citep{Rottgering1994} with a jet extending from the core to the northern hotspot. The UV continuum is elongated and aligned with the axis of the radio source \citep{Pentericci1999}.

The host galaxy has a maximum stellar mass of \textless3.98 $\times$ 10$^{11}$ M$_{\sun}$ \citep{Seymour2007}.
\citet{Ojik1997} found that the blue wing of the Ly$\alpha$ emission profile is absorbed by neutral gas. \citet{Nesvadba2009} found CO(3-2) emission in the halo of the radio structure.
This source has an intermediate continuum polarization of around 10\% \citep{Vernet2001}.

\citet{VM2002} discovered a giant halo ($\sim$130 kpc) of quiescent gas extending beyond the radio structures. 
They also detected two kinematic components in Ly$\alpha$, CIV and HeII. A kinematically quiescent component (FWHM \textless 400 km s$^{-1}$) extending across the whole object and beyond the radio structures, and a kinematically disturbed component (FWHM $\sim$1200 km s$^{-1}$) inside the radio structures. While the disturbed component traces an outflow triggered by the interaction between the ambient gas and the radio structures \citep{Humphrey2006}, \citet{Humphrey2007a} found evidence of infall for the quiescent gas. \citet{VM2002} constrained the extended gas metallicity to be close to solar.

\section{Observations and data reduction}\label{sec:Observations}

The new observations presented here were obtained using the Optical System for Imaging and low-Intermediate-Resolution Integrated Spectroscopy \citep[OSIRIS,][]{Cepa2000,Cepa2003} on the 10.4 m Gran Telescopio Canarias (GTC) as part of the observing proposal 2-GTC2/11BIACMEX. 
The instrument was operated in long-slit spectroscopy mode with 100 kHz CCD readout mode. The two targets were observed with the R1000B grism. Long-slit spectra were obtained for TXS 0211-122 and TXS 0828+193 with total integration time of $\sim$2 hours per target, split into 9 integrations of $\sim$800 seconds. A 0.41\arcsec \, wide slit was used for TXS 0211-122 and a 0.8\arcsec \, wide slit was used for TXS 0828+193. The spatial pixel scale is 0.254\arcsec\, per pixel. The slit was positioned perpendicularly to the major axis of the radio and optical emission. A summary of the observations is presented in Table \ref{table:Observation parameters}.
	
The data were reduced, calibrated and corrected for reddening caused by Galactic extinction using standard IRAF \citep[Image Reduction and Analysis Facility,][]{Tody1993} routines.

\begin{table*}
\caption{Journal of observations}{Journal of observations. (1) Object name;  (2) redshift; (3) date of the observations; (4) total exposure time; (5) slit position angle in degrees of North (i.e. N=0, E=90); (6) full width at half maximum of the seeing disc; (7) resolution; (8) sky transparency.} 
\label{table:Observation parameters}
\begin{tabular}{c c c c c c c c} 
\hline 
Object 	   &   z     &     Date obs. (2011) & Exp.  (s) &  PA ($^{\circ}$)   & Seeing (\arcsec)  & Spectral resolution (\AA) & Sky transparency \\
(1)                     &  (2)   &          (3)                    &(4)          &         (5)           &       (6)              &      (7)  &   (8) \\
\hline
TXS 0211-122   &  2.340  &  30 September     &    7200   &   22.5      & 0.92 $\pm$ 0.02        & 3.1 - 4.0 & Clear    \\
\hline
TXS 0828+193  &  2.572  &  28 November      &    7380   &    - 45     &  1.0 $\pm$ 0.03         & 6.2 - 9.3   &  Clear  \\
\hline
\end{tabular}
\end{table*}

\subsection{Keck II data}
	
In addition to the new observations we made use of existing observations of TXS 0211-122 and TXS 0828+193 obtained using the Low Resolution Imaging Spectrometer \citep[LRIS,][]{Oke1995} at the Keck II 10 m telescope.
A 300 line mm$^{-1}$ grating and a 1\arcsec \, wide slit were used, providing a dispersion of 2.4 \AA \, and an effective resolution of $\sim$10 \AA. The spectral range was $\sim$3900 - 9000 \AA. The spatial scale is 0.214\arcsec \, per pixel.
The slit was oriented along the radio axis of the galaxy. TXS 0211-122 was observed using a position angle (PA) of 104$^{\circ}$, and TXS 0828+193 using a PA of 44$^{\circ}$. During the observations the seeing varied between $\sim$0.5 and 1\arcsec. For observations and data reduction details see \citet{Vernet2001}. Results from this data have previously been published by \citet{Vernet2001, VM2002, VM2003, Humphrey2007a, Humphrey2007b, Humphrey2008a}.

\begin{figure}\centering
\includegraphics[trim=0 10 0 0mm, width=\columnwidth]{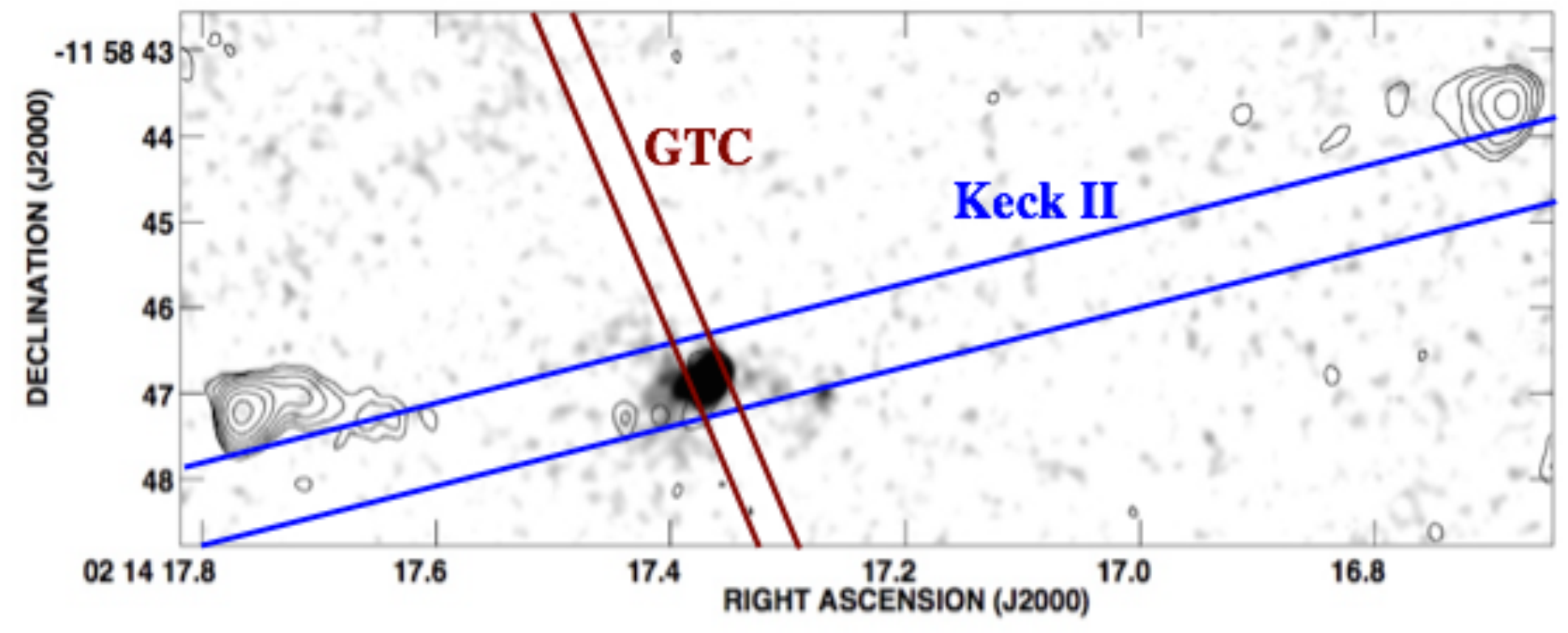}
\caption{Radio image of the full field of TXS 0211-122 \protect\citep{Pentericci1999} with the positions and widths of the long slits overplotted. The long slits are much longer that the dimension of the images.}\label{fig:TXS0211_slit}
\end{figure}

\begin{figure}\centering
\includegraphics[trim=0 10 0 0mm, width=6cm]{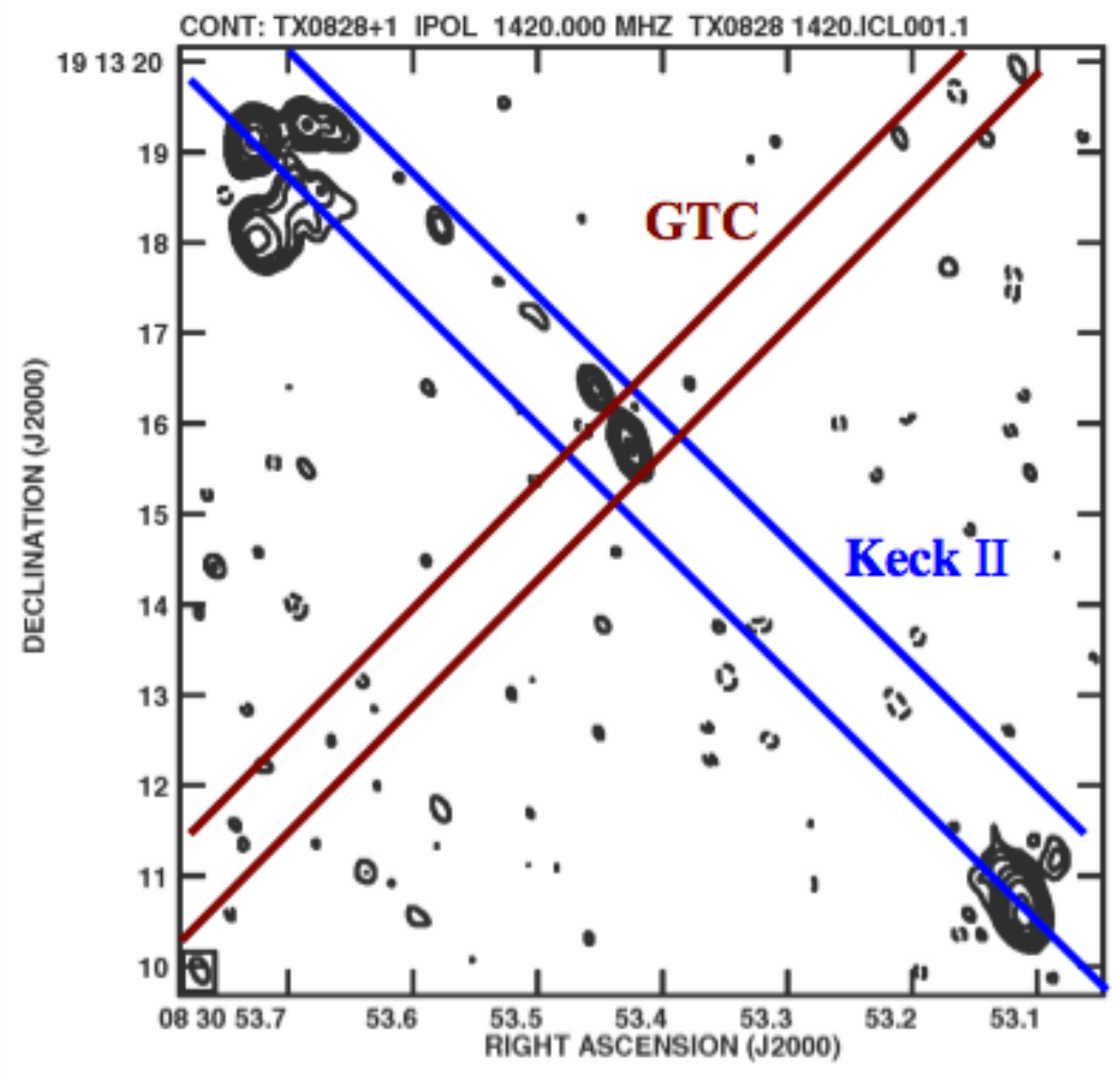}
\caption{MERLIN\protect\footnotemark radio image of TXS 0828+193. The positions and widths of the slits are overplotted. The long slits are much longer that the dimension of the images.}\label{fig:TXS0828_slit}
\end{figure}

The positions and widths of the slits are shown in Fig. \ref{fig:TXS0211_slit} and \ref{fig:TXS0828_slit}.

\footnotetext{MERLIN is a National Facility operated by the University of Manchester at Jodrell Bank Observatory on behalf of STFC.}

\subsection{Data Analysis}\label{sec:Fitting}

IRAF routines were used to measure the emission-line fluxes from simple line integrals.
Kinematic information was extracted from the two dimensional spectra by fitting single Gaussians to the spatial profiles of the emission lines. Gaussians were fitted at each spatial position along the slit, summing a number of spatial pixels to increase S/N.
The velocity and FWHM were measured from the Gaussians fitted to the emission lines. The FWHM was corrected in quadrature for instrumental broadening.
The systemic velocity was determined from the central wavelength of a Gaussian fit to the HeII emission line in the centre of the galaxy. The zero points of the spatial scales correspond to the location of the peak of the UV continuum which is assumed to be coincident with the location of the nucleus of the galaxy. This is assumed because, as already discussed by \citet{Pentericci2001} and \citet{VM2003} the radio core of the two radio galaxies is spatially coincident with the UV continuum peak. 
3$\sigma$ errors were estimated using the statistical properties of the spectra. 

Our main goal in this work was to study the properties of the extended gas, thus we needed to determine if the emission lines in the GTC spectra were extended. In order to do this we compared the spatial profile of the emission lines with the seeing. To determine the seeing profile in the observations the spatial profile of the seeing disc along the slit was rebuilt using a star present in the acquisition image. An aperture equal to the slit width in each observation was then used to extracted the stellar flux. Due to the slit widths, the stellar profiles obtained using this method have slightly narrower FWHMs than the seeing of the observations.

\subsection{Archival HST data}

Images of TXS 0828+193 were obtained from the HST science archive (PI: George Miley, Proposal ID 11738).
These images were taken with the Advanced Camera for Surveys (ACS) and the Wide Field Camera 3 (WFC3) of HST. TXS 0828+193 was observed on 2010 March 26$^{th}$ and on 2010 March 28$^{th}$ with the Wide Field Channel (WFC) of ACS. The WFC has a field of view of 202\arcsec\,$\times$ 202\arcsec\, and a pixel scale of 0\arcsec.049 per pixel. 
To carry out these observations the filters F814W and F606W were used . The total exposure time was 10173 seconds in F814W and 10173 seconds in F606W. 

TXS 0828+193 was also observed on 2010 March 10$^{th}$ and on 2011 May 8$^{th}$ with the IR Channel of the Wide Field Camera 3 (WFC3) using the filters F110W and F160W. The IR Channel has a total field of view of 123\arcsec\,$\times$ 136\arcsec\, with a pixel  scale of 0.13\arcsec. The FWHM resolution of HST/WFC3 and ACS during the observations was 0\arcsec.07 - 0\arcsec.14. Table \ref{table:HST_Observation} gives details of the observations, the spectral range covered by each of the filters and the rest frame of TXS 0828+193.
\begin{table*}
\caption{HST observations of TXS 0828+193 }{Journal of observations. (1) Instrument;  (2) filter; (3) integration time in seconds; (4) date of the observations; (5) minimum wavelength; (6) effective wavelength; (7) maximum wavelength; (8) rest frame wavelength range; (9) main emission lines; (10) galactic extinction.} 
\label{table:HST_Observation}
\begin{tabular}{c c c c c c c c c c} 
\hline 
Instrument	   &   Filter    &   Int. Time   &  Date obs. & Min. $\lambda$   &  Eff. $\lambda$  & Max. $\lambda$ & Rest frame & Main emission  & Galactic  \\
                    &                    &  (s)      & (TU)  &         (\AA)       &          (\AA)           &  (\AA)      &   (\AA)  &   lines   & extinction \\ 
                                        &     &                 &   &                &                     &        &   &    &  \\
(1)                     &  (2)   &          (3)         & (4)          &         (5)           &       (6)              &      (7)  &   (8) &   (9)  &  (10) \\

\hline
ACS   &  F606W      &   10173   &   2010     &  4633   &  5808       & 7180   &  1297.03 - 2010.08 & CIII] $\lambda$1909  & 0.085$\pm$0.01 \\
          &    &                  &  March 28$^{th}$   &           &                 &            & & HeII $\lambda$1640 &     \\
          &			  &                  &                                           &           &       &          &            & CIV $\lambda$1549 &     \\
\hline
ACS   &  F814W  &    10173   &  2010     & 6885    &                 & 8800 &  1927.49 - 2463.61   & CII] $\lambda$2326 & 0.058$\pm$0.02 \\
          &   &                &March 26$^{th}$   &           &                 &            & &  [NeIV] $\lambda$2424 &     \\
\hline
WFC3  & F110W 		  &   2470       &    2010      & 9319      &  11534  & 13749 & 2608.90 - 3849.10 & [OII] $\lambda$3727 & 0.029$\pm$0.005 \\ 
          &			  &                  &    March 10$^{th}$     &           &           &      &            & [NeV] $\lambda$3426 &     \\
          &			  &                  &                                           &           &        &         &            & MgII $\lambda$2800 &     \\
\hline
WFC3 & F160W                     &   2611        & 2011 	&  14027.5 &  15369 & 16710.5 & 3927.07 - 4678.19 & H$\gamma$ $\lambda$4342 & 0.02$\pm
$0.005 \\
          &			  &                  &      May 8$^{th}$  &           &                 &     &       & [OIII] $\lambda$4364  &     \\
\hline
\end{tabular}
\end{table*}

\section{Photoionization models}\label{sec:Models}
				
Our photoionization model runs were performed using the multipurpose code MAPPINGS 1e \citep{Binette1985, Ferruit1997}. In these models it is considered that an ionizing continuum of the form $f_{\nu} \propto \nu ^{\alpha}$ illuminates an isobaric, plane-parallel slab of gas. 
The model has several parameters: gas density of hydrogen (n$_H$), ionization parameter (U), spectral index of the ionizing continuum ($\alpha$), and gas metallicity (Z). The ionization parameter is defined as 
\begin{equation}
U=\frac{Q}{4 \pi r^2 n_H c} \,,  
\end{equation}
where $Q$ is the ionizing photon luminosity, $r$ is the distance of the ionized cloud from the ionizing source, $c$ is the speed of light, and $n_H$ is the hydrogen gas density. It indicates how intense the ionizing radiation is as felt by the cloud. In the grid of photoionization calculations, U increases from 0.00005 to 0.00005$\times$3$^{9}$ in multiplicative steps of times 3. 

Photoionization by a power law of index -1.0 was used because according to the results of \citet{VM1997} it is best able to reproduce the ratios observed in HzRGs. For the hydrogen density, a value of n$_{H}$=100 cm$^{-3}$ was chosen because previous studies have found, by various lines of argument, that the extended emission line gas has very approximately this density \citep{McCarthy1990, VM2002, VM2003}. 

Finally, the gas metallicity takes the values Z=0.2 Z$_{\sun}$, Z=Z$_{\sun}$ and Z=3 Z$_{\sun}$, where Z$_{\sun}$ is the solar metallicity \citep{Asplund2006}. The chemical abundances have been scaled with all metals scaled proportionally to oxygen, except nitrogen, which is scaled quadratically with oxygen \citep{Henry2000}. 

Besides photoionization by the AGN, ionizing shocks may also play a role in heating the gas \citep[e.g.][]{VM1999b, Tadhunter2000}, therefore shock and shock plus precursor models computed by \citet{Allen2008} using the code MAPPINGS III are also considered. 
In order to be consistent we chose n$_{H}$=100 cm$^{-3}$. The models available with this hydrogen density have solar metallicities. 
The shock velocities are in the range 100 km s$^{-1}$ $\leq$ v $\leq$ 1000 km s$^{-1}$. \citet{Overzier2005} derived a magnetic field strength of $\sim$100 - 200 $\mu$G for TXS 0828+193, thus we chose B=100 $\mu$G, and the magnetic parameter B/n$^{1/2}$=10 $\mu$G cm$^{3/2}$, where $n$ is the pre-shock number density.


\section{Results and Analysis}\label{sec:Results}
\subsection{TXS 0211-122}\label{sec:Results1}

\begin{figure}\centering
\textbf{TXS 0211-122}\par\medskip                                                             
\includegraphics[trim=60 180 60 45mm, width=\columnwidth]{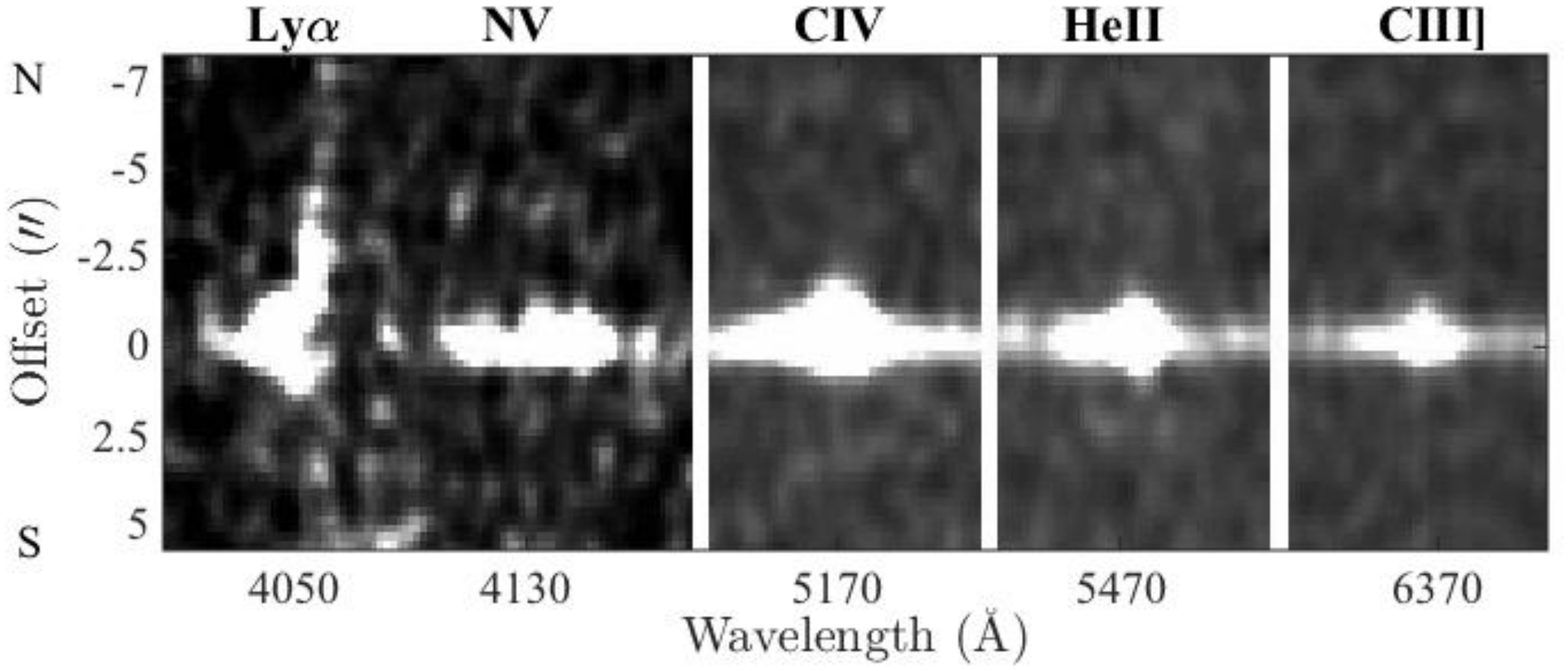}
\hfill
\includegraphics[trim=60 200 60 40mm,width=\columnwidth]{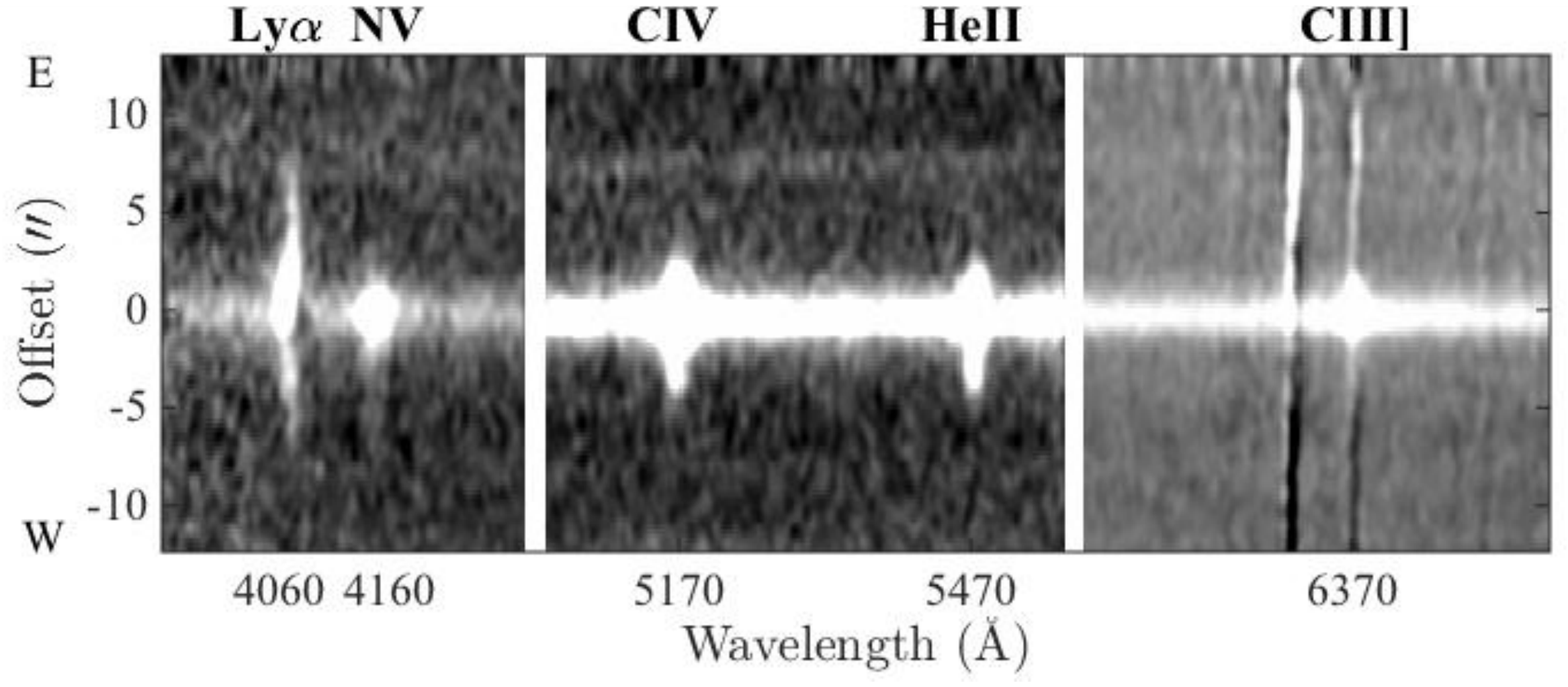}
\quad
\caption{Top panel: Sections of the two dimensional GTC spectrum of TXS 0211-122 obtained with the slit oriented perpendicularly to the radio axis (PA=22.5$^{\circ}$). The spectra are smoothed with a gaussian function with radius of 3 pixels to better show the emission line features. Bottom: Two dimensional Keck II spectrum of the main emission lines of TXS 0211-122, with the slit oriented along the radio axis (PA=104 $^{\circ}$). The position of the radio core (0) is identified 
with the brightest continuum emission.}\label{fig:2D_TXS0211}
\end{figure}

The two dimensional spectra of TXS 0211-122 are shown in Fig. \ref{fig:2D_TXS0211}. They show the spatial distribution of the brightest emission lines along the slits. Several emission lines show spatial asymmetry. The Ly$\alpha$ emission is spatially extended along both slit position angles. 

\begin{figure}
\centering
\textbf{TXS 0211-122}\par\medskip                                                             
\includegraphics[width=\linewidth]{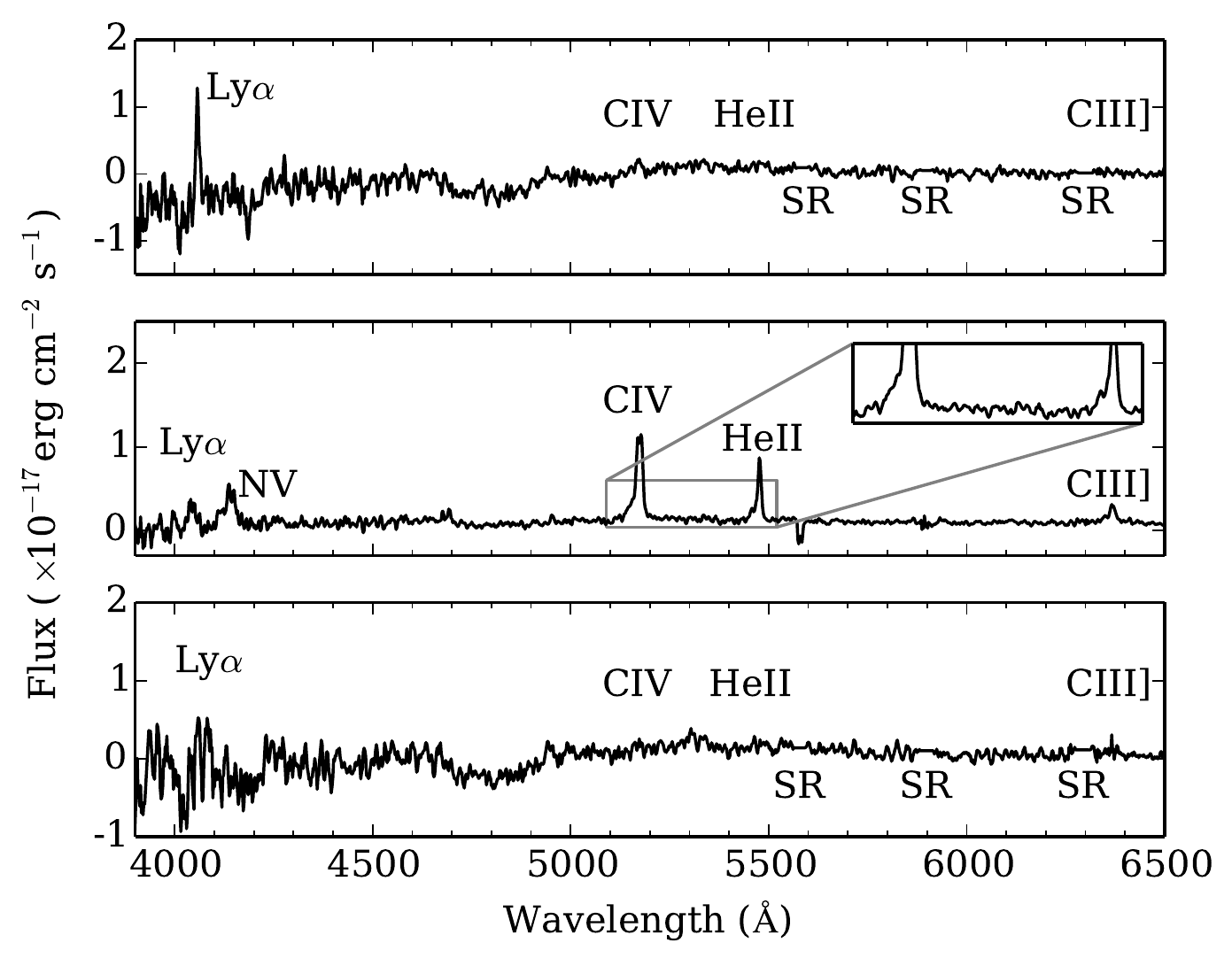}
\vspace*{-5mm}
\caption{One dimensional spectra of different regions of TXS 0211-122 observed with the slit positioned perpendicularly to the radio axis. The top spectrum shows the region between -7\arcsec \, and -1.3\arcsec \,, the middle figure shows the nuclear region (central 2\arcsec), and the bottom spectrum shows the region between 1.3\arcsec \, and 7.6\arcsec. The name of each emission line is marking the position of the line. The spectra have been smoothed to reduce noise and they show some skyline residuals (SR).}
\label{fig:GTC_TXS0211_regions}
\end{figure}

The one dimensional spectra showing different regions of TXS 0211-122 from the perpendicular slit (PA=22.5$^{\circ}$) are shown in Fig. \ref{fig:GTC_TXS0211_regions}. Ly$\alpha$ emission shows clear signs of absorption. It is asymmetric, has several peaks and dips and the flux drops below the continuum level in several regions. \\
The CIV$\lambda$1548.2 and CIV$\lambda$1550.8 emission lines are blended creating a double peaked profile. In the central region there is evidence for an excess of flux in the blue wing of HeII and CIV (see Fig. \ref{fig:GTC_TXS0211_regions}). \citet{Humphrey2006} proposed this is the signature of an outflow triggered by jet-gas interactions. 

\begin{figure}
\centering         
\textbf{TXS 0211-122}\par\medskip                                                 
\includegraphics[width=\columnwidth]{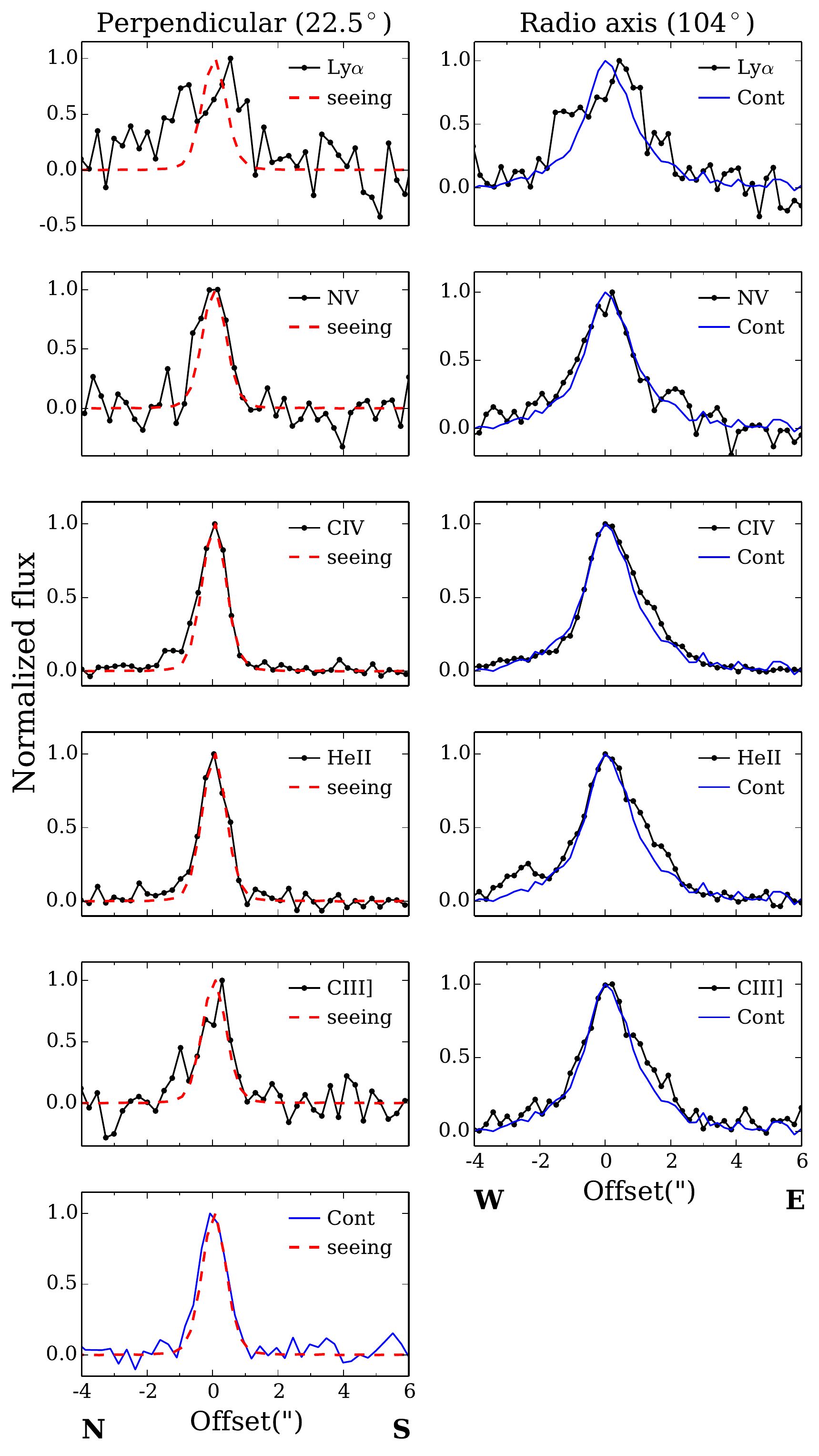}
\vspace*{-5mm}
\caption{Normalized flux of the different emission lines (black) compared with the seeing (red) and the continuum profile (blue). In the left we show the values observed perpendicularly to the radio axis and in the right the results observed along the radio axis. Zero in the spatial direction represents the peak of the continuum emission.}
\label{fig:TXS0211_SB}
\end{figure}

In the left panels of Fig. \ref{fig:TXS0211_SB} the spatial profiles of the emission lines and the spatial profile of the continuum are compared with the seeing profile in the perpendicular slit (PA=22.5$^{\circ}$). The profiles have been shifted in order to make the spatial centroid coincide with that of the stellar profile. In the right panels of Fig. \ref{fig:TXS0211_SB} we show the emission lines spatial profiles compared to the continuum profile in the observations with the slit placed along the radio axis (PA=104$^{\circ}$).

In the GTC spectrum the spatial distribution of the Ly$\alpha$, NV, CIV, HeII and CIII] emission lines is consistent with a Gaussian profile of FWHM(obs)=2.49$\pm$0.06\arcsec\, for Ly$\alpha$, FWHM(obs)=1.04$\pm$0.01\arcsec\, for NV, FWHM(obs)=1.02$\pm$0.01\arcsec\, for CIV, FWHM(obs)=1.02$\pm$0.03\arcsec\, for HeII and FWHM(obs)=1.1$\pm$0.2\arcsec\,for CIII]. All lines are spatially resolved compared with the seeing (FWHM=0.87$\pm$0.04\arcsec). The spatial variation of the kinematic properties of Ly$\alpha$, HeII and CIV in the direction perpendicular to the radio axis (Fig.\ref{fig:velocity_FWHM_TXS0211}) further confirms that the lines are spatially extended. 
Correcting for seeing broadening in quadrature, the intrinsic FWHM values are 2.33$\pm$0.07\arcsec\,(19.3$\pm$0.6 kpc) for Ly$\alpha$, 0.57$\pm$0.06\arcsec\,(4.7$\pm$0.5 kpc) for NV, 0.52$\pm$0.07\arcsec\,(4.3$\pm$0.6 kpc) for CIV, 0.52$\pm$0.87\arcsec\,(4.3$\pm$0.7 kpc) for HeII, and 0.7$\pm$0.2\arcsec\,(6$\pm$2 kpc) for CIII]. Because they are FWHM, these values should be considered lower limits to the actual extent of the emission lines.
The continuum is also resolved, with FWHM=0.99$\pm$0.01\arcsec. Correcting for seeing broadening the continuum shows FWHM=0.47$\pm$0.08\arcsec\,(3.9$\pm$0.7 kpc). \\
The Ly$\alpha$ emission shows an excess towards the North (see also Fig. \ref{fig:2D_TXS0211}). The excess of emission above the seeing wings shows that at $\sim$-2\arcsec\, Ly$\alpha$ is dominated by extended emission. \\
In the direction of the radio axis (Keck II, PA=104$^{\circ}$) the emission lines are more extended. As \citet{Ojik1994} and \citet{VM2003} have already shown Ly$\alpha$ has a total detected extent of $\sim$110 kpc. CIV and HeII are also extended, showing that the nebula is ionized and is significantly enriched in metals \citep{VM2003}.

\subsubsection{Offset UV Continuum Source}\label{sec:ContinuumSources}

\begin{figure}\centering
\textbf{TXS 0211-122}\par\medskip
\includegraphics[width=7cm]{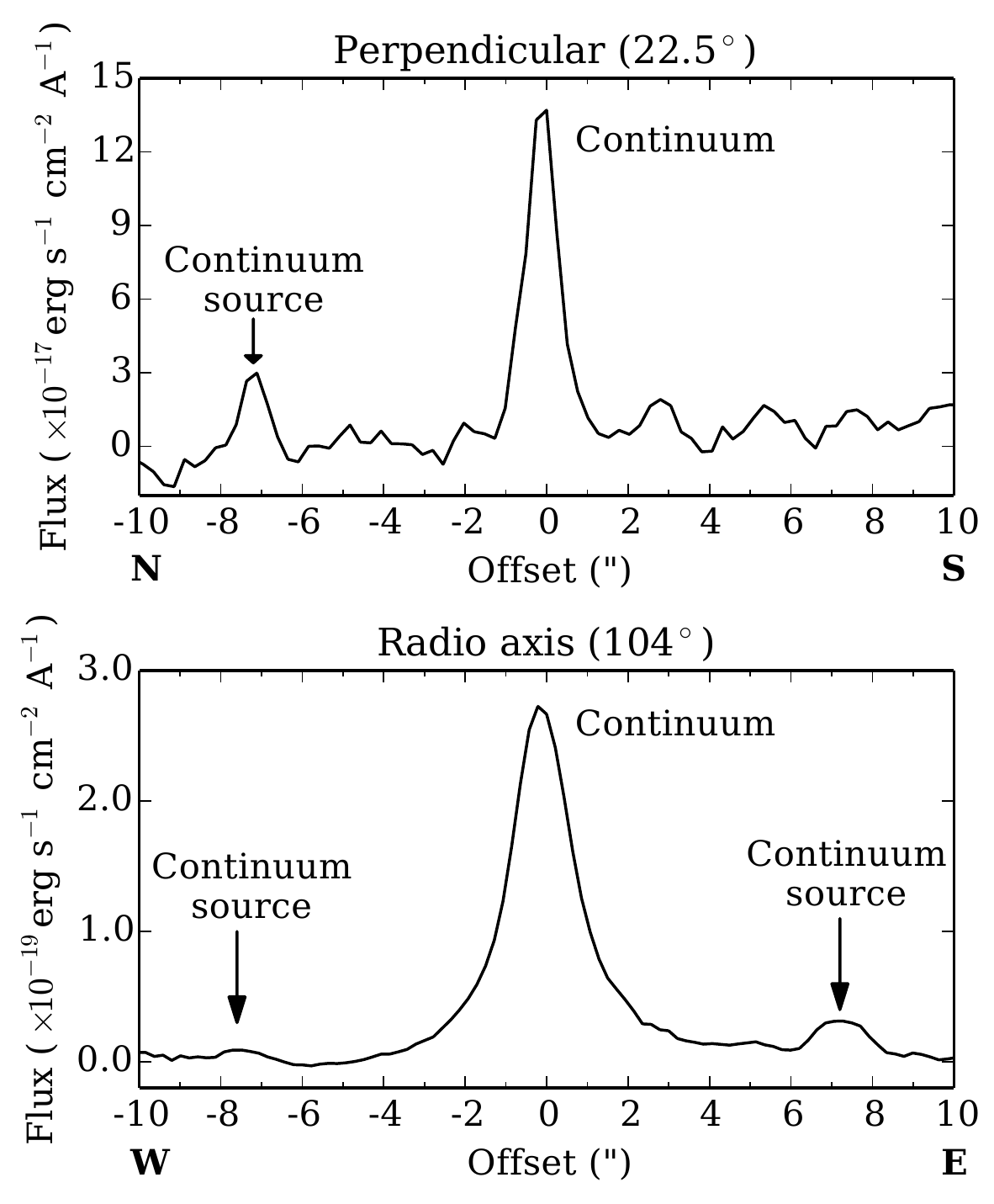}
\vspace*{-3mm}
\caption{Continuum sources observed in TXS 0211-122. Top: continuum source observed in the perpendicular slit. Bottom: continuum sources detected in the direction of the radio axis.}
 \label{fig:TXS0211_sources}
 \end{figure}

In the GTC spectrum (PA=22.5$^{\circ}$) a faint continuum source is detected at 7.1\arcsec$\pm$0.1\arcsec\,(59$\pm$1 kpc) from the centre of the galaxy (see top panel of Fig. \ref{fig:TXS0211_sources}). It is detected on the side to which Ly$\alpha$ is more extended. 

The source has a FWHM of 0.8\arcsec$\pm$0.3\arcsec \, and a signal to noise ratio of 4.8. Assuming that the source is associated with the radio galaxy, the rest-frame wavelength range where the source is detected is $\sim$1900 \AA \, - $\sim$2300 \AA. No emission lines are detected from this source.

This continuum source is reminiscent of the pair of similarly offset UV continuum sources detected previously by \citet{Humphrey2013TXS0211}, along PA=104$^{\circ}$. The sources are visible in the bottom panel of Fig. \ref{fig:TXS0211_sources} at -7.3\arcsec\,and 7.4\arcsec\,from the centre of the radio galaxy. This result will be discussed further in Section \ref{sec:Discussion}.

\subsubsection{Line ratios and comparison with models}

In order to better understand the properties and physical conditions in the large scale Ly$\alpha$ halo, emission line fluxes and ratios were measured from several spatial apertures, for comparison against the ionization models. 
The apertures are optimised to detect faint lines and low surface brightness extended emission. Table \ref{table:RatiosTXS0211_GTC} and Table \ref{table:RatiosTXS0211_Keck} show line ratios for GTC and Keck II data, respectively. 
The nuclear aperture shows a difference in line ratios between the Keck II data and GTC data but this could be due to a slight misalignment between the slits and/or to the different apertures chosen in each case.  
\begin{table*}\centering 
\vspace*{5mm}
\begin{tabular}{c c c c c c c c c c} 
\hline\hline 
Pos.  &  Ly$\alpha$/HeII & Ly$\alpha$/CIV  & Ly$\alpha$/NV      &      CIV/HeII    &        CIV/CIII]         & NV/HeII & NV/CIV \\
(1)     &   (2)                &     (3)            &             (4)          &          (5)                  &    (6)        &         (7)    &   (8)   \\
\hline
-3.3\arcsec : -1.8\arcsec   &  5.0$\pm$0.3   &    -     &               -            &    -            &     -                 &          -     &         -   \\
\hline
-1.8\arcsec : -0.8\arcsec   &  2.9$\pm$0.2    &  1.5$\pm$0.2      &              -            &    2$\pm$0.2              & 3.1$\pm$0.2  & - & -   \\
\hline
-0.8\arcsec : 0.8\arcsec     &  0.8$\pm$0.1  &  0.34$\pm$0.1   &  0.5$\pm$0.1  &  2.36$\pm$0.06   &  4.51$\pm$0.08  &  1.56$\pm$0.09 & 0.66$\pm$0.08  \\
\hline
0.8\arcsec : 1.8\arcsec     & 2.5$\pm$0.3     &   1.6$\pm$0.3  &          -  &               1.5$\pm$0.3           &  1.9$\pm$0.3   &  -        & -  \\
\hline 
\end{tabular}
\caption[Emission line ratios from several apertures in the TXS 0211-122 spectrum, observed in the direction perpendicular to the radio axis]{Emission line ratios from several apertures in the TXS 0211-122 spectrum observed in the direction perpendicular to the radio axis (PA=22.5$^{\circ}$). Columns are as follows: (1) position of the aperture along the slit in arcseconds; (2) the  Ly$\alpha$/HeII ratio; (3) the  Ly$\alpha$/CIV ratio; (4) the  Ly$\alpha$/NV ratio; (5) the CIV/HeII ratio; (6) the  CIV/CIII] ratio; (7) the  NV/HeII ratio; (8) the NV/CIV ratio.} 
\label{table:RatiosTXS0211_GTC} 
\end{table*}
\begin{table*}\centering 
\vspace*{5mm}
\begin{tabular}{c c c c c c c c c c} 
\hline\hline 
Pos.  &  Ly$\alpha$/HeII & Ly$\alpha$/CIV  & Ly$\alpha$/NV      &      CIV/HeII    &        CIV/CIII]         & NV/HeII & NV/CIV \\
(1)     &   (2)                &          (3)            &             (4)          &          (5)                &    (6)        &         (7)    &   (8)         \\
\hline
-5.8\arcsec\, : -3.9\arcsec  &    5.4$\pm$0.3  &      3.6$\pm$0.3   &      -  &    1.5$\pm$0.4    &      1.7$\pm$0.4           &   -  & -  \\
\hline
-3.9\arcsec\, : -1.9\arcsec  &    1.3$\pm$0.1  &  1.4$\pm$0.1   &    1.4$\pm$0.1  &    1.0$\pm$0.1    &   3.3$\pm$0.2       &     0.9$\pm$0.2  & 1.0$\pm$0.2  \\
\hline
-1.9\arcsec\, : 1.3\arcsec   &  0.65$\pm$0.05    &   0.33$\pm$0.04   &  0.67$\pm$0.04 & 1.94$\pm$0.03  &  3.43$\pm$0.05  &  0.96$\pm$0.05 &  0.49$\pm$ 0.04 \\
\hline
1.3\arcsec\, : 4.1\arcsec &    4.4$\pm$0.1    &   2.52$\pm$0.07   &    4.58$\pm$0.09  &    1.8$\pm$0.1    &    2.8$\pm$0.12        &  1.0$\pm$0.2        & 0.5$\pm$ 0.2  \\
\hline
\end{tabular}
\caption{Emission line ratio measurements from the TXS 0211-122 spectrum, observed along the radio axis (PA=104$^{\circ}$). Columns are as in \ref{table:RatiosTXS0211_GTC}.} 
\label{table:RatiosTXS0211_Keck} 
\end{table*}

\begin{figure}\centering
\textbf{TXS 0211-122}\par\medskip                                                             
\includegraphics[width=\columnwidth]{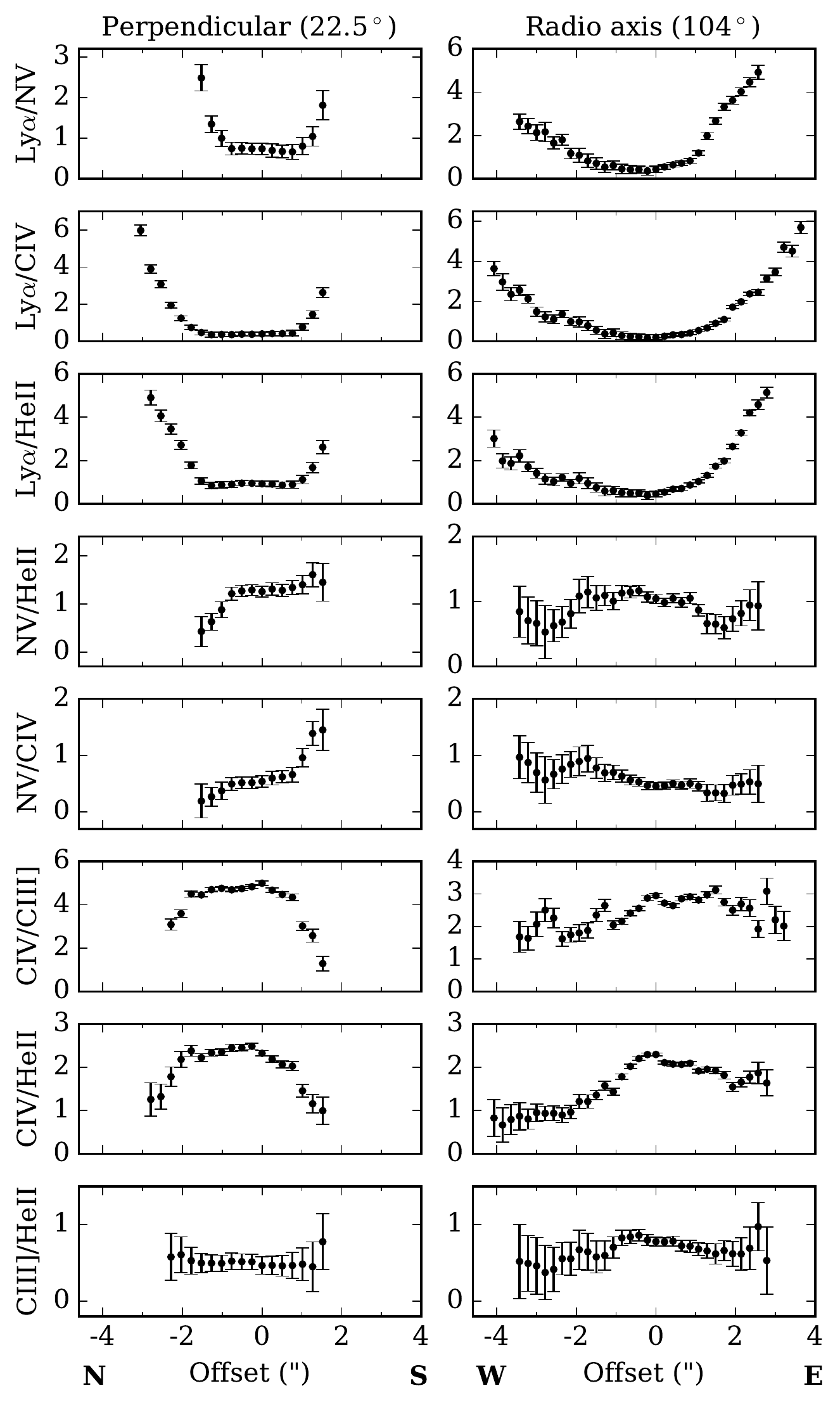}
\vspace*{-5mm}
\caption[Ratios between the emission lines in TXS 0211-122 as a function of distance in the galaxy]{Ratios between the emission lines of TXS 0211-122 as a function of distance in the galaxy. In the left we have the flux ratios in the direction perpendicular to the radio axis (PA=22.5$^{\circ}$) and in the right in the direction of the radio axis (PA=104$^{\circ}$).}
\label{fig:TXS0211_ratios}
\end{figure}

Fig. \ref{fig:TXS0211_ratios} shows how the line ratios vary along the slits. The emission line ratios of this radio galaxy show strong NV and weak Ly$\alpha$ relative to the other emission lines \citep[compared with typical values in HzRGs, e.g.][]{McCarthy1993, Humphrey2008a}. 
The ratio between Ly$\alpha$ and the other emission lines presents a "U shape" along both slits. This provides evidence for a variation in Ly$\alpha$ absorption, with strong absorption in the central regions, and progressively less absorption at larger radii. 

HzRGs showing strong Ly$\alpha$ absorption often show significant CIV absorption \citep[e.g.][]{Binette2000, Humphrey2013a}. We find no evidence for CIV absorption in the form of unusually low CIV flux ratios or a U-shaped CIV line ratio curve reminiscent of Ly$\alpha$. However, without higher spectral resolution (FWHM $\la$ 10 km s$^{-1}$) we are unable to definitively rule out the presence of CIV absorption.

\begin{figure}\centering
\textbf{TXS 0211-122} \\ Perpendicular (22.5$^{\circ}$)\par\medskip                                                          
\includegraphics[width=\columnwidth]{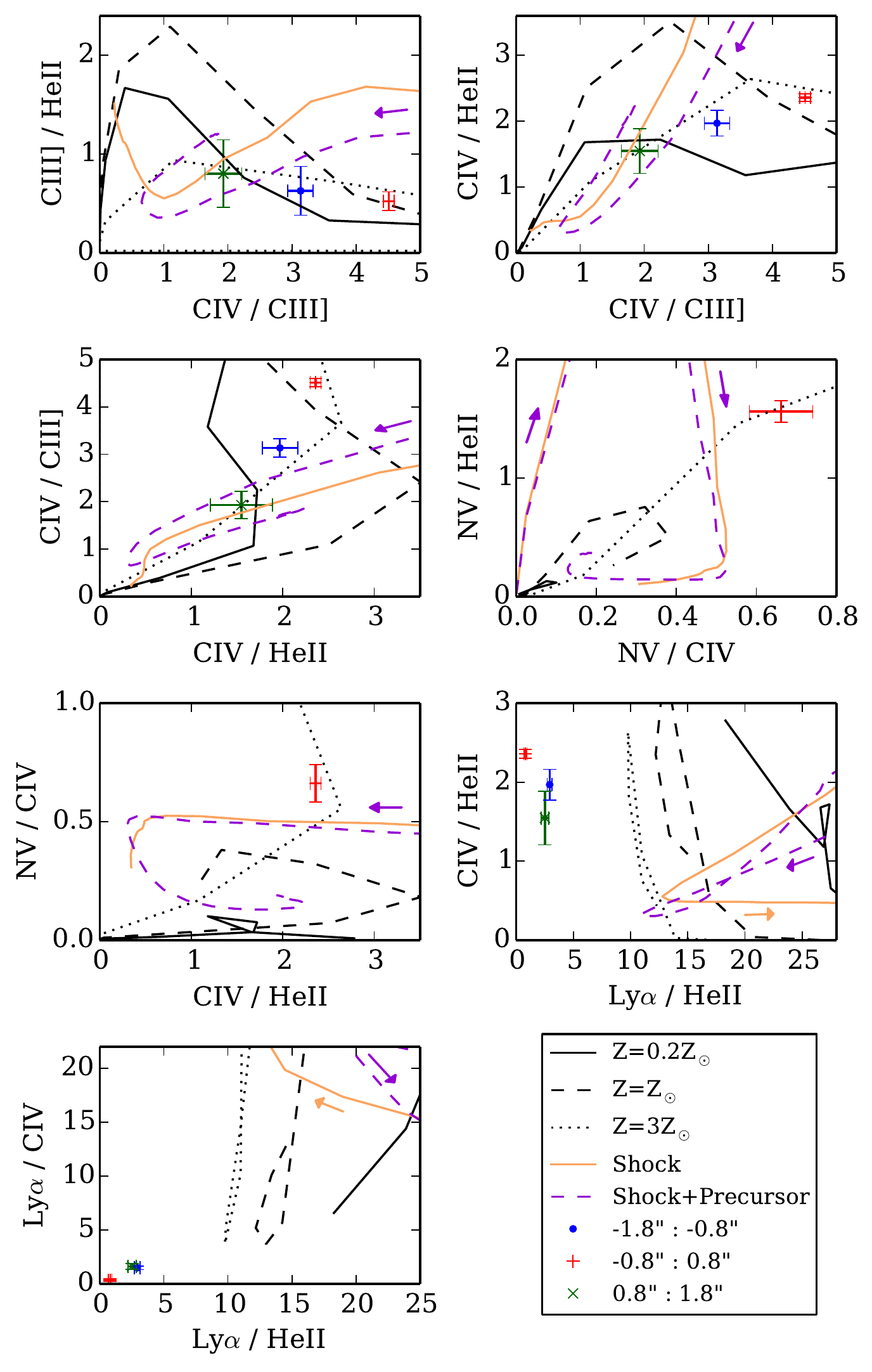}
\vspace*{-5mm}
\caption{Diagnostic diagrams involving the Ly$\alpha$, NV, HeII, CIV and CIII] emission lines. The black lines represent sequences of photoionization models for three different metallicities. U varies along each metallicity line. Shock models by \protect\citet{Allen2008} are also shown. Orange solid lines denote the predictions of pure shock models, and violet-dotted lines denote the prediction of shock plus precursor models. The arrows indicate the direction of increasing shock velocity.}
\label{fig:GTC_TXS0211_grids}
\end{figure}

Line ratios between the different emission lines are now compared to the ionization models. Diagnostic diagrams were constructed in order to visualize the comparison. \\
According to the diagnostic diagrams involving CIV, HeII and CIII] in Fig. \ref{fig:GTC_TXS0211_grids} a photoionization model with Z=3Z$_{\sun}$ and a spatial variation in ionization parameter is able to reproduce our flux ratio measurements. 
However, the region from 0.8\arcsec\, to 1.8\arcsec \, can be reproduced by a photoionization model with Z=3Z$_{\sun}$ or by a photoionization model with Z=0.2Z$_{\sun}$. This illustrates the strong degeneracy between metallicity solutions when using the CIV, HeII and CIII] line ratios. \\
In the direction perpendicular to the radio axis NV is only detected in the centre of the galaxy. The NV strength in relation to CIV and HeII suggests high levels of chemical enrichment of the gas. The ratio NV/HeII versus NV/CIV suggests a metallicity of Z=3Z$_{\sun}$. The small variation in CIII]/HeII suggests little or no variation in C/He and, C/H. 

\begin{figure}\centering                        
\textbf{TXS 0211-122} \\ Radio axis (104$^{\circ}$)\par\medskip                                 
\includegraphics[width=\columnwidth]{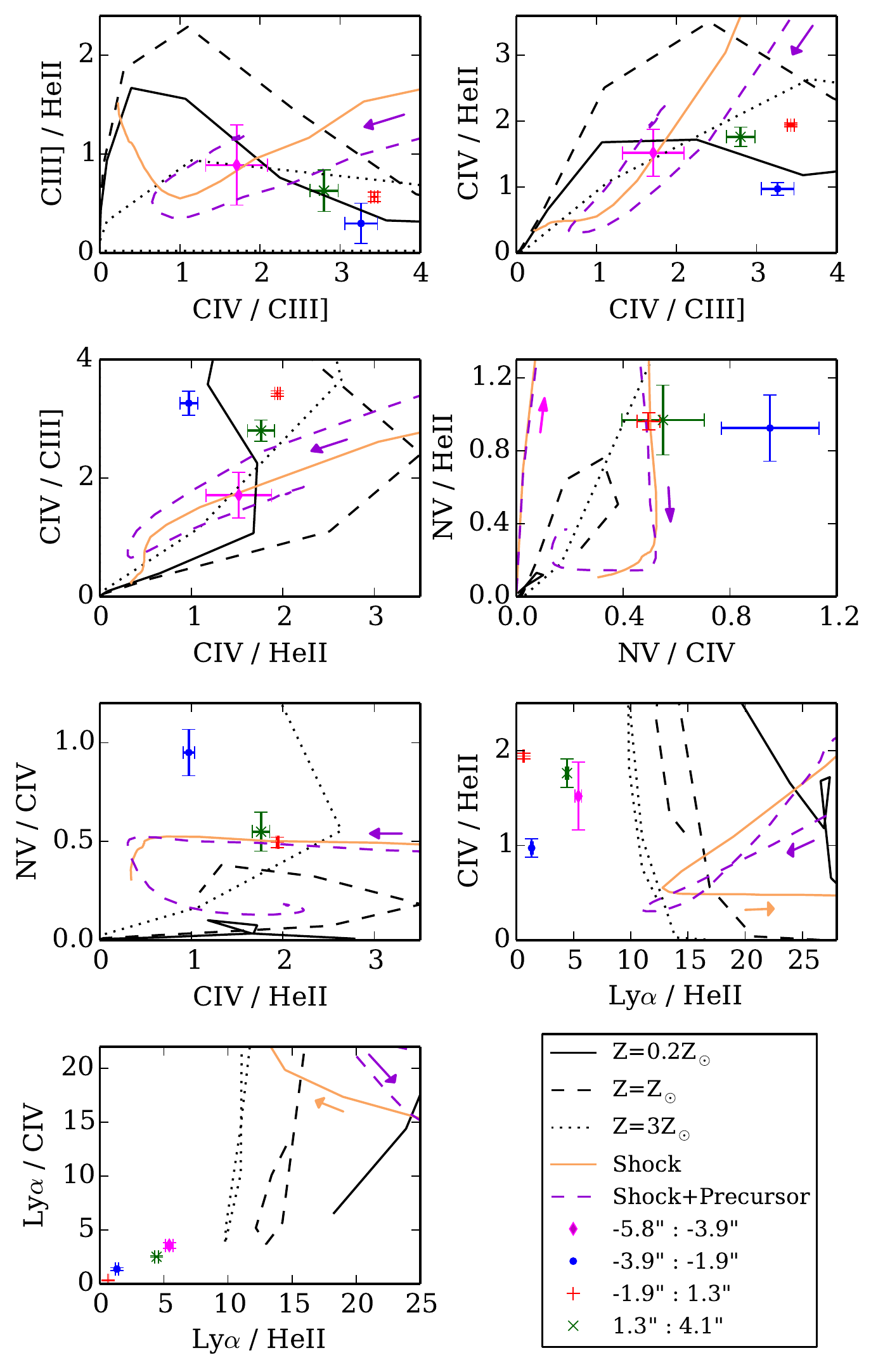}
\vspace*{-5mm}
\caption{Flux ratios of TXS 0211-122 observed with the slit placed along the radio axis plotted on diagnostic diagrams involving Ly$\alpha$, NV, CIV, HeII and CIII]. The diagrams are the same as in Fig. \ref{fig:GTC_TXS0211_grids}.}
\label{fig:grid_Keck_TXS0211}
\end{figure}

Next, we compare the observed and model line ratios for PA=104$^{\circ}$ (along the radio axis). The diagnostic diagrams are shown in Fig. \ref{fig:grid_Keck_TXS0211}.  
In aperture -5.8\arcsec\,to -3.9\arcsec\,the degeneracy due to the use of CIV, HeII and CIII] line ratios is clearly apparent. 
The line ratios observed in aperture -3.9\arcsec\ to -1.9\arcsec\,lie far from the models in most of the diagrams.
Both a photoionization model with Z=0.2Z$_{\sun}$ and Z=3Z$_{\sun}$ and a variation in the ionization parameter are able to reproduce our results. 

The models are not able to reproduce the Ly$\alpha$/HeII and Ly$\alpha$/CIV line ratios. The photoionization models, shock and shock plus precursor models all predict higher Ly$\alpha$/HeII ratios than those measured. The cause of this discrepancy is likely to be strong absorption of Ly$\alpha$ by HI \citep[and possibly dust,][]{Ojik1994}. 
The diagnostic diagrams suggest supersolar metallicities, even in regions far from the nucleus. These results are in agreement with the results from \citet{VM1999c, Vernet2001, VM2001} who found that the metallicities of the extended emission line regions of powerful HzRGs are close to solar or supersolar. However, some HzRGs show evidence for lower metallicities \citep[e.g.][]{Overzier2001, Iwamuro2003}.

\subsubsection{Kinematics}\label{sec:TXS 0211-122 Kinematics}

\begin{figure*}\centering
\textbf{TXS 0211-122}\par\medskip                                                             
\includegraphics[width=\columnwidth]{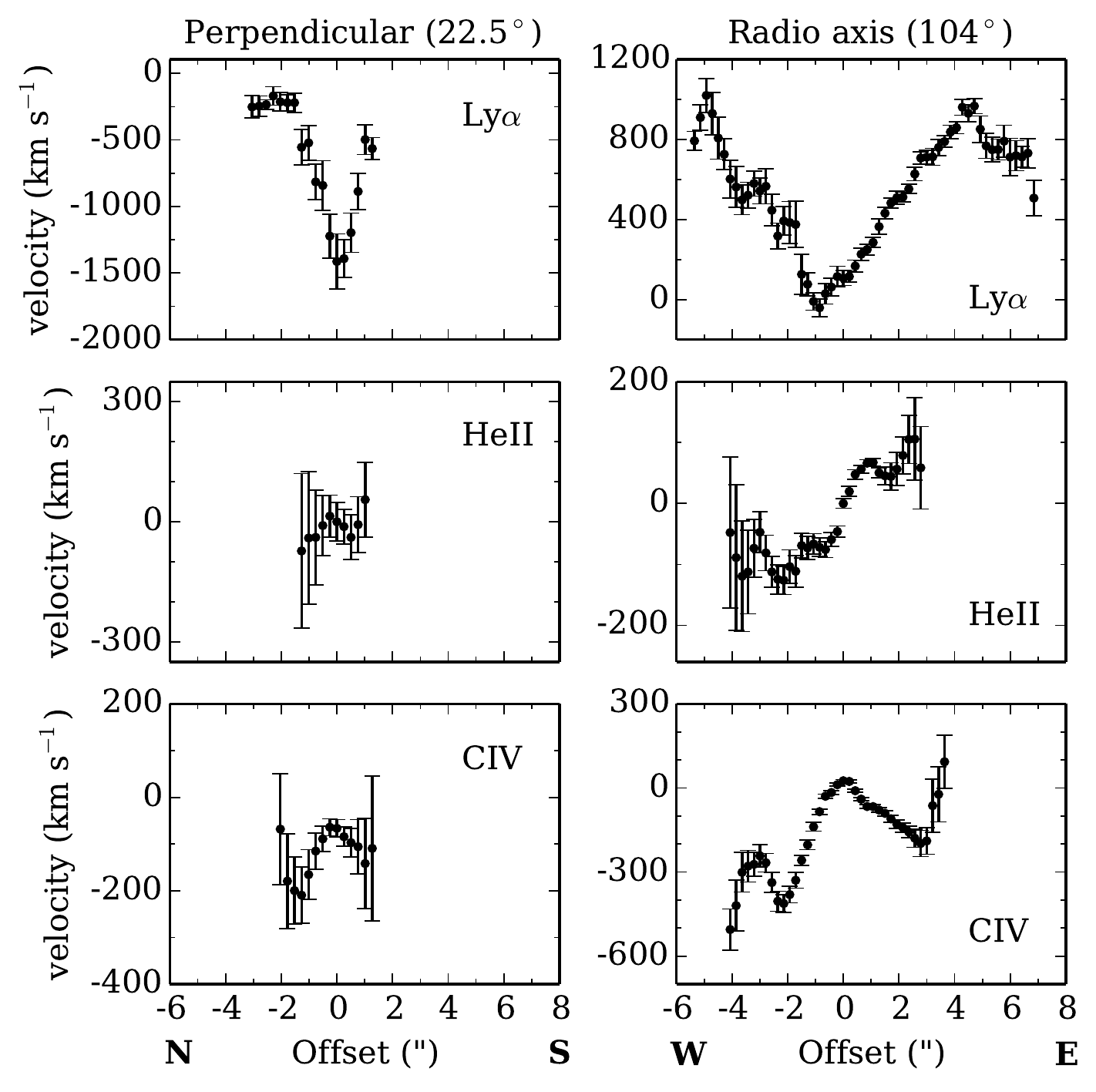}
\hfill
\quad
\includegraphics[width=\columnwidth]{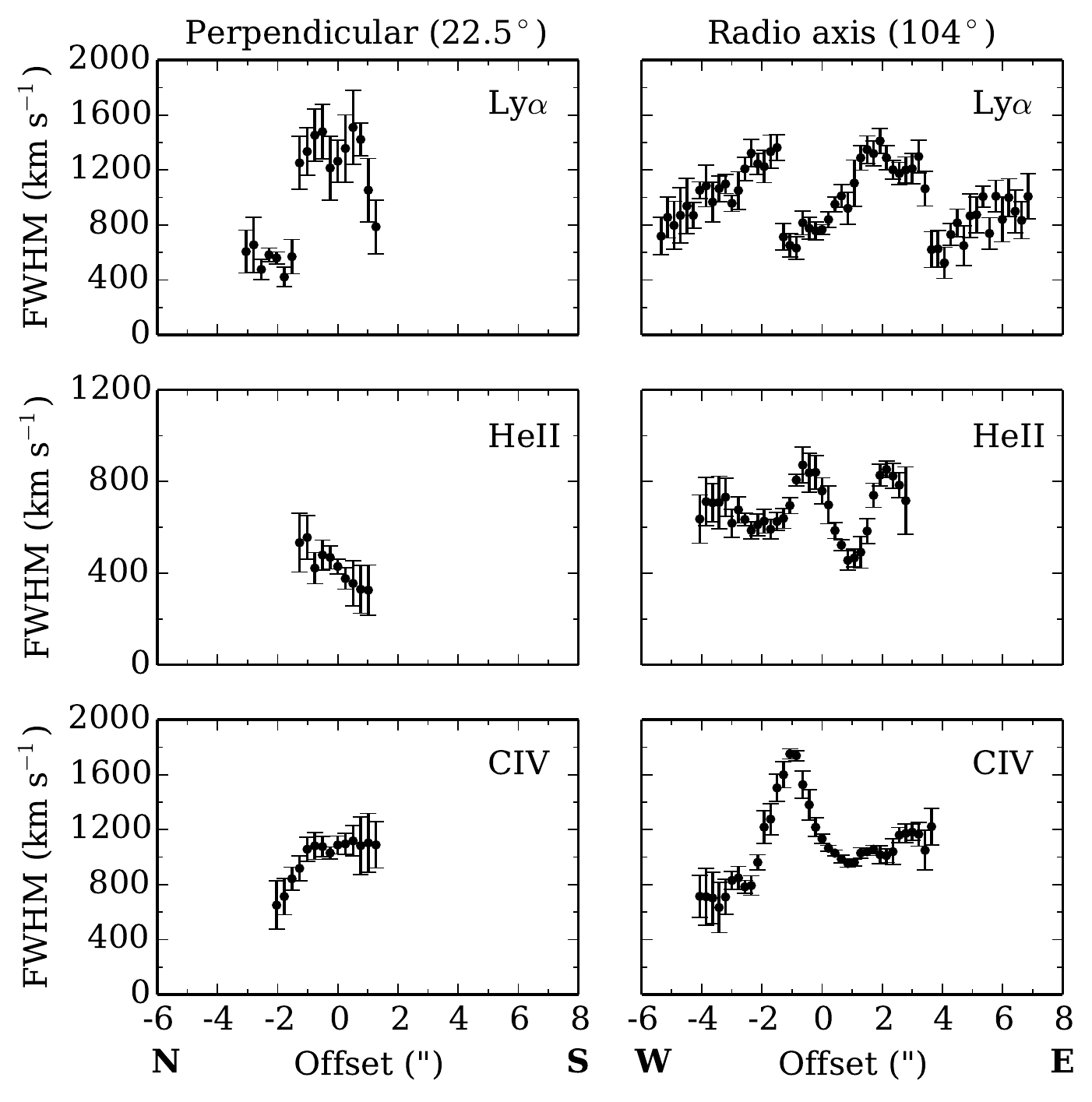}
\caption{Left: Velocity profiles of Ly$\alpha$ (top), HeII (middle) and CIV (bottom). Right: FWHM of Ly$\alpha$, HeII and CIV. The figures in the left represent the variation in the direction perpendicular to the radio axis, while the figures in the right represent the variation along the radio axis.}
\label{fig:velocity_FWHM_TXS0211}
\end{figure*}

In the left panel of Fig. \ref{fig:velocity_FWHM_TXS0211} we show the overall velocity centroids of Ly$\alpha$, HeII and CIV emission along the two long slits for TXS 0211-122. The emission lines show strong spatial variation in their velocity centroids.

The maximum difference of the Ly$\alpha$ emission line velocity and the systemic velocity of the galaxy observed perpendicularly to the radio axis (PA=22.5$^{\circ}$) is $\sim$1400 km s$^{-1}$. The maximum velocity shift of HeII and CIV across the nebula are $\sim$130 km s$^{-1}$ and $\sim$150 km s$^{-1}$, respectively.
In the direction perpendicular to the radio axis HeII and CIV show similar velocity curves to each other, but both differ greatly from the velocity curve of Ly$\alpha$, suggesting that the measured kinematics of Ly$\alpha$ emission are strongly affected by HI absorption and/or scattering.

In the direction of the radio axis (PA=104$^{\circ}$) the maximum difference between Ly$\alpha$ velocity and the systemic velocity is $\sim$1000 km s$^{-1}$. HeII shows a maximum velocity shift of $\sim$230 km s$^{-1}$ and CIV has a maximum velocity shift of $\sim$600 km s$^{-1}$. In this direction HeII shows a rather constant velocity curve. The velocity curve of Ly$\alpha$ shows a large range in velocity. Once again this could be due to HI absorption or scattering. The velocity curve of CIV also differs from the velocity curve of HeII, and CIV appears to be blueshifted relative to HeII. This result is further explored later in this section. 

The right panel of Fig. \ref{fig:velocity_FWHM_TXS0211} shows how the FWHM of Ly$\alpha$, HeII and CIV vary along the two slits. In the direction perpendicular to the radio axis there is a sudden jump in the FWHM and velocity curve of Ly$\alpha$ at -1.5\arcsec. This jump is not seen in the other UV lines, suggesting a change in the impact of absorption or resonant scattering on Ly$\alpha$.   

In the direction of the radio axis (PA=104$^{\circ}$) Ly$\alpha$ shows several changes in the FWHM. \citet{VM2003} found narrow and broad components in the Ly$\alpha$ emission of this radio source. The differences in the FHWM(Ly$\alpha$) observed in Fig. \ref{fig:velocity_FWHM_TXS0211} may be due to the different components. 
In regions where the broad component is affecting more the width of the line broad lines are produced, and in regions where the narrow component dominates the FWHMs are lower. 

The velocity curve of HeII displays perturbed kinematics\footnote{In this work highly perturbed emission refers to emission lines with FWHM$\geq$1000 km s$^{-1}$ and perturbed emission refers to gas with FWHM$\leq$1000 km s$^{-1}$.} (FWHM\textless 1000 km s$^{-1}$) along both PAs. The observed FWHM are higher than what we would expect from gravitational motions \citep[e.g.][]{VM2002}. In addition, Ly$\alpha$ and CIV show high FWHM in both PAs, suggesting that there are highly perturbed motions along both axes.

\subsubsection{Correlations between kinematics and line ratios}

Looking at the results obtained for the Keck II data (PA=104$^{\circ}$) and comparing the figure of CIV/HeII (Fig. \ref{fig:TXS0211_ratios}) and the velocity of CIV (Fig. \ref{fig:velocity_FWHM_TXS0211}) it can be seen that the lowest values of CIV/HeII have the highest blueshift. 
This result is consistent with the presence of ionized gas outflows along the radio axis, with a relatively lower ionization state, as has been found in a number of powerful radio galaxies at low to high redshifts \citep[e.g.][]{VM1999b, Humphrey2006}. 

In order to investigate possible correlations between the CIV/HeII ratio and the velocities of the different emission lines the \citet{Spearman1904} rank correlation coefficient, $\rho$, was calculated.

\begin{figure}\centering                                                       
\textbf{TXS 0211-122}\par\medskip                                                             
\includegraphics[width=\columnwidth]{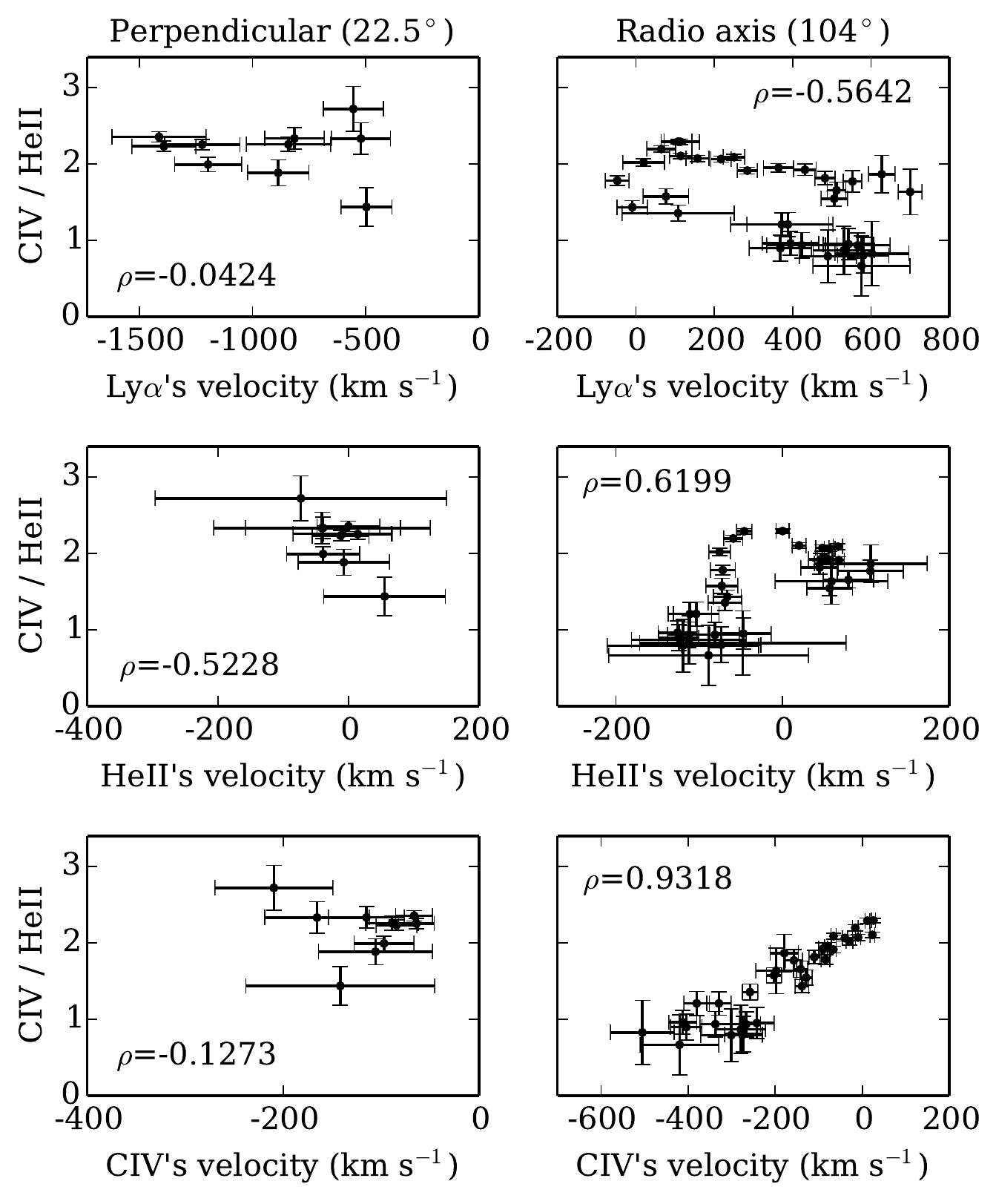}
\vspace*{-5mm}
\caption{Variation of CIV/HeII ratio with the velocity of different emission lines. The top left panel shows the ratio CIV/HeII as a function of the velocity of Ly$\alpha$ in the perpendicular slit, the top right panel shows the variation of CIV/HeII as a function of the velocity of Ly$\alpha$ along the radio axis, the middle left panel shows the variation of CIV/HeII as a function of the velocity of HeII in the perpendicular slit, the middle right panel shows the variation of CIV/HeII as a function of the velocity of HeII in the slit placed along the radio axis, the bottom left figure shows how the CIV/HeII varies with the velocity of CIV in the perpendicular slit and the bottom right figure shows how the CIV/HeII varies with the velocity of CIV along the radio axis.
$\rho$ represents the Spearman's rank correlation coefficient.}\label{fig:TXS0211_CIVHeII_velocity}
\end{figure} 

The results are presented in Fig. \ref{fig:TXS0211_CIVHeII_velocity}. There is a strong correlation ($\rho$=0.932) between CIV's velocity and the ratio CIV/HeII in the direction of the radio axis. The figure also shows a correlation ($\rho$=0.620) between HeII's velocity and CIV/HeII. 
The significance of the values 0.932 and 0.620 was tested by comparing them to critical values for Spearman's rank correlation coefficient. Both correlations are significant at a confidence level of more than 99\%. \\
In the direction perpendicular to the radio axis no correlation is apparent, suggesting that the ionized outflowing gas is preferentially situated along the radio axis.

\subsection{TXS 0828+193}

\begin{figure}\centering
\textbf{TXS 0828+193}\par\medskip                                                                                                                   
\includegraphics[trim=60 180 60 45mm, width=\columnwidth]{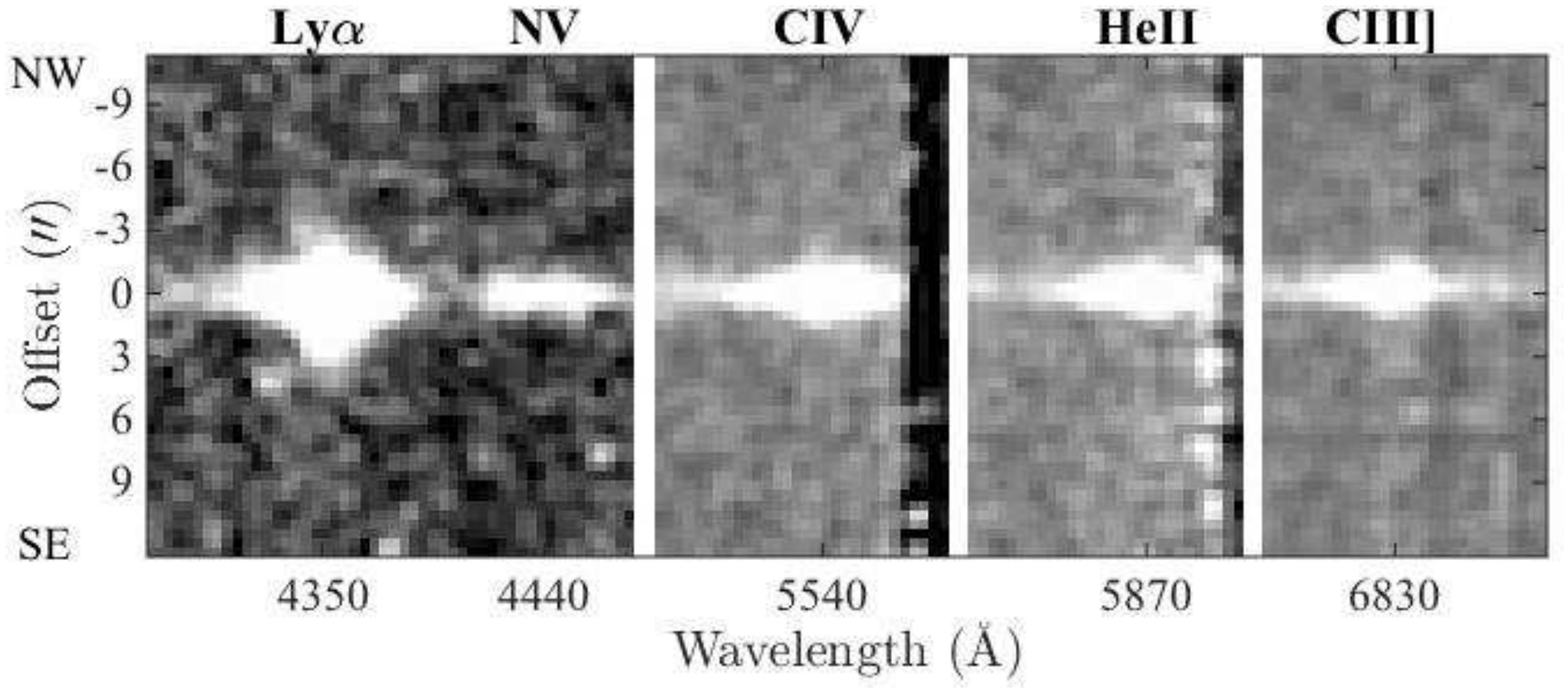}
\hfill                                                      
\includegraphics[trim=60 200 60 40mm, width=\columnwidth]{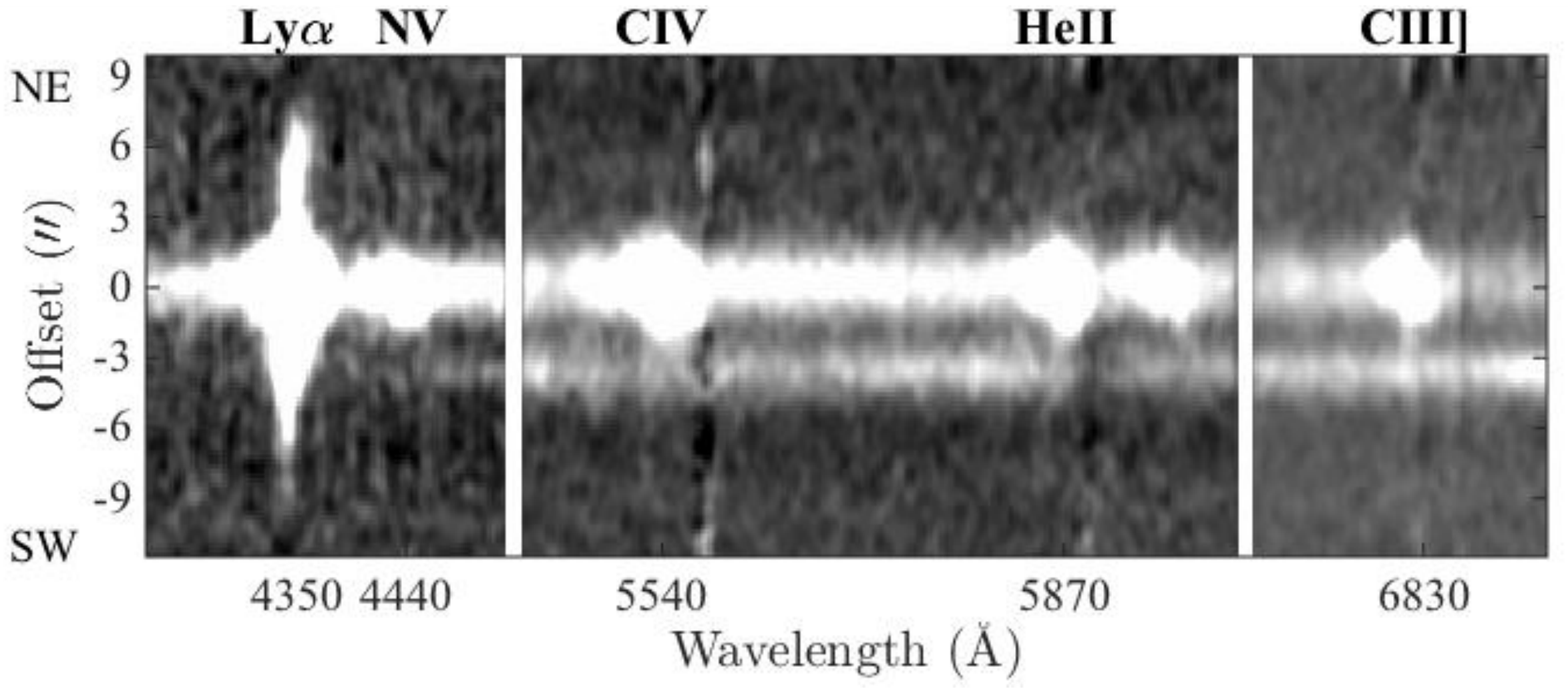}
\quad
\caption{Top: 2D spectrum of the main emission lines for TXS 0828+193, with the slit oriented perpendicularly to the radio axis.  Bottom: 2D spectrum of the main emission lines for TXS 0828+193, with the slit oriented in the direction of the radio axis. The zero in the spatial direction corresponds to the peak of the continuum emission.}
\label{fig:figure_2D_TXS0828}
\end{figure}

The two dimensional spectra of TXS 0828+193 are shown in Fig. \ref{fig:figure_2D_TXS0828}. Ly$\alpha$ is the brightest and most extended emission line along both PAs.  

\begin{figure}\centering      
\textbf{TXS 0828+193}\par\medskip                                                                                                               
\includegraphics[width=\columnwidth]{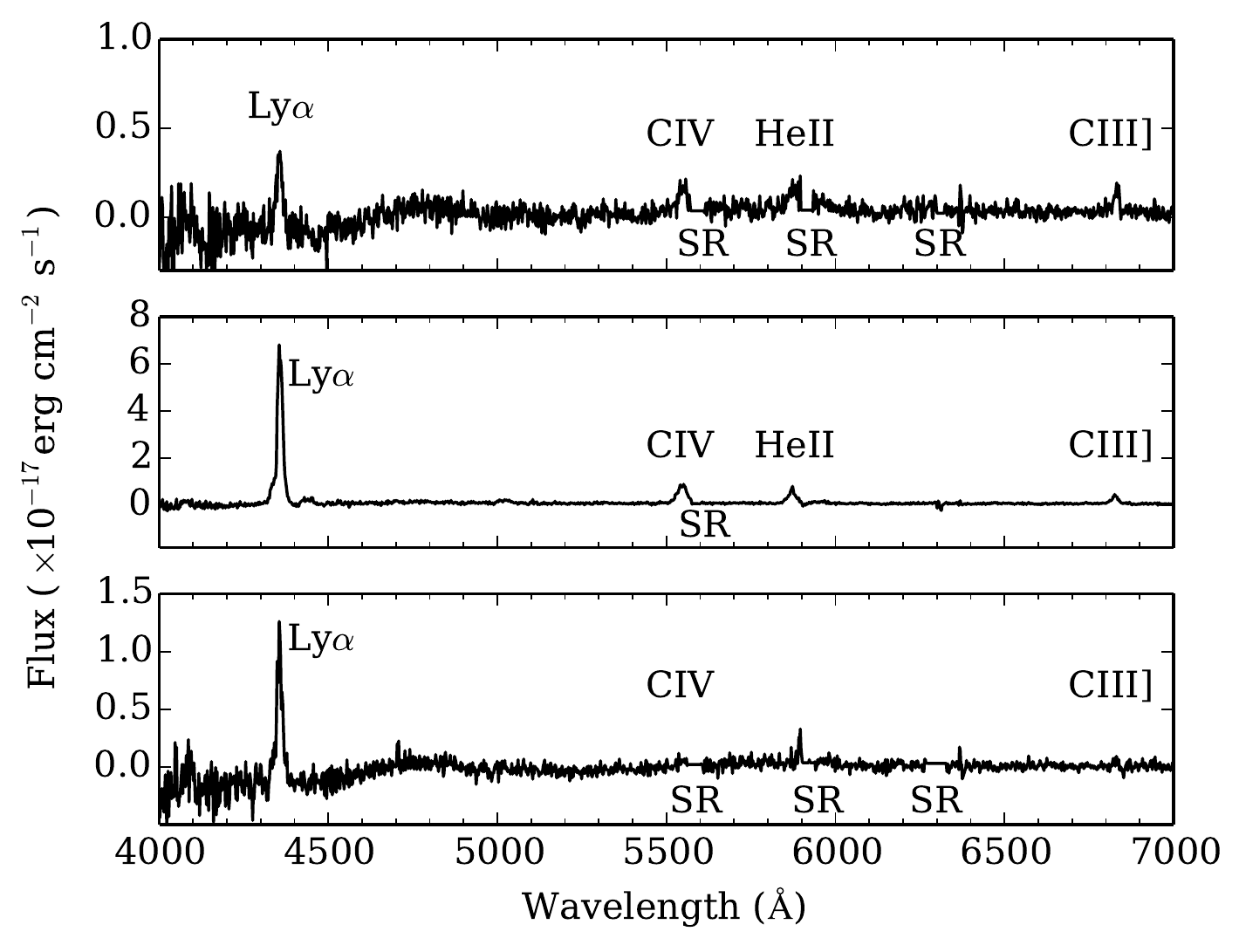}
\vspace*{-5mm}
\caption{One dimensional spectra of different regions of TXS 0828+193 observed with the slit oriented perpendicularly to the radio axis. The top spectrum shows the region between -4.3\arcsec \, and -1.3\arcsec\,, the spectrum in the middle shows the nuclear region (central 2\arcsec) and the bottom spectrum shows the region between 1.3\arcsec \, and 4.6\arcsec. Ly$\alpha$, NV, CIV, HeII and CIII] positions are indicated in the spectra. NV is only detected in the central region. There are some skyline residuals (SR) affecting the emission lines. CIV emission line is affected by the 5577 \AA\, (OI) skyline, HeII is affected by a skyline at 5889 \AA\, (Na D) and CIII] is affected by a skyline at $\sim$6828 \AA\,(OH).}
\label{fig:GTC_TXS0828_regions}
\end{figure}

One dimensional spectra showing different regions of TXS 0828+193 observed with the slit positioned perpendicularly to the radio axis are shown in Fig. \ref{fig:GTC_TXS0828_regions}. An excess of emission in the blue wing of Ly$\alpha$ is seen in the middle and bottom panels of Fig. \ref{fig:GTC_TXS0828_regions}.

\begin{figure}\centering   
\textbf{TXS 0828+193}\par\medskip                                                                                                                   
\includegraphics[width=\columnwidth]{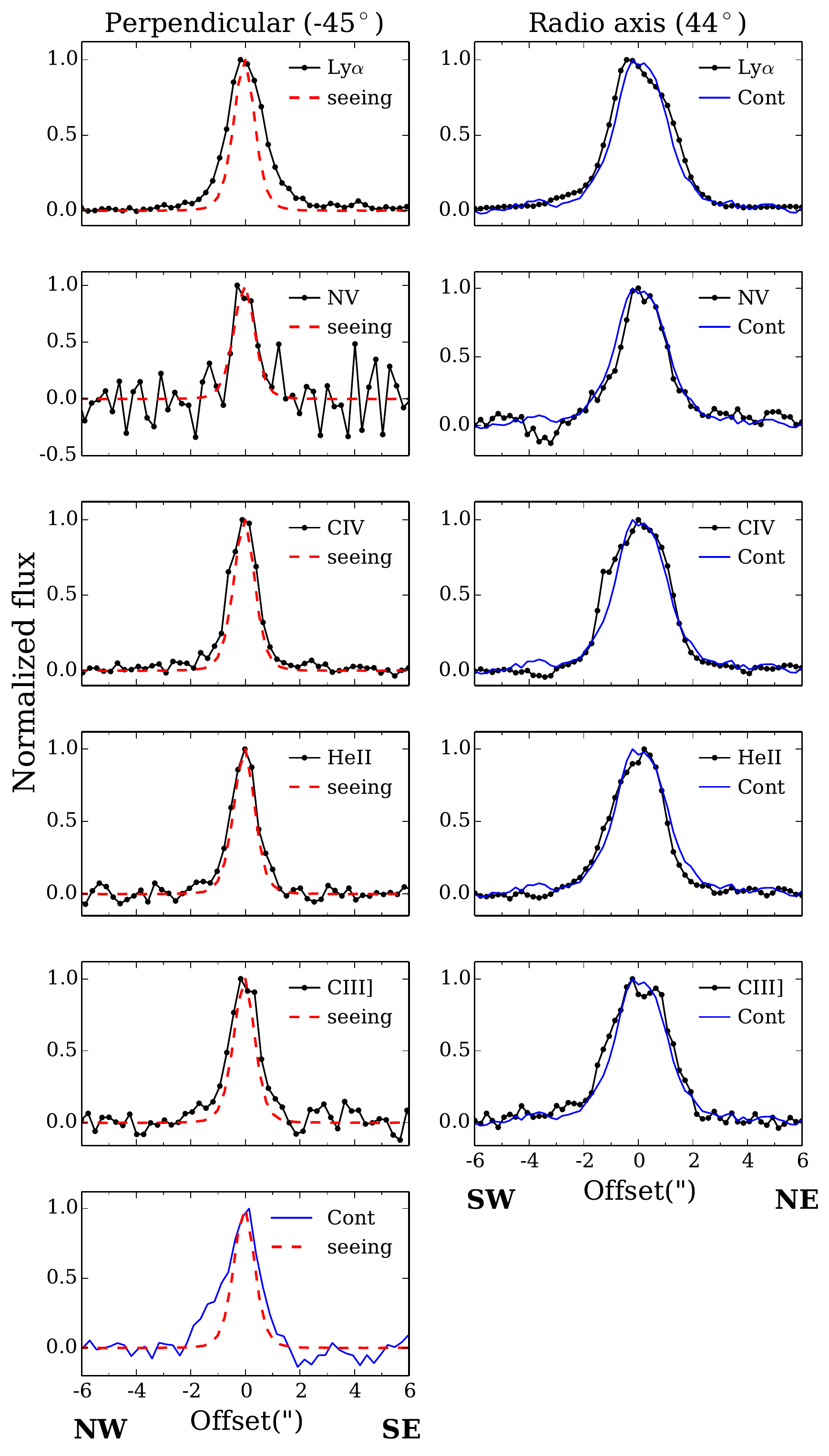}
\vspace*{-5mm}
\caption{Normalized flux of the different emission lines (black) compared with the seeing (red) and the continuum profile (blue). In the left we have the values observed perpendicularly to the radio axis and in the right the normalized flux along the radio axis. Zero in the spatial direction represents the peak of the continuum emission.}\label{fig:TXS0828_SB}
\end{figure}

In Fig. \ref{fig:TXS0828_SB} we show the spatial profiles of the different emission lines and the spatial profile of the continuum for both PAs. For the perpendicular slit we also show the seeing (FWHM=0.93$\pm$0.04\arcsec). 
In the GTC spectrum (PA=$-45^{\circ}$) Ly$\alpha$ has an extent of 6.9\arcsec\,($\sim$56 kpc), much less than the $\sim$130 kpc extent \textit{along} the radio axis measured by \citet{VM2002}. \\
The spatial distribution of the Ly$\alpha$, CIV, HeII and CIII] emission lines is consistent with a Gaussian profile of FWHM(obs)=1.62$\pm$0.02\arcsec\,for Ly$\alpha$, FWHM(obs)=1.21$\pm$0.01\arcsec\,for CIV, FWHM(obs)=1.16$\pm$0.03\arcsec\,for HeII and FWHM(obs)=1.33$\pm$0.01\arcsec\,for CIII]. Correcting for seeing broadening in quadrature, the intrinsic FWHM values are 1.33$\pm$0.04\arcsec\,(10.8$\pm$0.3 kpc) for Ly$\alpha$, 0.77$\pm$0.05\arcsec\,(6.3$\pm$0.4 kpc) for CIV, 0.69$\pm$0.07\arcsec\, (5.6$\pm$0.6 kpc) for HeII, and 0.95$\pm$0.04\arcsec\,(7.7$\pm$0.3 kpc) for CIII]. 
NV is consistent with the seeing, we determine an upper limit of 0.97\arcsec for the FWHM. After correcting for seeing broadening the intrinsic FWHM is $\sim$0.27\arcsec\, (2.2 kpc). The continuum is resolved, with an intrinsic FWHM of 1.24$\pm$0.08\arcsec\,(10.1$\pm$0.7 kpc). It is asymmetric and shows stronger emission towards the NW.\\
In this source, all the emission lines (except NV) are spatially extended in the direction perpendicular to the radio axis. This is further supported by the fact that the kinematic properties of Ly$\alpha$, CIV and HeII vary along the slit (Fig. \ref{fig:TXS0828_velocity_FWHM}). \\ 
The emission lines are more extended in the direction of the radio axis, as is also the case in TXS 0211-122.

\subsubsection{Line ratios and comparison with models}
	
\begin{figure}\centering                                                        
\textbf{TXS 0828+193}\par\medskip                                                             
\includegraphics[width=\columnwidth]{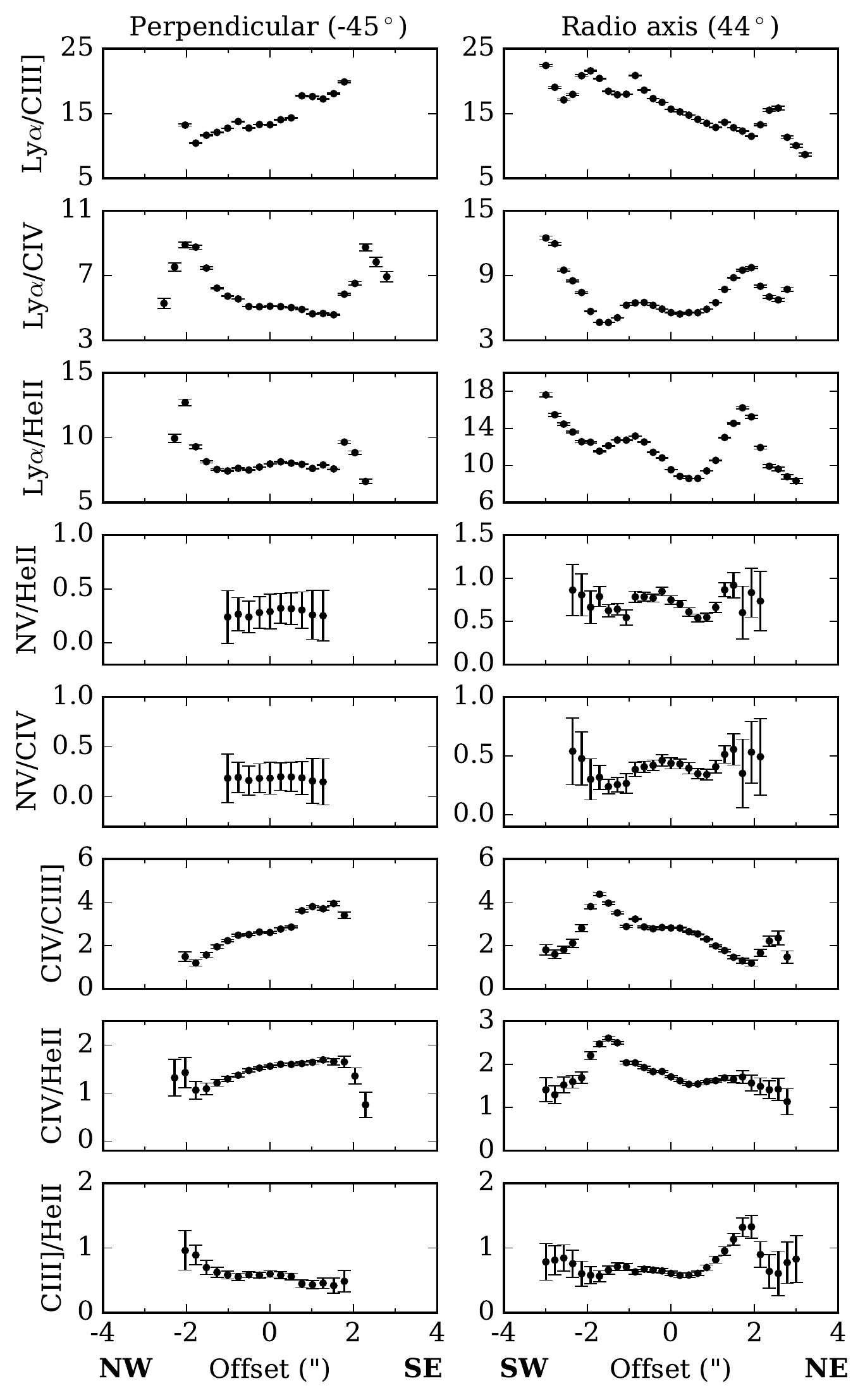}
\vspace*{-5mm}
\caption{Ratios between the emission lines in TXS 0828+193 as a function of distance in the galaxy. In the left are shown the flux ratios in the direction perpendicular to the radio axis (PA=-45$^{\circ}$) and in the right in the direction of the radio axis (PA=44$^{\circ}$).}
\label{fig:TXS0828_ratios}
\end{figure}

Fig. \ref{fig:TXS0828_ratios} shows how the different emission line ratios vary along the slit. 
The emission line ratios in TXS 0828+193 are typical of HzRGs \citep[e.g.][]{McCarthy1993, Humphrey2008a}. In the direction perpendicular to the radio axis, (Fig. \ref{fig:TXS0828_ratios} left panels), the emission line ratios Ly$\alpha$/HeII and Ly$\alpha$/CIV show a peak around $\sim$-2\arcsec. Another local maximum is seen for the same line ratios at $\sim$2\arcsec.
This may suggest that Ly$\alpha$ emission is enhanced in these regions. However, it is not immediately clear whether this is due to enhanced Ly$\alpha$ or lower fluxes for the other lines. 
In the direction of the radio axis (Fig. \ref{fig:TXS0828_ratios} right) Ly$\alpha$/CIV, Ly$\alpha$/HeII and CIII]/HeII show a local maximum at $\sim$2\arcsec, but this time with only two corresponding features at $\sim$-2\arcsec (CIV/CIII] and CIV/HeII). 

\begin{figure}\centering                                                       
\textbf{TXS 0828+193} \\ Perpendicular (-45$^{\circ}$)\par\medskip                                     
\includegraphics[width=\columnwidth]{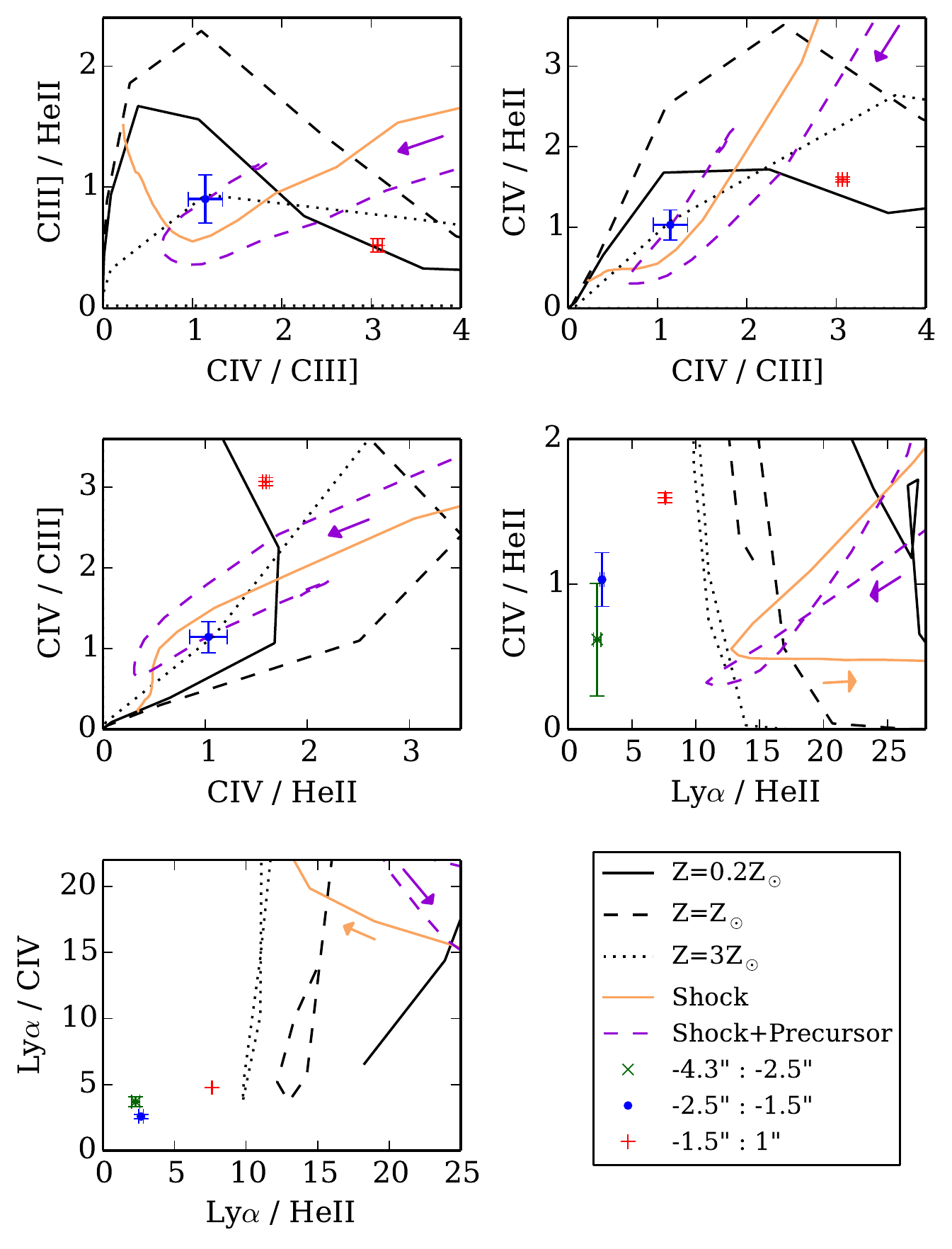}
\vspace*{-5mm}
\caption{Flux ratios of TXS 0828+193 observed with the slit oriented perpendicularly to the radio axis plotted on diagnostic diagrams involving Ly$\alpha$, CIV, HeII and CIII]. The black lines represent sequences of photoionization models for constant metallicities. U varies along each metallicity line.
Shock models by \protect\citet{Allen2008} are also shown. Orange solid lines denote the predictions of pure shock models, and violet-dotted lines denote the prediction of shock plus precursor models. The arrows indicate the direction of increasing velocity.}
\label{fig:GTC_TXS0828_grid}
\end{figure}

In order to understand the physical conditions and properties of the extended emission line regions, a comparison is made between the results of our observations and photoionization and shock models. For this purpose, several apertures were extracted along the slits. 
Table \ref{table:RatiosTXS0828GTC} shows the different apertures and the flux ratios for the different emission lines in the direction perpendicular to the radio axis (PA=-45$^{\circ}$). \\
To help distinguish between ionization mechanisms in the gas line-ratio diagrams were used. The models are the same as used for TXS 0211-122. In Fig. \ref{fig:GTC_TXS0828_grid} diagnostic diagrams showing photoionization and shock models for the slit perpendicular to the radio axis are shown. NV is consistent with the seeing thus we did not include it in the diagrams.
\begin{table*}\centering 
\vspace*{5mm}
\begin{tabular}{c c c c c c c c c c} 
\hline\hline 
Pos.  &  Ly$\alpha$/HeII & Ly$\alpha$/CIV  & Ly$\alpha$/CIII]     &      CIV/HeII    &        CIV/CIII]      \\   
(1)     &   (2)                &          (3)            &             (4)          &          (5)                &    (6)    \\    
\hline
-4.3\arcsec\, : -2.5\arcsec   &  2.3$\pm$0.3    &      3.7$\pm$0.4  &    -  &  0.6$\pm$0.4    \\  
\hline
-2.5\arcsec\, : -1.5\arcsec   &  2.7$\pm$0.3    &      2.6$\pm$0.2  & 3.0$\pm$0.2   &  1.0$\pm$0.2      &    1.14$\pm$0.2  \\  
\hline
-1.5\arcsec\, : 1\arcsec      &    7.61$\pm$0.03 &   4.77$\pm$0.02 &  14.7$\pm$0.05 & 1.60$\pm$0.03 &    3.07$\pm$0.05 \\
\hline
1\arcsec\, : 2\arcsec   &                -                   &  24.6$\pm$0.3   &  36.7$\pm$0.3  &  -   & 1.5$\pm$0.4 \\ 
\hline 
\end{tabular}
\caption[Emission line ratios from the TXS 0828+193 spectrum observed in the direction perpendicular to the radio axis]{Emission line ratio measurements from the TXS 0828+193 spectrum, observed perpendicularly to the radio axis (PA=-45$^{\circ}$). Columns are as follows: (1) position of the aperture along the slit in arcseconds; (2) the  Ly$\alpha$/HeII ratio; (3) the Ly$\alpha$/CIV ratio; (4) the Ly$\alpha$/CIII] ratio; (5) the CIV/HeII ratio; (6) the  CIV/CIII] ratio.} \label{table:RatiosTXS0828GTC} 
\end{table*}
\begin{table*}\centering 
\vspace*{5mm}
\begin{tabular}{c c c c c c c c c c} 
\hline\hline 
Pos.  &  Ly$\alpha$/HeII & Ly$\alpha$/CIV  & Ly$\alpha$/CIII]     &      CIV/HeII    &        CIV/CIII]         & NV/HeII & NV /CIV\\
(1)     &   (2)                &          (3)            &             (4)          &          (5)                &    (6)        &         (7)    &   (8)    \\
\hline
-3.6\arcsec\, : -1.3\arcsec &   19.39$\pm$0.08 &   8.94$\pm$0.04 & 16.33$\pm$0.06 & 2.2$\pm$0.1   & 1.83$\pm$0.07   &  0.6$\pm$0.2  & 0.3$\pm$0.2 & \\
\hline
-1.3\arcsec\, : 0.9\arcsec   & 10.85$\pm$0.01  & 5.90$\pm$0.01  & 19.91$\pm$0.02 &  1.84$\pm$0.01   & 3.37$\pm$0.02 &  0.56$\pm$0.04 & 0.30$\pm$0.04 \\
\hline
0.9\arcsec\, : 2.8\arcsec &    8.3$\pm$0.03   &  5.22$\pm$0.02 &   20.92$\pm$0.06  &  1.60$\pm$0.04 &  4.01$\pm$0.06   &  0.3$\pm$0.2   & 0.3$\pm$0.2 \\
\hline
2.8\arcsec\, : 3.6\arcsec   &  6.5$\pm$0.2      & 5.0$\pm$0.2   &     -     & 1.3$\pm$0.2       &   - & -&-    \\
\hline 
\end{tabular}
\caption[Emission line ratios from the TXS 0828+193 spectrum observed along the radio axis]{Emission line ratio measurements from the TXS 0828+193 spectrum, observed along the radio axis (PA=44$^{\circ}$). Columns are as in Table \ref{table:RatiosTXS0828GTC}.}\label{table:KeckTXS0828_ratios} 
\end{table*}

In Table \ref{table:KeckTXS0828_ratios} it can be seen that in the outer regions of the nebula Ly$\alpha$ emission is very bright in comparison to HeII, CIV and CIII]. The variation in the ratio CIV/CIII] suggests that the state of ionization of the gas is not constant across the nebula.
The ratio CIV/HeII suggests an increase in the ionization state of the gas, from NE ($\sim$3.6\arcsec) to SW ($\sim$-3.6\arcsec). This result is in agreement with results from \citet{Humphrey2007b}. 

It is apparent from Fig. \ref{fig:GTC_TXS0828_grid} that a photoionization model with Z=3Z$_{\sun}$ is able to reproduce the ratios for the region -2.5\arcsec\, to -1.5\arcsec\, in some diagrams. However the ratios are also reproduced by shock and shock plus precursor models.
Due to the low Ly$\alpha$/HeII ratio we observe the models are not able to reproduce our results. 
The detection of strong HeII emission strongly disfavours photoionization by young stars and resonant scattering as main powering mechanisms.
\begin{figure}
\centering                               
\textbf{TXS 0828+193} \\ Radio axis (44$^{\circ}$)\par\medskip                                                             
\includegraphics[width=\columnwidth]{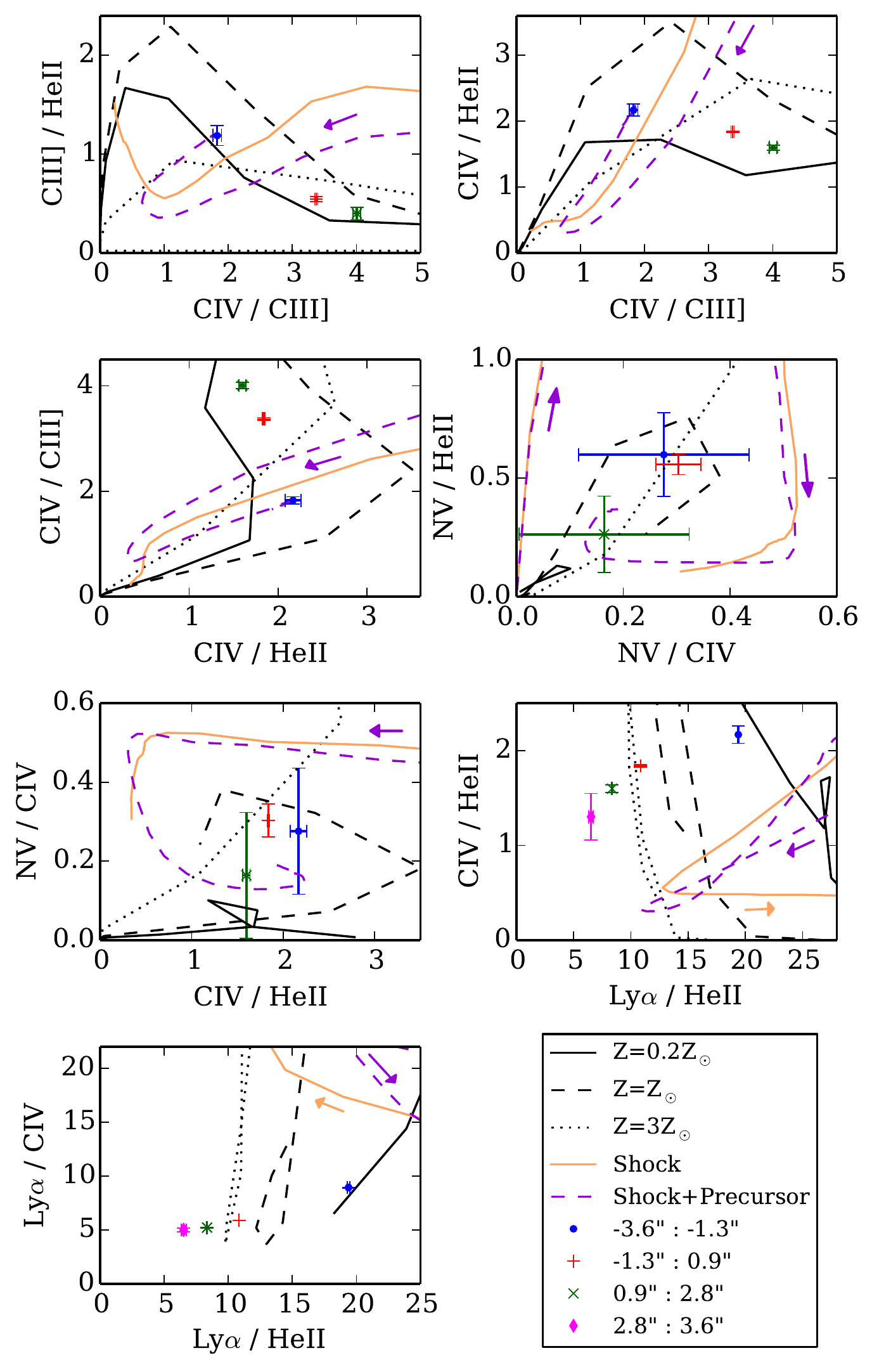}
\vspace*{-5mm}
\caption{Flux ratios observed with the slit placed along the radio axis plotted on diagnostic diagrams involving Ly$\alpha$, NV, CIV, HeII and CIII]. The models are the same as in Figure\ref{fig:GTC_TXS0828_grid}.}
\label{fig:Keck_TXS0828_grids}
\end{figure}

In Fig. \ref{fig:Keck_TXS0828_grids} diagrams showing photoionization and shock models for the slit along the radio axis are presented.
The diagram of NV/HeII versus NV/CIV suggests that the gas metallicity is high. Photoionization models with metallicities between Z$_{\sun}$ and 3Z$_{\sun}$ give the best reproduction of the results, however, due to the overlapping between the photoionization and the shock models we cannot rule out the possibility that both mechanisms are ionizing the gas.
This is in agreement with the results of \citet{Humphrey2008a} who showed that AGN photoionization is the main ionization mechanism in radio galaxies but shocks also make a small contribution to the ionization of the gas.

\subsubsection{Kinematics}

\begin{figure*}\centering
\textbf{TXS 0828+193}\par\medskip                                                             
\includegraphics[width=\columnwidth]{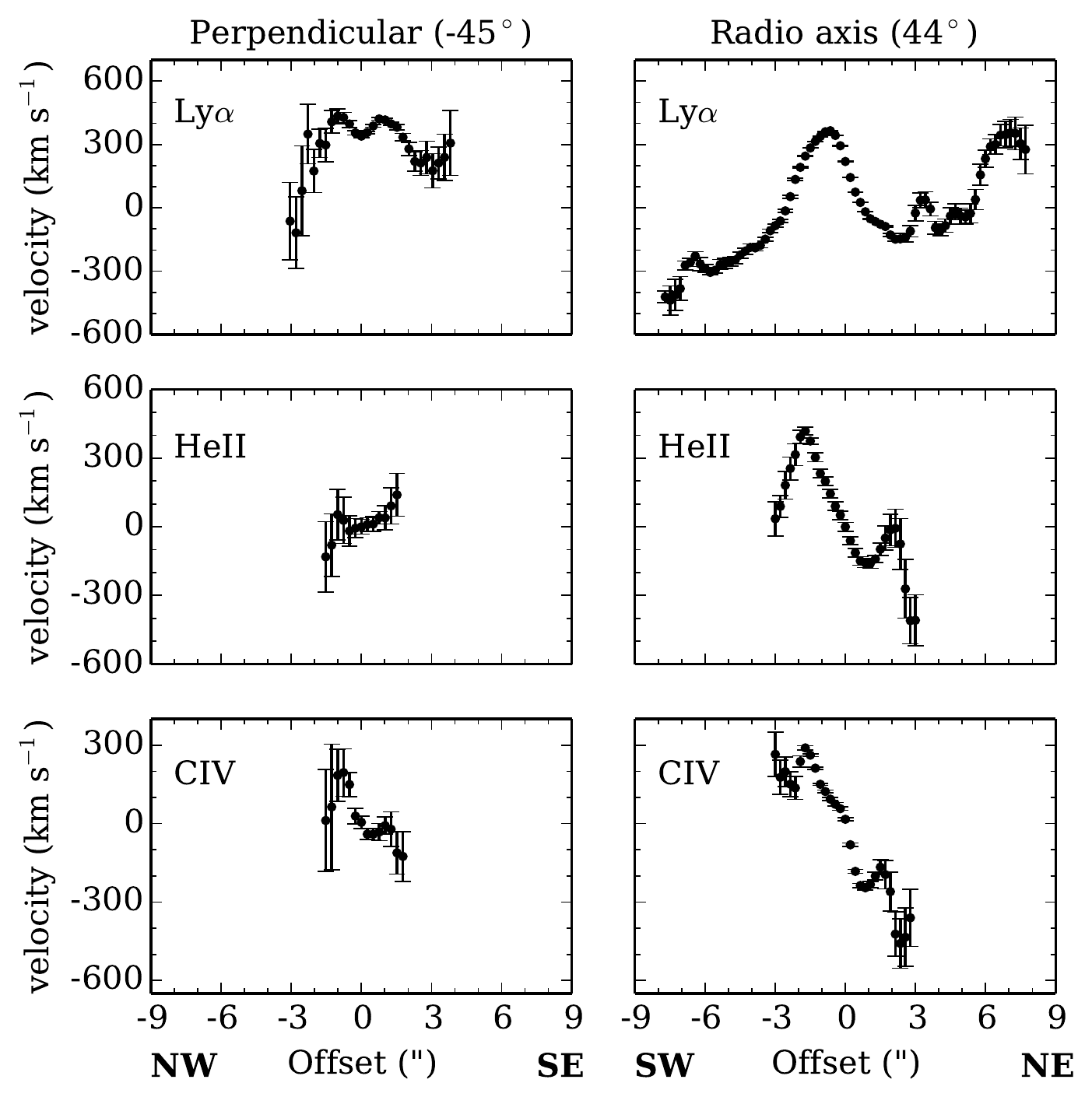}
\hfill
\includegraphics[width=\columnwidth]{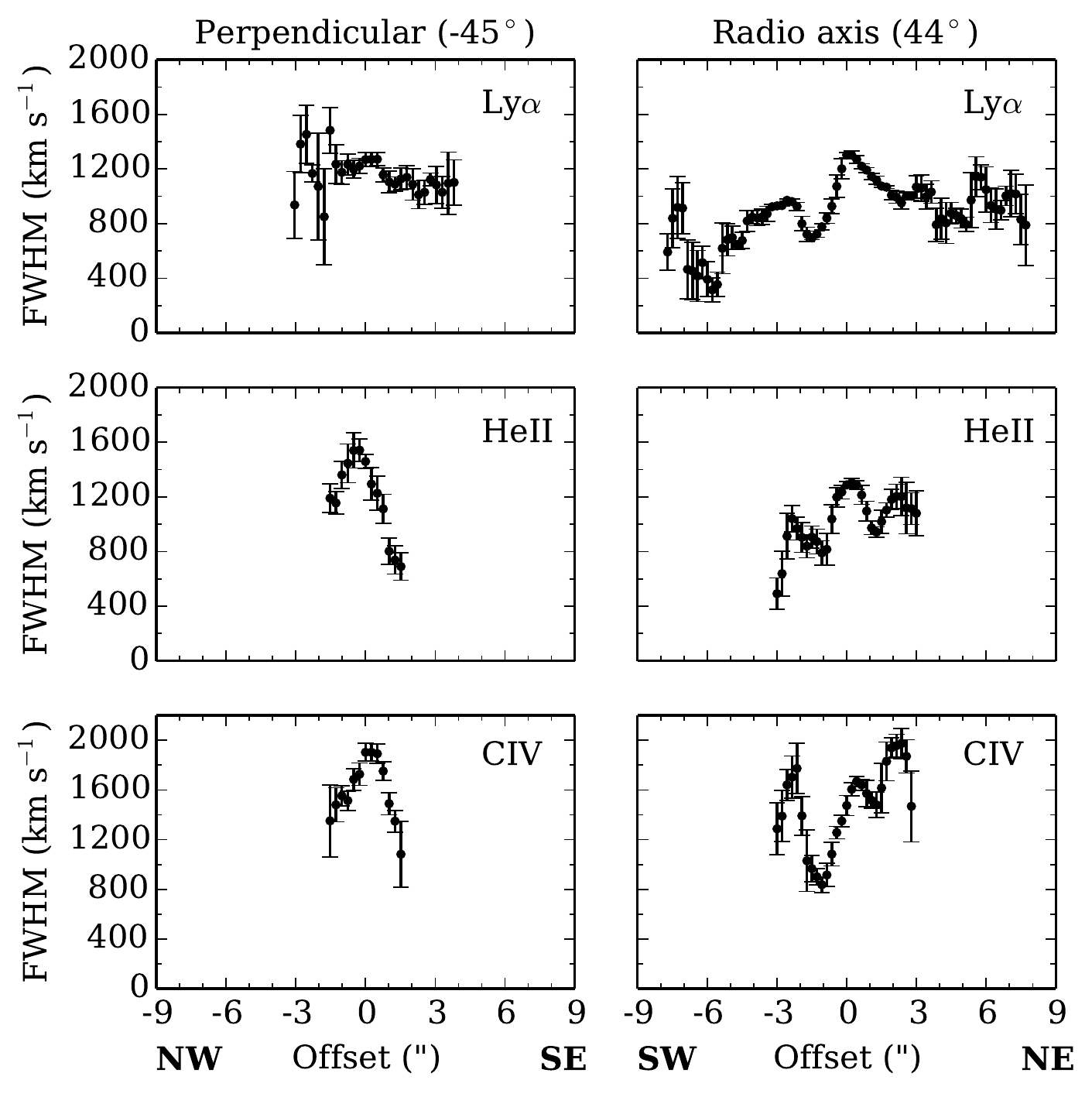}
\quad
\caption{Left: Ly$\alpha$ (top), HeII (middle) and CIV (bottom) velocity profiles for TXS 0828+193. Right: FWHM for Ly$\alpha$, HeII and CIV. The figures in the left represent the variation in the direction perpendicular to the radio axis, while the figures in the right represent the variation along the radio axis.}
\label{fig:TXS0828_velocity_FWHM}
\end{figure*}

The velocity curves of Ly$\alpha$, CIV and HeII are shown in the left panel of Fig. \ref{fig:TXS0828_velocity_FWHM}. 
Perpendicularly to the radio axis HeII shows a near systemic velocity curve (within the error bars). In the same region CIV also presents near systemic velocities. The CIV velocity curve resembles that of Ly$\alpha$. \\
In the centre of the galaxy Ly$\alpha$ emission is redshifted (v$_{Ly\alpha}$$\sim$340 km s$^{-1}$) relatively to HeII.
The modulus of the maximum difference in radial velocity of HeII is $\Delta$v(HeII)$\sim$220 km s$^{-1}$, for CIV the maximum difference in radial velocity $\Delta$v(CIV)$\sim$310 km s$^{-1}$.

There is a velocity shift ($\sim$340 km s$^{-1}$ for PA=-45$^{\circ}$, and $\sim$220 km s$^{-1}$ for PA=44$^{\circ}$) between Ly$\alpha$ and HeII. There are at least two possible reasons for this. First, Ly$\alpha$ and HeII may originate in gas from different locations, and thus different kinematic properties, in the overall nebulosity. Secondly, Ly$\alpha$ can be shifted from other lines as a consequence of partial absorption of the Ly$\alpha$ emission by HI \citep[e.g.][]{Rottgering1995, Ojik1997}. In this case, Ly$\alpha$ must be absorbed in the central regions because Ly$\alpha$/HeII and Ly$\alpha$/CIV are well below the photoionization model predictions.

In the direction perpendicular to the radio axis (PA=-45$^{\circ}$) Ly$\alpha$ emission (Fig. \ref{fig:TXS0828_velocity_FWHM}) from the central regions show an approximately constant redshifted velocity relatively to HeII.
In this direction there are some changes in the velocity pattern of Ly$\alpha$ at $\sim$-2\arcsec\, and $\sim$2\arcsec. \\
In the direction of the radio axis (PA=44$^{\circ}$) we see that HeII and CIV show similar velocity curves. At $\sim$-2\arcsec\, and $\sim$2\arcsec \, CIV and HeII show local maxima in their velocity curves.

The right panel of Fig. \ref{fig:TXS0828_velocity_FWHM} shows the FWHM of the emission lines as a function of distance from the nucleus.
In the direction perpendicular to the radio axis (PA=-45$^{\circ}$) the emission lines show large FWHMs (\textgreater 1000 km s$^{-1}$).

\subsubsection{HST observations}

\begin{figure}\centering
\textbf{TXS 0828+193}\par\medskip       
\includegraphics[width=4cm]{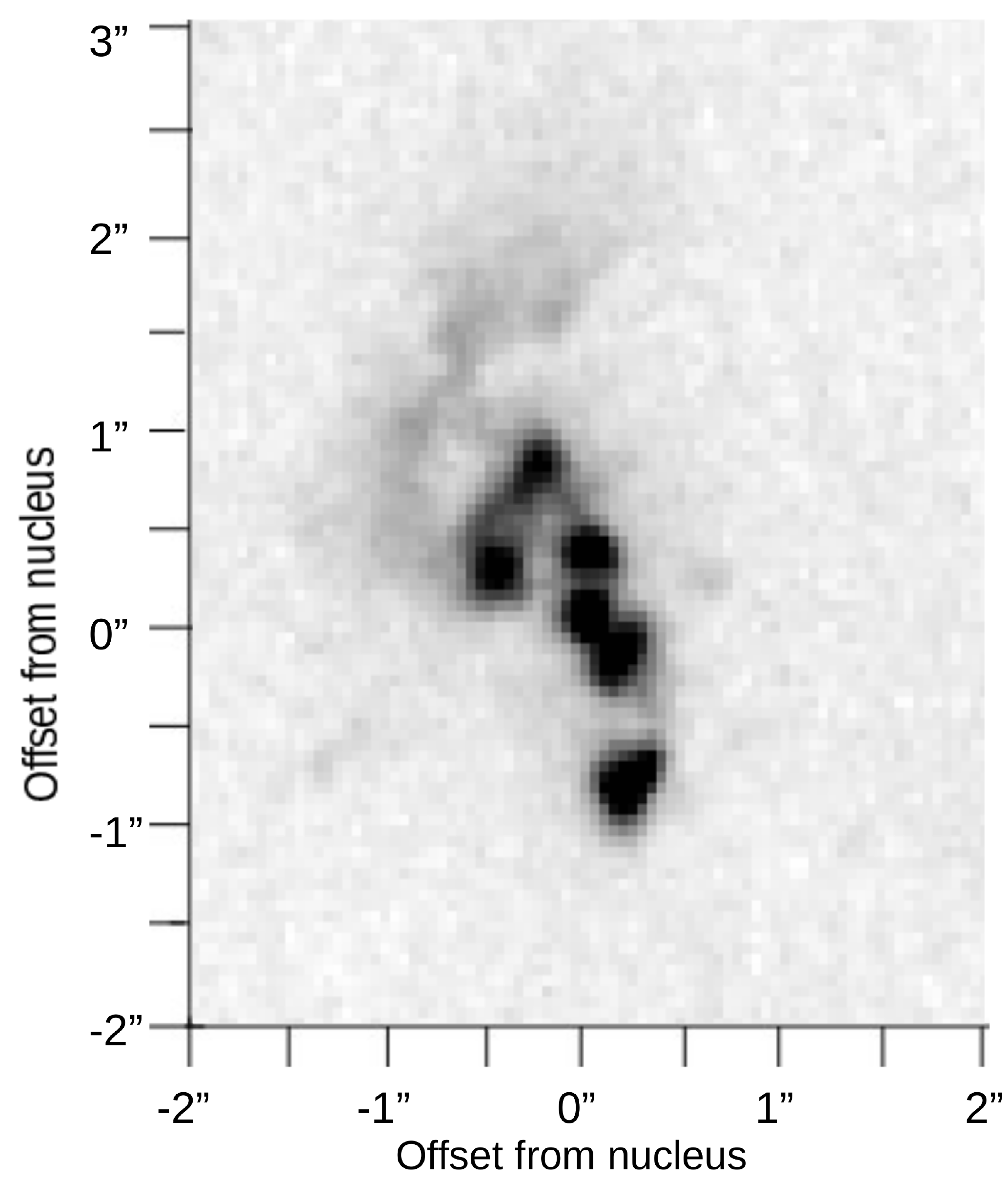}
\hfill
\includegraphics[width=4cm]{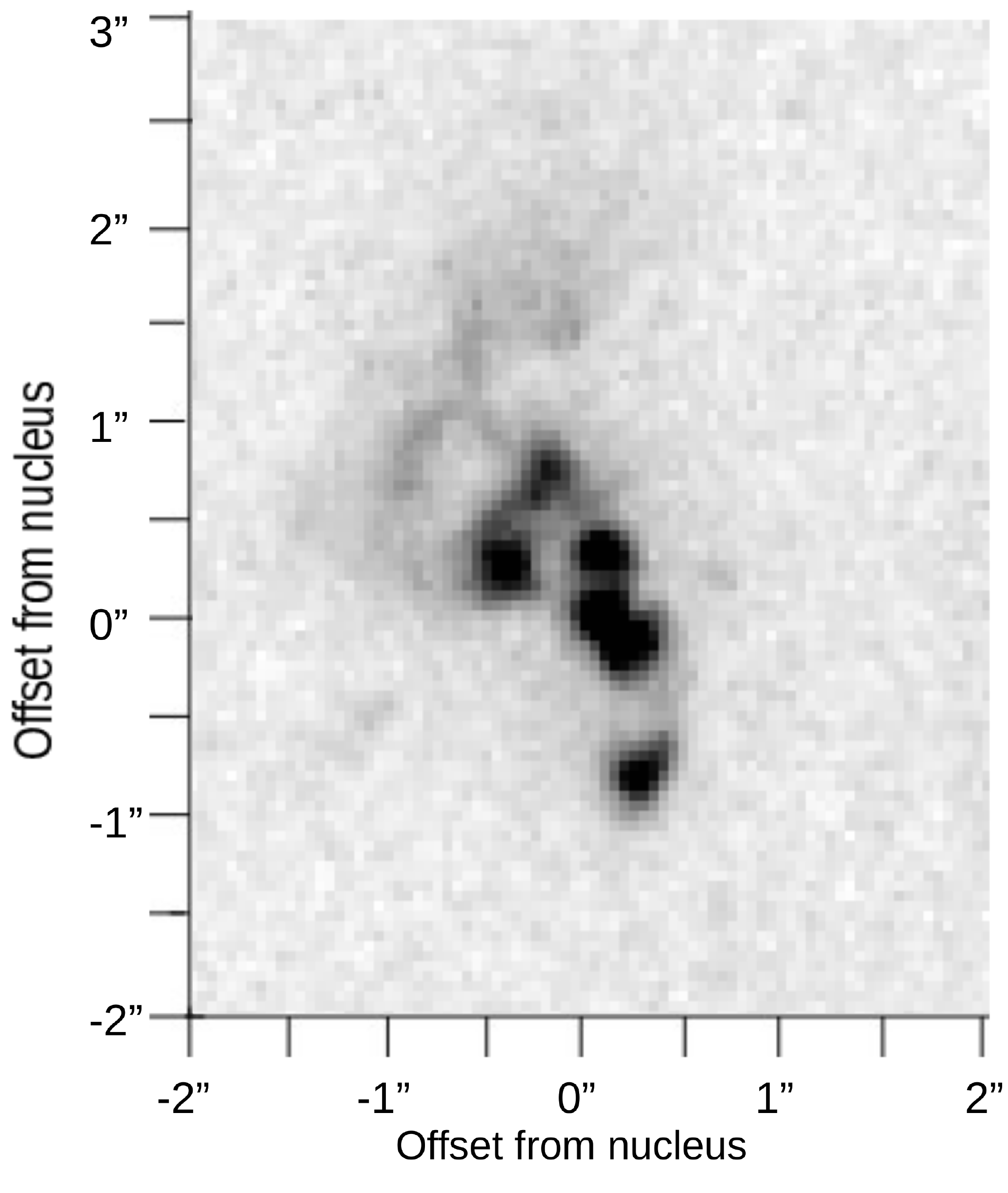}
\quad
\includegraphics[width=4cm]{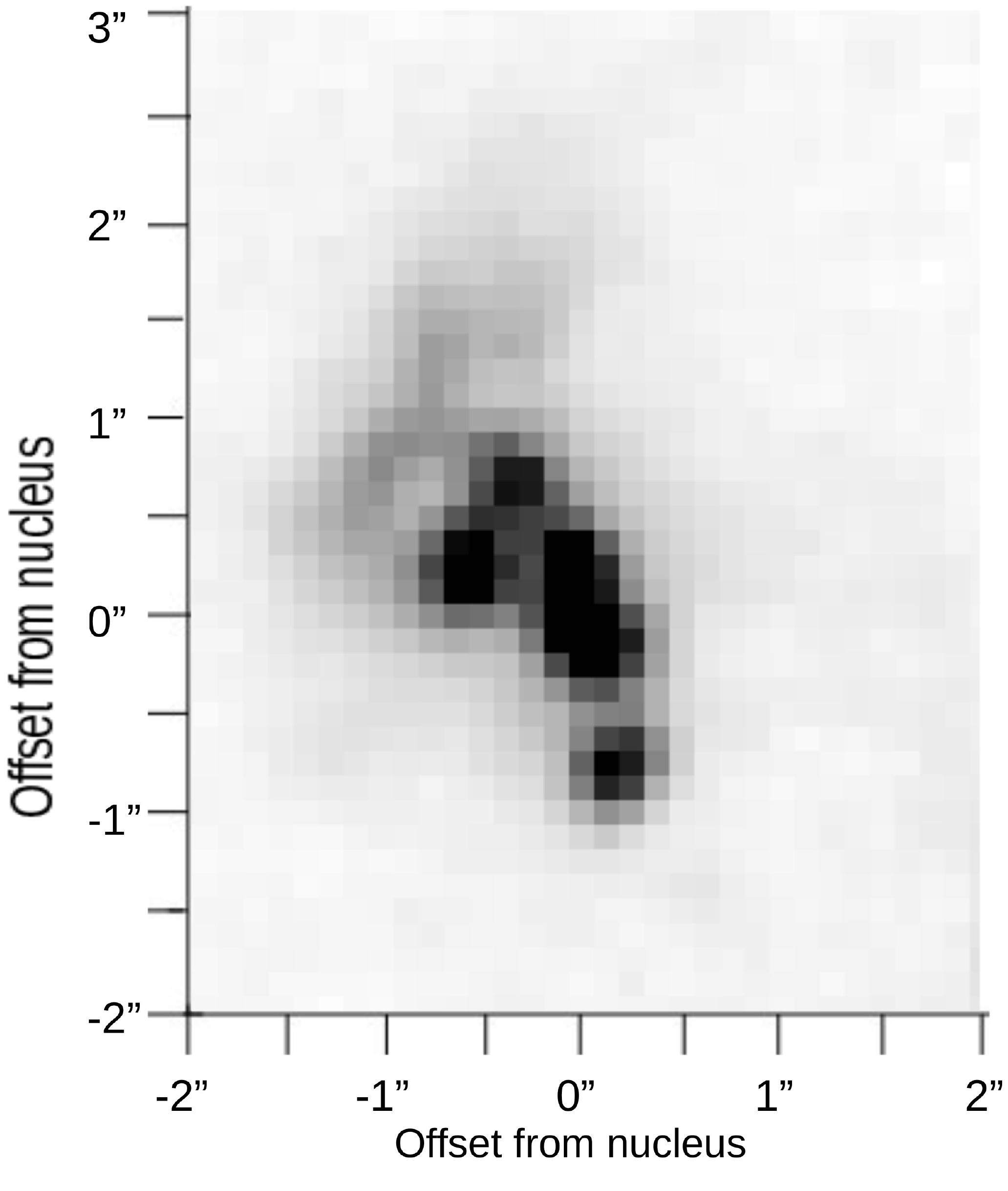}
\hfill 
\includegraphics[width=4cm]{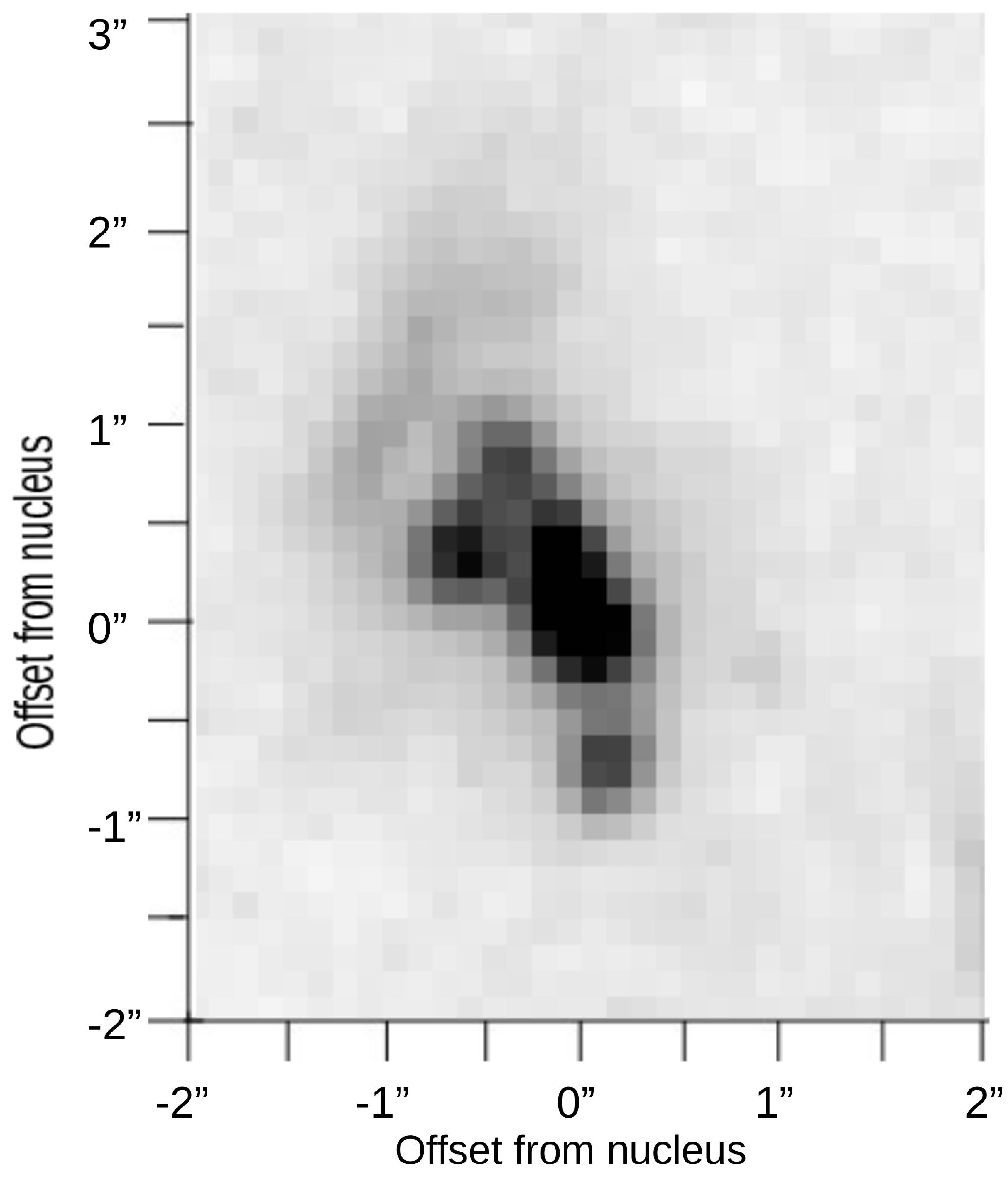}
\caption{HST observations of TXS 0828+193 using different filters. Top left: F606W image. Top right: F814W image. Bottom left: F110W image. Bottom right: F160W image.}\label{fig:TXS0828_HST}
\end{figure}

Fig. \ref{fig:TXS0828_HST} shows the HST images of TXS 0828+193. As already noted by \citet{Pentericci1999} using a relatively shallower WFPC2 image, in the central $\sim$2\arcsec\,of the galaxy the rest-frame UV and optical emission shows a triangular morphology and is closely aligned with the position angle of the radio source, with a number of relatively bright and compact clumps embedded within a diffuse low surface brightness component. 

The new HST images reveal additional, faint morphological features in TXS 0828+193. Thanks to the greater depth of our images, the diffuse UV emission on the NE side of the galaxy, also detected in the WFPC2 image of \citet{Pentericci1999}, is revealed to have a double-shell/loop morphology. This feature is detected in all 4 of the $HST$ images, and is reminiscent of the super-bubble morphology seen in $HST$ images of MRC 0406-244 \citep{Rush1997, Pentericci2001, Taniguchi2001, Humphrey2009,Hatch2013}. 

In the F160W image, the morphology in the central $\sim$2\arcsec\,of the galaxy shows 6 relatively compact sources with an X-shaped distribution about the central source.

\section{Discussion}  \label{sec:Discussion}

\subsection{Outflows}

Perturbed/extreme kinematics are known to be present along the radio axis of several HzRGs \citep[e.g.][]{Tadhunter1991,McCarthy1996,VM1999a}, including TXS 0211-122 and TXS 0828+193 \citep{VM2002, Humphrey2006}. In the case of TXS 0211-122, we find further evidence for this in the form of an anticorrelation between the CIV/HeII flux ratio and magnitude of the blueshift of CIV and HeII from the systemic velocity, but only along the radio axis. No such (anti)correlation is seen along the perpendicular slit, suggesting that the ionized outflow is not galaxy-wide, and is located preferentially along the radio axis. Moreover, the FWHM of the HeII emission along the perpendicular slit is consistent with gravitational gas kinematics (\textless 600 km s$^{-1}$). These results are consistent with  a scenario in which the radio structures are disturbing the gas kinematics along the radio axis \citep[e.g.][]{Best1998,VM1999b,Christensen2006}. 

In the case of TXS 0828+193, the long-slit observations show that Ly$\alpha$, CIV, and HeII are relatively broad along both slit PAs. The large FWHM seen in the direction perpendicular to the radio axis is evidence that the disturbed emission line kinematics are not confined to the radio axis, with relatively broad Ly$\alpha$ (FWHM\textgreater1000 km s$^{-1}$) being detected out to a radial distance of $\sim$30 kpc from the nucleus. In this galaxy there seems to be an enhancement in the emission of several lines and a change in the velocity pattern at $\pm$2\arcsec\, from the nucleus, suggesting the presence of a mechanism operating at r$\sim$2\arcsec\,that is disturbing the gas and altering its excitation properties. 

\begin{figure}\centering
\textbf{TXS 0828+193}\par\medskip                                                             
\includegraphics[width=7 cm]{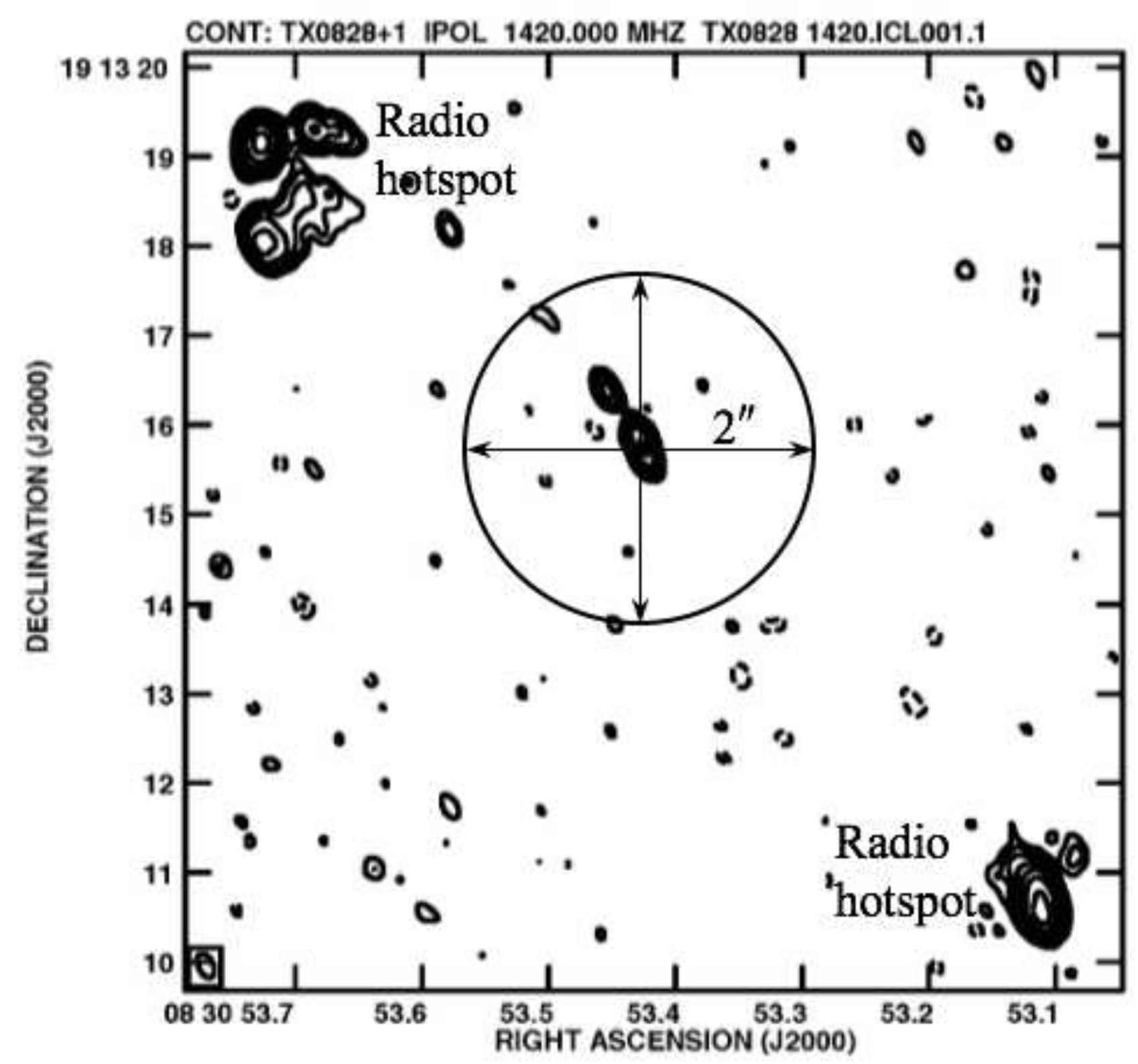}
\caption[Expanding bubble scheme overplotted in the image of the full field of TXS 0828+193]{Expanding bubble scheme overplotted in the image of the full field of TXS 0828+193.}\label{fig:TXS0828_bubble}
\end{figure}

We argue that the above results from TXS 0828+193 can be naturally explained by a scenario consisting of an expanding bubble of gas with a radius of 
$\sim$16 kpc ($\sim$2\arcsec\,, see Fig. \ref{fig:TXS0828_bubble}), inside which the most kinematically perturbed gas is located, and outside of which the gas kinematics are less extreme. At the edge of the bubble, one might expect there to be a shock-ionized shell of interstellar medium that has been swept up as the outflow progresses through the galaxy, although the enhancement in local gas density may also alter the ionization conditions there independently of the ionization mechanism. Interestingly, the HST images (Fig. \ref{fig:TXS0828_HST}) reveal an apparent bubble or loop structure located at $\sim$1.4\arcsec (11 kpc) to the NE along the radio axis, which we suggest may be part of this bubble.

Based on timescale arguments, we find it plausible that the expanding bubble and radio-loud activity were triggered simultaneously. In the case of the radio source, we assume an average speed of hotspot advance of $\sim$0.1\textit{c} \citep[e.g.][]{Carilli1991, Liu1992, Best1995, Arshakian2000}. For a hotspot to core radius of $\sim$50 kpc, we thus obtain an approximate age of $\sim 2 \times 10^{7}$ years. 

For the expanding bubble, we assume an expansion velocity of $\sim$1000 km s$^{-1}$ \citep[e.g.][]{Bland1988, Heckman2003}. Taken together with its radius of $\sim$16 kpc, this gives an age of $\sim 2 \times 10^{7}$ years. Despite the calculated ages being rather approximate, their similarity shows that both phenomena could plausibly have been triggered at (or near) the same point in time, presumably by the AGN activity (for the jets) and feedback activity (for the bubble). A possible alternative means to explain their apparently simultaneous triggering is through a merger event, during which gas is driven into the nuclear regions of the host galaxy to trigger a massive starburst and a resultant galaxy-wide super wind, in addition to triggering radio-loud AGN activity. 

Interestingly, from its velocity curve along the perpendicular slit (Fig. \ref{fig:TXS0828_velocity_FWHM}) we find that Ly$\alpha$ is almost always redshifted in relation to the systemic velocity. Taking this result at face value (ignoring line transfer effects), one might conclude that the extended Ly$\alpha$ emitting gas is infalling towards the centre of the host galaxy \citep[e.g.][]{Humphrey2007a, Humphrey2013a}. However, the transfer of Ly$\alpha$ photons through an outflowing neutral medium may also result in the Ly$\alpha$ profile being shifted to the red \citep[e.g.][]{Dijkstra2006}. As such, this result cannot be unambiguously interpreted. 

\subsection{Offset continuum sources in TXS 0211-122}

Using the same Keck II long slit spectrum of TXS 0211-122 (PA$=$104$^{\circ}$), \citet{Humphrey2013TXS0211} detected a pair of diametrically opposed UV continuum sources positioned $\sim$60 kpc either side of the radio galaxy, at the extreme outer edge of the Ly$\alpha$ emission halo. The brighter of the two UV continuum sources was found to be linearly polarized, implying illumination by the central AGN and the presence of dust. Its detection in H$\alpha$ confirmed its similar redshift to the radio galaxy. \citet{Humphrey2013TXS0211} proposed to have detected in emission (for the first time) a giant shell of gas of the type that is thought to be responsible for the giant HI and CIV absorbers detected in many HzRGs \citep[e.g.][]{Ojik1997}. 

Our detection of a further UV continuum source, also located at $\sim$60 kpc from the AGN and along a very different PA, corroborates the result of \citet{Humphrey2013TXS0211} and provides further information about the nature of this structure(s). The detection of three (compared to two) equidistant offset continuum sources now makes it even more improbable that these are merely fortuitously positioned companion galaxies. If our two slits are intersecting a single continuous structure, then it would need to cover an angle of at least 180$^{\circ}$, and be approximately circular/semi-circular.  Interestingly, our GTC spectrum shows the Ly$\alpha$ emission to be brightest/more extended on the side of the galaxy where the offset UV continuum source is positioned, which suggests the presence or enhancement of extended Ly$\alpha$ may be closely related to the offset continuum source (or vice versa). Our favoured interpretation is that this is a large-scale shell of gas and dust, produced by a powerful feedback event in the host galaxy of the HzRG, similar to the interpretation of \citet{Humphrey2013TXS0211}. 

A potential alternative explanation is that we may have detected the observational manifestation of an accretion shock, produced during the infall of cold gas into the host galaxy of the HzRG \citep[e.g.][]{Barkana2003, Barkana2004, Humphrey2008b}. 
In this scenario, infalling gas would be shocked and ionized upon reaching the virial radius, with the density of gas experiencing a local enhancement there. Unfortunately, the radius at which this effect would occur is not well constrained, but is not incompatible with the 60 kpc offsets measured for the UV continuum sources we observe \citep[see also][]{Humphrey2013TXS0211}. However, the apparent presence of dust in at least one of the offset UV sources would seem to disfavour an external origin.

\section{Conclusions}\label{sec:Conclusions}

Using long-slit spectroscopy from the 10.4 m GTC and the 10 m Keck II telescopes, we have studied the kinematics and ionization properties of the extended emission line nebulae associated with two HzRGs, TXS 0211-122 and TXS 0828+193. We have also presented previously unpublished HST images of TXS 0828+193.

In addition to the large spatial extent of Ly$\alpha$ emission along the radio axis reported by previous works \citep{VM2002,VM2003}, we find that both radio galaxies also show spatially very extended Ly$\alpha$ emission in the direction perpendicular to the radio axis. In the case of TXS 0211-122, the flux and velocity profiles of Ly$\alpha$ are strongly affected by HI absorption/scattering. Moreover, we find evidence for outflowing gas along the radio axis which may be the result of jet-gas interactions, in agreement with previous studies. In contrast, in our slit placed perpendicularly to the radio axis, we find less perturbed gas kinematics, which suggests outflows of ionized gas in this object are focused along the radio jet axis, rather than being galaxy-wide. We also find evidence for a large-scale UV-emitting arc or shell-like feature circumscribing the radio galaxy and the Ly$\alpha$ halo, possibly resulting from feedback activity. 

For TXS 0828+193, extended Ly$\alpha$ emission ($\sim$56 kpc) is detected in the direction perpendicular to the radio axis, in addition to the previously reported large extent along the radio axis \citep[130 kpc,][]{VM2002}. Along both slit position angles we find evidence for gas with highly perturbed kinematics, which we argue is part of a roughly spherical, expanding bubble or shell of gas powered by feedback activity in the central regions of the radio galaxy. Our results suggest there is a diversity in the spatial distribution of ionized outflows in powerful radio galaxies at z$\sim$2.5.

\section*{Acknowledgments}

We would like to thank Laura Pentericci for allowing us to reproduce the radio/optical image of TXS 0211-122. We thank Bjorn Emonts for the useful comments that helped improve this work. We also thank Bob Fosbury, Andrea Cimatti and Marshall Cohen for the important contribution they have made to the HzRG Keck II spectropolarimetry project from which some of the data herein comes. 
AH and PL acknowledge Funda\c{c}\~{a}o para a Ci\^{e}ncia e a Tecnologia (FCT) support through UID/FIS/04434/2013, and through project FCOMP-01-0124-FEDER-029170 (Reference FCT PTDC/FIS-AST/3214/2012) funded by FCT-MEC (PIDDAC) and FEDER (COMPETE), in addition to FP7 project PIRSES-GA-2013-612701. AH also acknowledges a Marie Curie Fellowship co-funded by the FP7 and the FCT (DFRH/WIIA/57/2011), FP7 / FCT Complementary Support grant SFRH/BI/52155/2013, and FCT grant SFRH/BPD/107919/2015. PL is supported by a post-doctoral grant SFRH/BPD/72308/2010, funded by the FCT. Support for RO was provided by CNPq programs 459040/2014-6 and 400738/2014-7.


\bibliographystyle{mnras}
\bibliography{refs} 

\begin{thebibliography}{}
\makeatletter
\relax
\def\mn@urlcharsother{\let\do\@makeother \do\$\do\&\do\#\do\^\do\_\do\%\do\~}
\def\mn@doi{\begingroup\mn@urlcharsother \@ifnextchar [ {\mn@doi@}
  {\mn@doi@[]}}
\def\mn@doi@[#1]#2{\def\@tempa{#1}\ifx\@tempa\@empty \href
  {http://dx.doi.org/#2} {doi:#2}\else \href {http://dx.doi.org/#2} {#1}\fi
  \endgroup}
\def\mn@eprint#1#2{\mn@eprint@#1:#2::\@nil}
\def\mn@eprint@arXiv#1{\href {http://arxiv.org/abs/#1} {{\tt arXiv:#1}}}
\def\mn@eprint@dblp#1{\href {http://dblp.uni-trier.de/rec/bibtex/#1.xml}
  {dblp:#1}}
\def\mn@eprint@#1:#2:#3:#4\@nil{\def\@tempa {#1}\def\@tempb {#2}\def\@tempc
  {#3}\ifx \@tempc \@empty \let \@tempc \@tempb \let \@tempb \@tempa \fi \ifx
  \@tempb \@empty \def\@tempb {arXiv}\fi \@ifundefined
  {mn@eprint@\@tempb}{\@tempb:\@tempc}{\expandafter \expandafter \csname
  mn@eprint@\@tempb\endcsname \expandafter{\@tempc}}}

\bibitem[\protect\citeauthoryear{{Adams}, {Hill}  \& {MacQueen}}{{Adams}
  et~al.}{2009}]{Adams2009}
{Adams} J.~J.,  {Hill} G.~J.,   {MacQueen} P.~J.,  2009, \mn@doi [\apj]
  {10.1088/0004-637X/694/1/314}, \href
  {http://adsabs.harvard.edu/abs/2009ApJ...694..314A} {694, 314}

\bibitem[\protect\citeauthoryear{{Allen}, {Groves}, {Dopita}, {Sutherland}  \&
  {Kewley}}{{Allen} et~al.}{2008}]{Allen2008}
{Allen} M.~G.,  {Groves} B.~A.,  {Dopita} M.~A.,  {Sutherland} R.~S.,
  {Kewley} L.~J.,  2008, \mn@doi [\apjs] {10.1086/589652}, \href
  {http://adsabs.harvard.edu/abs/2008ApJS..178...20A} {178, 20}

\bibitem[\protect\citeauthoryear{{Arshakian} \& {Longair}}{{Arshakian} \&
  {Longair}}{2000}]{Arshakian2000}
{Arshakian} T.~G.,  {Longair} M.~S.,  2000, \mn@doi [\mnras]
  {10.1046/j.1365-8711.2000.03098.x}, \href
  {http://adsabs.harvard.edu/abs/2000MNRAS.311..846A} {311, 846}

\bibitem[\protect\citeauthoryear{{Asplund}, {Grevesse}  \& {Jacques
  Sauval}}{{Asplund} et~al.}{2006}]{Asplund2006}
{Asplund} M.,  {Grevesse} N.,   {Jacques Sauval} A.,  2006, \mn@doi [Nuclear
  Physics A] {10.1016/j.nuclphysa.2005.06.010}, \href
  {http://adsabs.harvard.edu/abs/2006NuPhA.777....1A} {777, 1}

\bibitem[\protect\citeauthoryear{{Barkana}}{{Barkana}}{2004}]{Barkana2004}
{Barkana} R.,  2004, \mn@doi [\mnras] {10.1111/j.1365-2966.2004.07177.x}, \href
  {http://adsabs.harvard.edu/abs/2004MNRAS.347...59B} {347, 59}

\bibitem[\protect\citeauthoryear{{Barkana} \& {Loeb}}{{Barkana} \&
  {Loeb}}{2003}]{Barkana2003}
{Barkana} R.,  {Loeb} A.,  2003, \nat, \href
  {http://adsabs.harvard.edu/abs/2003Natur.421..341B} {421, 341}

\bibitem[\protect\citeauthoryear{{Best}, {Bailer}, {Longair}  \&
  {Riley}}{{Best} et~al.}{1995}]{Best1995}
{Best} P.~N.,  {Bailer} D.~M.,  {Longair} M.~S.,   {Riley} J.~M.,  1995,
  \mn@doi [\mnras] {10.1093/mnras/275.4.1171}, \href
  {http://adsabs.harvard.edu/abs/1995MNRAS.275.1171B} {275, 1171}

\bibitem[\protect\citeauthoryear{{Best}, {Carilli}, {Garrington}, {Longair}  \&
  {Rottgering}}{{Best} et~al.}{1998}]{Best1998}
{Best} P.~N.,  {Carilli} C.~L.,  {Garrington} S.~T.,  {Longair} M.~S.,
  {Rottgering} H.~J.~A.,  1998, \mn@doi [\mnras]
  {10.1046/j.1365-8711.1998.01676.x}, \href
  {http://adsabs.harvard.edu/abs/1998MNRAS.299..357B} {299, 357}

\bibitem[\protect\citeauthoryear{{Bicknell}, {Sutherland}, {van Breugel},
  {Dopita}, {Dey}  \& {Miley}}{{Bicknell} et~al.}{2000}]{Bicknell2000}
{Bicknell} G.~V.,  {Sutherland} R.~S.,  {van Breugel} W.~J.~M.,  {Dopita}
  M.~A.,  {Dey} A.,   {Miley} G.~K.,  2000, \mn@doi [\apj] {10.1086/309343},
  \href {http://adsabs.harvard.edu/abs/2000ApJ...540..678B} {540, 678}

\bibitem[\protect\citeauthoryear{{Binette}, {Dopita}  \& {Tuohy}}{{Binette}
  et~al.}{1985}]{Binette1985}
{Binette} L.,  {Dopita} M.~A.,   {Tuohy} I.~R.,  1985, \mn@doi [\apj]
  {10.1086/163544}, \href {http://adsabs.harvard.edu/abs/1985ApJ...297..476B}
  {297, 476}

\bibitem[\protect\citeauthoryear{{Binette}, {Kurk}, {Villar-Mart{\'{\i}}n}  \&
  {R{\"o}ttgering}}{{Binette} et~al.}{2000}]{Binette2000}
{Binette} L.,  {Kurk} J.~D.,  {Villar-Mart{\'{\i}}n} M.,   {R{\"o}ttgering}
  H.~J.~A.,  2000, \aap, \href
  {http://adsabs.harvard.edu/abs/2000A%26A...356...23B} {356, 23}

\bibitem[\protect\citeauthoryear{{Bland} \& {Tully}}{{Bland} \&
  {Tully}}{1988}]{Bland1988}
{Bland} J.,  {Tully} B.,  1988, \mn@doi [\nat] {10.1038/334043a0}, \href
  {http://adsabs.harvard.edu/abs/1988Natur.334...43B} {334, 43}

\bibitem[\protect\citeauthoryear{{Carilli}, {Perley}, {Dreher}  \&
  {Leahy}}{{Carilli} et~al.}{1991}]{Carilli1991}
{Carilli} C.~L.,  {Perley} R.~A.,  {Dreher} J.~W.,   {Leahy} J.~P.,  1991,
  \mn@doi [\apj] {10.1086/170813}, \href
  {http://adsabs.harvard.edu/abs/1991ApJ...383..554C} {383, 554}

\bibitem[\protect\citeauthoryear{{Carilli}, {R{\"o}ttgering}, {van Ojik},
  {Miley}, {Breugel}  \& {W.~J.~M.~van}}{{Carilli} et~al.}{1997}]{Carilli1997}
{Carilli} C.~L.,  {R{\"o}ttgering} H.~J.~A.,  {van Ojik} R.,  {Miley} G.~K.,
  {Breugel}  {W.~J.~M.~van} 1997, \mn@doi [\apjs] {10.1086/312973}, \href
  {http://adsabs.harvard.edu/abs/1997ApJS..109....1C} {109, 1}

\bibitem[\protect\citeauthoryear{{Cepa} et~al.,}{{Cepa}
  et~al.}{2000}]{Cepa2000}
{Cepa} J.,  et~al., 2000, in {Iye} M.,  {Moorwood} A.~F.,  eds,  \procspie Vol.
  4008, Optical and IR Telescope Instrumentation and Detectors. pp 623--631

\bibitem[\protect\citeauthoryear{{Cepa} et~al.,}{{Cepa}
  et~al.}{2003}]{Cepa2003}
{Cepa} J.,  et~al., 2003, in {Iye} M.,  {Moorwood} A.~F.~M.,  eds,  \procspie
  Vol. 4841, Instrument Design and Performance for Optical/Infrared
  Ground-based Telescopes. pp 1739--1749, \mn@doi{10.1117/12.460913}

\bibitem[\protect\citeauthoryear{{Christensen}, {Jahnke}, {Wisotzki},
  {S{\'a}nchez}, {Exter}  \& {Roth}}{{Christensen}
  et~al.}{2006}]{Christensen2006}
{Christensen} L.,  {Jahnke} K.,  {Wisotzki} L.,  {S{\'a}nchez} S.~F.,  {Exter}
  K.,   {Roth} M.~M.,  2006, \mn@doi [\aap] {10.1051/0004-6361:20054578}, \href
  {http://adsabs.harvard.edu/abs/2006A%26A...452..869C} {452, 869}

\bibitem[\protect\citeauthoryear{{Dannerbauer} et~al.,}{{Dannerbauer}
  et~al.}{2014}]{Dannerbauer2014}
{Dannerbauer} H.,  et~al., 2014, \mn@doi [\aap] {10.1051/0004-6361/201423771},
  \href {http://adsabs.harvard.edu/abs/2014A%26A...570A..55D} {570, A55}

\bibitem[\protect\citeauthoryear{{De Breuck}, {van Breugel}, {Stanford},
  {R{\"o}ttgering}, {Miley}  \& {Stern}}{{De Breuck} et~al.}{2002}]{Breuck2002}
{De Breuck} C.,  {van Breugel} W.,  {Stanford} S.~A.,  {R{\"o}ttgering} H.,
  {Miley} G.,   {Stern} D.,  2002, \mn@doi [\aj] {10.1086/324632}, \href
  {http://adsabs.harvard.edu/abs/2002AJ....123..637D} {123, 637}

\bibitem[\protect\citeauthoryear{{De Breuck} et~al.,}{{De Breuck}
  et~al.}{2010}]{Breuck2010}
{De Breuck} C.,  et~al., 2010, \mn@doi [\apj] {10.1088/0004-637X/725/1/36},
  \href {http://adsabs.harvard.edu/abs/2010ApJ...725...36D} {725, 36}

\bibitem[\protect\citeauthoryear{{Dijkstra}, {Haiman}  \& {Spaans}}{{Dijkstra}
  et~al.}{2006}]{Dijkstra2006}
{Dijkstra} M.,  {Haiman} Z.,   {Spaans} M.,  2006, \mn@doi [\apj]
  {10.1086/506243}, \href {http://adsabs.harvard.edu/abs/2006ApJ...649...14D}
  {649, 14}

\bibitem[\protect\citeauthoryear{{Douglas}, {Bash}, {Torrence}  \&
  {Wolfe}}{{Douglas} et~al.}{1980}]{Douglas1980}
{Douglas} J.~N.,  {Bash} F.~N.,  {Torrence} G.~W.,   {Wolfe} C.,  1980,
  University of Texas Publications in Astronomy, \href
  {http://adsabs.harvard.edu/abs/1980PAUTx..17....1D} {17, 1}

\bibitem[\protect\citeauthoryear{{Dubinski}}{{Dubinski}}{1998}]{Dubinski1998}
{Dubinski} J.,  1998, \mn@doi [\apj] {10.1086/305901}, \href
  {http://adsabs.harvard.edu/abs/1998ApJ...502..141D} {502, 141}

\bibitem[\protect\citeauthoryear{{Emonts} et~al.,}{{Emonts}
  et~al.}{2014}]{Emonts2014}
{Emonts} B.~H.~C.,  et~al., 2014, \mn@doi [\mnras] {10.1093/mnras/stt2398},
  \href {http://adsabs.harvard.edu/abs/2014MNRAS.438.2898E} {438, 2898}

\bibitem[\protect\citeauthoryear{{Ferruit}, {Binette}, {Sutherland}  \&
  {Pecontal}}{{Ferruit} et~al.}{1997}]{Ferruit1997}
{Ferruit} P.,  {Binette} L.,  {Sutherland} R.~S.,   {Pecontal} E.,  1997, \aap,
  \href {http://adsabs.harvard.edu/abs/1997A%26A...322...73F} {322, 73}

\bibitem[\protect\citeauthoryear{{Fosbury} et~al.,}{{Fosbury}
  et~al.}{1982}]{Fosbury1982}
{Fosbury} R.~A.~E.,  et~al., 1982, \mnras, \href
  {http://adsabs.harvard.edu/abs/1982MNRAS.201..991F} {201, 991}

\bibitem[\protect\citeauthoryear{{Galametz} et~al.,}{{Galametz}
  et~al.}{2012}]{Galametz2012}
{Galametz} A.,  et~al., 2012, \mn@doi [\apj] {10.1088/0004-637X/749/2/169},
  \href {http://adsabs.harvard.edu/abs/2012ApJ...749..169G} {749, 169}

\bibitem[\protect\citeauthoryear{{Hatch}, {Overzier}, {Kurk}, {Miley},
  {R{\"o}ttgering}  \& {Zirm}}{{Hatch} et~al.}{2009}]{Hatch2009}
{Hatch} N.~A.,  {Overzier} R.~A.,  {Kurk} J.~D.,  {Miley} G.~K.,
  {R{\"o}ttgering} H.~J.~A.,   {Zirm} A.~W.,  2009, \mn@doi [\mnras]
  {10.1111/j.1365-2966.2009.14525.x}, \href
  {http://adsabs.harvard.edu/abs/2009MNRAS.395..114H} {395, 114}

\bibitem[\protect\citeauthoryear{{Hatch} et~al.,}{{Hatch}
  et~al.}{2013}]{Hatch2013}
{Hatch} N.~A.,  et~al., 2013, \mn@doi [\mnras] {10.1093/mnras/stt1734}, \href
  {http://adsabs.harvard.edu/abs/2013MNRAS.436.2244H} {436, 2244}

\bibitem[\protect\citeauthoryear{{Heckman}}{{Heckman}}{2003}]{Heckman2003}
{Heckman} T.~M.,  2003, in {Avila-Reese} V.,  {Firmani} C.,  {Frenk} C.~S.,
  {Allen} C.,  eds,  Revista Mexicana de Astronomia y Astrofisica Conference
  Series Vol. 17, Revista Mexicana de Astronomia y Astrofisica Conference
  Series. pp 47--55

\bibitem[\protect\citeauthoryear{{Henry}, {Edmunds}  \& {K{\"o}ppen}}{{Henry}
  et~al.}{2000}]{Henry2000}
{Henry} R.~B.~C.,  {Edmunds} M.~G.,   {K{\"o}ppen} J.,  2000, \mn@doi [\apj]
  {10.1086/309471}, \href {http://adsabs.harvard.edu/abs/2000ApJ...541..660H}
  {541, 660}

\bibitem[\protect\citeauthoryear{{Humphrey}, {Villar-Mart{\'{\i}}n}, {Fosbury},
  {Vernet}  \& {di Serego Alighieri}}{{Humphrey} et~al.}{2006}]{Humphrey2006}
{Humphrey} A.,  {Villar-Mart{\'{\i}}n} M.,  {Fosbury} R.,  {Vernet} J.,   {di
  Serego Alighieri} S.,  2006, \mn@doi [\mnras]
  {10.1111/j.1365-2966.2006.10224.x}, \href
  {http://adsabs.harvard.edu/abs/2006MNRAS.369.1103H} {369, 1103}

\bibitem[\protect\citeauthoryear{{Humphrey}, {Villar-Mart{\'{\i}}n}, {Fosbury},
  {Binette}, {Vernet}, {De Breuck}  \& {di Serego Alighieri}}{{Humphrey}
  et~al.}{2007a}]{Humphrey2007a}
{Humphrey} A.,  {Villar-Mart{\'{\i}}n} M.,  {Fosbury} R.,  {Binette} L.,
  {Vernet} J.,  {De Breuck} C.,   {di Serego Alighieri} S.,  2007a, \mn@doi
  [\mnras] {10.1111/j.1365-2966.2006.11344.x}, \href
  {http://adsabs.harvard.edu/abs/2007MNRAS.375..705H} {375, 705}

\bibitem[\protect\citeauthoryear{{Humphrey}, {Iwamuro}, {Villar-Mart{\'{\i}}n},
  {Binette}, {Fosbury}  \& {di Serego Alighieri}}{{Humphrey}
  et~al.}{2007b}]{Humphrey2007b}
{Humphrey} A.,  {Iwamuro} F.,  {Villar-Mart{\'{\i}}n} M.,  {Binette} L.,
  {Fosbury} R.,   {di Serego Alighieri} S.,  2007b, \mn@doi [\mnras]
  {10.1111/j.1365-2966.2007.12463.x}, \href
  {http://adsabs.harvard.edu/abs/2007MNRAS.382.1729H} {382, 1729}

\bibitem[\protect\citeauthoryear{{Humphrey}, {Villar-Mart{\'{\i}}n}, {Vernet},
  {Fosbury}, {di Serego Alighieri}  \& {Binette}}{{Humphrey}
  et~al.}{2008a}]{Humphrey2008a}
{Humphrey} A.,  {Villar-Mart{\'{\i}}n} M.,  {Vernet} J.,  {Fosbury} R.,  {di
  Serego Alighieri} S.,   {Binette} L.,  2008a, \mn@doi [\mnras]
  {10.1111/j.1365-2966.2007.12506.x}, \href
  {http://adsabs.harvard.edu/abs/2008MNRAS.383...11H} {383, 11}

\bibitem[\protect\citeauthoryear{{Humphrey} et~al.,}{{Humphrey}
  et~al.}{2008b}]{Humphrey2008b}
{Humphrey} A.,  et~al., 2008b, \mn@doi [\mnras]
  {10.1111/j.1365-2966.2008.13826.x}, \href
  {http://adsabs.harvard.edu/abs/2008MNRAS.390.1505H} {390, 1505}

\bibitem[\protect\citeauthoryear{{Humphrey}, {Iwamuro}, {Villar-Mart{\'{\i}}n},
  {Binette}  \& {Sung}}{{Humphrey} et~al.}{2009}]{Humphrey2009}
{Humphrey} A.,  {Iwamuro} F.,  {Villar-Mart{\'{\i}}n} M.,  {Binette} L.,
  {Sung} E.~C.,  2009, \mn@doi [\mnras] {10.1111/j.1745-3933.2009.00719.x},
  \href {http://adsabs.harvard.edu/abs/2009MNRAS.399L..34H} {399, L34}

\bibitem[\protect\citeauthoryear{{Humphrey}, {Binette}, {Villar-Mart{\'{\i}}n},
  {Aretxaga}  \& {Papaderos}}{{Humphrey} et~al.}{2013a}]{Humphrey2013a}
{Humphrey} A.,  {Binette} L.,  {Villar-Mart{\'{\i}}n} M.,  {Aretxaga} I.,
  {Papaderos} P.,  2013a, \mn@doi [\mnras] {10.1093/mnras/sts055}, \href
  {http://adsabs.harvard.edu/abs/2013MNRAS.428..563H} {428, 563}

\bibitem[\protect\citeauthoryear{{Humphrey}, {Vernet}, {Villar-Mart{\'{\i}}n},
  {di Serego Alighieri}, {Fosbury}  \& {Cimatti}}{{Humphrey}
  et~al.}{2013b}]{Humphrey2013TXS0211}
{Humphrey} A.,  {Vernet} J.,  {Villar-Mart{\'{\i}}n} M.,  {di Serego Alighieri}
  S.,  {Fosbury} R.~A.~E.,   {Cimatti} A.,  2013b, \mn@doi [\apjl]
  {10.1088/2041-8205/768/1/L3}, \href
  {http://adsabs.harvard.edu/abs/2013ApJ...768L...3H} {768, L3}

\bibitem[\protect\citeauthoryear{{Iwamuro} et~al.,}{{Iwamuro}
  et~al.}{2003}]{Iwamuro2003}
{Iwamuro} F.,  et~al., 2003, \mn@doi [\apj] {10.1086/378845}, \href
  {http://adsabs.harvard.edu/abs/2003ApJ...598..178I} {598, 178}

\bibitem[\protect\citeauthoryear{{Jarvis} et~al.,}{{Jarvis}
  et~al.}{2001a}]{Jarvis2001a}
{Jarvis} M.~J.,  et~al., 2001a, \mn@doi [\mnras]
  {10.1111/j.1365-2966.2001.04726.x}, \href
  {http://adsabs.harvard.edu/abs/2001MNRAS.326.1563J} {326, 1563}

\bibitem[\protect\citeauthoryear{{Jarvis}, {Rawlings}, {Eales}, {Blundell},
  {Bunker}, {Croft}, {McLure}  \& {Willott}}{{Jarvis}
  et~al.}{2001b}]{Jarvis2001b}
{Jarvis} M.~J.,  {Rawlings} S.,  {Eales} S.,  {Blundell} K.~M.,  {Bunker}
  A.~J.,  {Croft} S.,  {McLure} R.~J.,   {Willott} C.~J.,  2001b, \mn@doi
  [\mnras] {10.1111/j.1365-2966.2001.04730.x}, \href
  {http://adsabs.harvard.edu/abs/2001MNRAS.326.1585J} {326, 1585}

\bibitem[\protect\citeauthoryear{{Jarvis}, {Wilman}, {R{\"o}ttgering}  \&
  {Binette}}{{Jarvis} et~al.}{2003}]{Jarvis2003}
{Jarvis} M.~J.,  {Wilman} R.~J.,  {R{\"o}ttgering} H.~J.~A.,   {Binette} L.,
  2003, \mn@doi [\mnras] {10.1046/j.1365-8711.2003.06053.x}, \href
  {http://adsabs.harvard.edu/abs/2003MNRAS.338..263J} {338, 263}

\bibitem[\protect\citeauthoryear{{Klamer}, {Ekers}, {Sadler}  \&
  {Hunstead}}{{Klamer} et~al.}{2004}]{Klamer2004}
{Klamer} I.~J.,  {Ekers} R.~D.,  {Sadler} E.~M.,   {Hunstead} R.~W.,  2004,
  \mn@doi [\apjl] {10.1086/424843}, \href
  {http://adsabs.harvard.edu/abs/2004ApJ...612L..97K} {612, L97}

\bibitem[\protect\citeauthoryear{{Legrand}, {Kunth}, {Mas-Hesse}  \&
  {Lequeux}}{{Legrand} et~al.}{1997}]{Legrand1997}
{Legrand} F.,  {Kunth} D.,  {Mas-Hesse} J.~M.,   {Lequeux} J.,  1997, \aap,
  \href {http://adsabs.harvard.edu/abs/1997A%26A...326..929L} {326, 929}

\bibitem[\protect\citeauthoryear{{Liu}, {Pooley}  \& {Riley}}{{Liu}
  et~al.}{1992}]{Liu1992}
{Liu} R.,  {Pooley} G.,   {Riley} J.~M.,  1992, \mn@doi [\mnras]
  {10.1093/mnras/257.4.545}, \href
  {http://adsabs.harvard.edu/abs/1992MNRAS.257..545L} {257, 545}

\bibitem[\protect\citeauthoryear{{Mayo}, {Vernet}, {De Breuck}, {Galametz},
  {Seymour}  \& {Stern}}{{Mayo} et~al.}{2012}]{Mayo2012}
{Mayo} J.~H.,  {Vernet} J.,  {De Breuck} C.,  {Galametz} A.,  {Seymour} N.,
  {Stern} D.,  2012, \mn@doi [\aap] {10.1051/0004-6361/201118254}, \href
  {http://adsabs.harvard.edu/abs/2012A%26A...539A..33M} {539, A33}

\bibitem[\protect\citeauthoryear{{McCarthy}}{{McCarthy}}{1993}]{McCarthy1993}
{McCarthy} P.~J.,  1993, \mn@doi [\araa] {10.1146/annurev.aa.31.090193.003231},
  \href {http://adsabs.harvard.edu/abs/1993ARA%26A..31..639M} {31, 639}

\bibitem[\protect\citeauthoryear{{McCarthy}, {van Breugel}, {Spinrad}  \&
  {Djorgovski}}{{McCarthy} et~al.}{1987}]{McCarthy1987}
{McCarthy} P.~J.,  {van Breugel} W.,  {Spinrad} H.,   {Djorgovski} S.,  1987,
  \mn@doi [\apjl] {10.1086/185000}, \href
  {http://adsabs.harvard.edu/abs/1987ApJ...321L..29M} {321, L29}

\bibitem[\protect\citeauthoryear{{McCarthy}, {Spinrad}, {Dickinson}, {van
  Breugel}, {Liebert}, {Djorgovski}  \& {Eisenhardt}}{{McCarthy}
  et~al.}{1990}]{McCarthy1990}
{McCarthy} P.~J.,  {Spinrad} H.,  {Dickinson} M.,  {van Breugel} W.,  {Liebert}
  J.,  {Djorgovski} S.,   {Eisenhardt} P.,  1990, \mn@doi [\apj]
  {10.1086/169503}, \href {http://adsabs.harvard.edu/abs/1990ApJ...365..487M}
  {365, 487}

\bibitem[\protect\citeauthoryear{{McCarthy}, {Spinrad}  \& {van
  Breugel}}{{McCarthy} et~al.}{1995}]{McCarthy1995}
{McCarthy} P.~J.,  {Spinrad} H.,   {van Breugel} W.,  1995, \mn@doi [\apjs]
  {10.1086/192178}, \href {http://adsabs.harvard.edu/abs/1995ApJS...99...27M}
  {99, 27}

\bibitem[\protect\citeauthoryear{{McCarthy}, {Baum}  \& {Spinrad}}{{McCarthy}
  et~al.}{1996}]{McCarthy1996}
{McCarthy} P.~J.,  {Baum} S.~A.,   {Spinrad} H.,  1996, \mn@doi [\apjs]
  {10.1086/192339}, \href {http://adsabs.harvard.edu/abs/1996ApJS..106..281M}
  {106, 281}

\bibitem[\protect\citeauthoryear{{Miley} \& {De Breuck}}{{Miley} \& {De
  Breuck}}{2008}]{Miley2008}
{Miley} G.,  {De Breuck} C.,  2008, \mn@doi [\aapr]
  {10.1007/s00159-007-0008-z}, \href
  {http://adsabs.harvard.edu/abs/2008A%26ARv..15...67M} {15, 67}

\bibitem[\protect\citeauthoryear{{Nesvadba} et~al.,}{{Nesvadba}
  et~al.}{2009}]{Nesvadba2009}
{Nesvadba} N.~P.~H.,  et~al., 2009, \mn@doi [\mnras]
  {10.1111/j.1745-3933.2009.00631.x}, \href
  {http://adsabs.harvard.edu/abs/2009MNRAS.395L..16N} {395, L16}

\bibitem[\protect\citeauthoryear{{Oke} et~al.,}{{Oke} et~al.}{1995}]{Oke1995}
{Oke} J.~B.,  et~al., 1995, \mn@doi [\pasp] {10.1086/133562}, \href
  {http://adsabs.harvard.edu/abs/1995PASP..107..375O} {107, 375}

\bibitem[\protect\citeauthoryear{{Overzier}, {R{\"o}ttgering}, {Kurk}  \& {De
  Breuck}}{{Overzier} et~al.}{2001}]{Overzier2001}
{Overzier} R.~A.,  {R{\"o}ttgering} H.~J.~A.,  {Kurk} J.~D.,   {De Breuck} C.,
  2001, \mn@doi [\aap] {10.1051/0004-6361:20010041}, \href
  {http://adsabs.harvard.edu/abs/2001A%26A...367L...5O} {367, L5}

\bibitem[\protect\citeauthoryear{{Overzier}, {Harris}, {Carilli}, {Pentericci},
  {R{\"o}ttgering}  \& {Miley}}{{Overzier} et~al.}{2005}]{Overzier2005}
{Overzier} R.~A.,  {Harris} D.~E.,  {Carilli} C.~L.,  {Pentericci} L.,
  {R{\"o}ttgering} H.~J.~A.,   {Miley} G.~K.,  2005, \mn@doi [\aap]
  {10.1051/0004-6361:2004165}, \href
  {http://adsabs.harvard.edu/abs/2005A%26A...433...87O} {433, 87}

\bibitem[\protect\citeauthoryear{{Pentericci}, {R{\"o}ttgering}, {Miley},
  {McCarthy}, {Spinrad}, {van Breugel}  \& {Macchetto}}{{Pentericci}
  et~al.}{1999}]{Pentericci1999}
{Pentericci} L.,  {R{\"o}ttgering} H.~J.~A.,  {Miley} G.~K.,  {McCarthy} P.,
  {Spinrad} H.,  {van Breugel} W.~J.~M.,   {Macchetto} F.,  1999, \aap, \href
  {http://adsabs.harvard.edu/abs/1999A%26A...341..329P} {341, 329}

\bibitem[\protect\citeauthoryear{{Pentericci}, {Van Reeven}, {Carilli},
  {R{\"o}ttgering}  \& {Miley}}{{Pentericci} et~al.}{2000}]{Pentericci2000}
{Pentericci} L.,  {Van Reeven} W.,  {Carilli} C.~L.,  {R{\"o}ttgering}
  H.~J.~A.,   {Miley} G.~K.,  2000, \mn@doi [\aaps] {10.1051/aas:2000104},
  \href {http://adsabs.harvard.edu/abs/2000A%26AS..145..121P} {145, 121}

\bibitem[\protect\citeauthoryear{{Pentericci}, {McCarthy}, {R{\"o}ttgering},
  {Miley}, {van Breugel}  \& {Fosbury}}{{Pentericci}
  et~al.}{2001}]{Pentericci2001}
{Pentericci} L.,  {McCarthy} P.~J.,  {R{\"o}ttgering} H.~J.~A.,  {Miley} G.~K.,
   {van Breugel} W.~J.~M.,   {Fosbury} R.,  2001, \mn@doi [\apjs]
  {10.1086/321781}, \href {http://adsabs.harvard.edu/abs/2001ApJS..135...63P}
  {135, 63}

\bibitem[\protect\citeauthoryear{{Reuland} et~al.,}{{Reuland}
  et~al.}{2003}]{Reuland2003}
{Reuland} M.,  et~al., 2003, \mn@doi [\apj] {10.1086/375619}, \href
  {http://adsabs.harvard.edu/abs/2003ApJ...592..755R} {592, 755}

\bibitem[\protect\citeauthoryear{{Rocca-Volmerange}, {Le Borgne}, {De Breuck},
  {Fioc}  \& {Moy}}{{Rocca-Volmerange} et~al.}{2004}]{Rocca2004}
{Rocca-Volmerange} B.,  {Le Borgne} D.,  {De Breuck} C.,  {Fioc} M.,   {Moy}
  E.,  2004, \mn@doi [\aap] {10.1051/0004-6361:20031717}, \href
  {http://adsabs.harvard.edu/abs/2004A%26A...415..931R} {415, 931}

\bibitem[\protect\citeauthoryear{{R{\"o}ttgering}}{{R{\"o}ttgering}}{1993}]{Rottgering1993}
{R{\"o}ttgering} H.~J.~A.,  1993, PhD thesis, Ph.~D.~thesis, University of
  Leiden (1993)

\bibitem[\protect\citeauthoryear{{R{\"o}ttgering}, {Lacy}, {Miley}, {Chambers}
  \& {Saunders}}{{R{\"o}ttgering} et~al.}{1994}]{Rottgering1994}
{R{\"o}ttgering} H.~J.~A.,  {Lacy} M.,  {Miley} G.~K.,  {Chambers} K.~C.,
  {Saunders} R.,  1994, \aaps, \href
  {http://adsabs.harvard.edu/abs/1994A%26AS..108...79R} {108, 79}

\bibitem[\protect\citeauthoryear{{R{\"o}ttgering}, {Hunstead}, {Miley}, {van
  Ojik}  \& {Wieringa}}{{R{\"o}ttgering} et~al.}{1995}]{Rottgering1995}
{R{\"o}ttgering} H.~J.~A.,  {Hunstead} R.~W.,  {Miley} G.~K.,  {van Ojik} R.,
  {Wieringa} M.~H.,  1995, \mnras, \href
  {http://adsabs.harvard.edu/abs/1995MNRAS.277..389R} {277, 389}

\bibitem[\protect\citeauthoryear{{Rush}, {McCarthy}, {Athreya}  \&
  {Persson}}{{Rush} et~al.}{1997}]{Rush1997}
{Rush} B.,  {McCarthy} P.~J.,  {Athreya} R.~M.,   {Persson} S.~E.,  1997, \apj,
  \href {http://adsabs.harvard.edu/abs/1997ApJ...484..163R} {484, 163}

\bibitem[\protect\citeauthoryear{{Seymour} et~al.,}{{Seymour}
  et~al.}{2007}]{Seymour2007}
{Seymour} N.,  et~al., 2007, \mn@doi [\apjs] {10.1086/517887}, \href
  {http://adsabs.harvard.edu/abs/2007ApJS..171..353S} {171, 353}

\bibitem[\protect\citeauthoryear{{Spearman}}{{Spearman}}{1904}]{Spearman1904}
{Spearman} C.,  1904, The Proof and Measurement of Association between Two
  Things.
University of Illinois Press, \mn@doi{10.2307/1412159}

\bibitem[\protect\citeauthoryear{{Swinbank} et~al.,}{{Swinbank}
  et~al.}{2015}]{Swinbank2015}
{Swinbank} A.~M.,  et~al., 2015, \mn@doi [\mnras] {10.1093/mnras/stv366}, \href
  {http://adsabs.harvard.edu/abs/2015MNRAS.449.1298S} {449, 1298}

\bibitem[\protect\citeauthoryear{{Tadhunter}}{{Tadhunter}}{1991}]{Tadhunter1991}
{Tadhunter} C.~N.,  1991, \mnras, \href
  {http://adsabs.harvard.edu/abs/1991MNRAS.251P..46T} {251, 46P}

\bibitem[\protect\citeauthoryear{{Tadhunter}, {Villar-Martin}, {Morganti},
  {Bland-Hawthorn}  \& {Axon}}{{Tadhunter} et~al.}{2000}]{Tadhunter2000}
{Tadhunter} C.~N.,  {Villar-Martin} M.,  {Morganti} R.,  {Bland-Hawthorn} J.,
  {Axon} D.,  2000, \mn@doi [\mnras] {10.1046/j.1365-8711.2000.03416.x}, \href
  {http://adsabs.harvard.edu/abs/2000MNRAS.314..849T} {314, 849}

\bibitem[\protect\citeauthoryear{{Taniguchi} et~al.,}{{Taniguchi}
  et~al.}{2001}]{Taniguchi2001}
{Taniguchi} Y.,  et~al., 2001, \mn@doi [\apjl] {10.1086/323652}, \href
  {http://adsabs.harvard.edu/abs/2001ApJ...559L...9T} {559, L9}

\bibitem[\protect\citeauthoryear{{Tody}}{{Tody}}{1993}]{Tody1993}
{Tody} D.,  1993, in {Hanisch} R.~J.,  {Brissenden} R.~J.~V.,   {Barnes} J.,
  eds,  Astronomical Society of the Pacific Conference Series Vol. 52,
  Astronomical Data Analysis Software and Systems II. p.~173

\bibitem[\protect\citeauthoryear{{Venemans} et~al.,}{{Venemans}
  et~al.}{2004}]{Venemans2004}
{Venemans} B.~P.,  et~al., 2004, \mn@doi [\aap] {10.1051/0004-6361:200400041},
  \href {http://adsabs.harvard.edu/abs/2004A%26A...424L..17V} {424, L17}

\bibitem[\protect\citeauthoryear{{Vernet}, {Fosbury}, {Villar-Mart{\'{\i}}n},
  {Cohen}, {Cimatti}, {di Serego Alighieri}  \& {Goodrich}}{{Vernet}
  et~al.}{2001}]{Vernet2001}
{Vernet} J.,  {Fosbury} R.~A.~E.,  {Villar-Mart{\'{\i}}n} M.,  {Cohen} M.~H.,
  {Cimatti} A.,  {di Serego Alighieri} S.,   {Goodrich} R.~W.,  2001, \mn@doi
  [\aap] {10.1051/0004-6361:20000076}, \href
  {http://adsabs.harvard.edu/abs/2001A%26A...366....7V} {366, 7}

\bibitem[\protect\citeauthoryear{{Villar-Mart{\'{\i}}n}, {Tadhunter}  \&
  {Clark}}{{Villar-Mart{\'{\i}}n} et~al.}{1997}]{VM1997}
{Villar-Mart{\'{\i}}n} M.,  {Tadhunter} C.,   {Clark} N.,  1997, \aap, \href
  {http://adsabs.harvard.edu/abs/1997A%26A...323...21V} {323, 21}

\bibitem[\protect\citeauthoryear{{Villar-Mart{\'{\i}}n}, {Tadhunter},
  {Morganti}, {Axon}  \& {Koekemoer}}{{Villar-Mart{\'{\i}}n}
  et~al.}{1999a}]{VM1999b}
{Villar-Mart{\'{\i}}n} M.,  {Tadhunter} C.,  {Morganti} R.,  {Axon} D.,
  {Koekemoer} A.,  1999a, \mn@doi [\mnras] {10.1046/j.1365-8711.1999.02603.x},
  \href {http://adsabs.harvard.edu/abs/1999MNRAS.307...24V} {307, 24}

\bibitem[\protect\citeauthoryear{{Villar-Mart{\'{\i}}n}, {Binette}  \&
  {Fosbury}}{{Villar-Mart{\'{\i}}n} et~al.}{1999b}]{VM1999a}
{Villar-Mart{\'{\i}}n} M.,  {Binette} L.,   {Fosbury} R.~A.~E.,  1999b, \aap,
  \href {http://adsabs.harvard.edu/abs/1999A%26A...346....7V} {346, 7}

\bibitem[\protect\citeauthoryear{{Villar-Mart{\'{\i}}n}, {Fosbury}, {Binette},
  {Tadhunter}  \& {Rocca-Volmerange}}{{Villar-Mart{\'{\i}}n}
  et~al.}{1999c}]{VM1999c}
{Villar-Mart{\'{\i}}n} M.,  {Fosbury} R.~A.~E.,  {Binette} L.,  {Tadhunter}
  C.~N.,   {Rocca-Volmerange} B.,  1999c, \aap, \href
  {http://adsabs.harvard.edu/abs/1999A%26A...351...47V} {351, 47}

\bibitem[\protect\citeauthoryear{{Villar-Mart{\'{\i}}n}, {Fosbury}, {Vernet},
  {Cohen}, {Cimatti}  \& {di Serego Alighieri}}{{Villar-Mart{\'{\i}}n}
  et~al.}{2001}]{VM2001}
{Villar-Mart{\'{\i}}n} M.,  {Fosbury} R.,  {Vernet} J.,  {Cohen} M.,  {Cimatti}
  A.,   {di Serego Alighieri} S.,  2001, \mn@doi [Astrophysics and Space
  Science Supplement] {10.1023/A:1012764309451}, \href
  {http://adsabs.harvard.edu/abs/2001ApSSS.277..571V} {277, 571}

\bibitem[\protect\citeauthoryear{{Villar-Mart{\'{\i}}n}, {Vernet}, {di Serego
  Alighieri}, {Fosbury}, {Pentericci}, {Cohen}, {Goodrich}  \&
  {Humphrey}}{{Villar-Mart{\'{\i}}n} et~al.}{2002}]{VM2002}
{Villar-Mart{\'{\i}}n} M.,  {Vernet} J.,  {di Serego Alighieri} S.,  {Fosbury}
  R.,  {Pentericci} L.,  {Cohen} M.,  {Goodrich} R.,   {Humphrey} A.,  2002,
  \mn@doi [\mnras] {10.1046/j.1365-8711.2002.05751.x}, \href
  {http://adsabs.harvard.edu/abs/2002MNRAS.336..436V} {336, 436}

\bibitem[\protect\citeauthoryear{{Villar-Mart{\'{\i}}n}, {Vernet}, {di Serego
  Alighieri}, {Fosbury}, {Humphrey}  \& {Pentericci}}{{Villar-Mart{\'{\i}}n}
  et~al.}{2003}]{VM2003}
{Villar-Mart{\'{\i}}n} M.,  {Vernet} J.,  {di Serego Alighieri} S.,  {Fosbury}
  R.,  {Humphrey} A.,   {Pentericci} L.,  2003, \mn@doi [\mnras]
  {10.1046/j.1365-2966.2003.07090.x}, \href
  {http://adsabs.harvard.edu/abs/2003MNRAS.346..273V} {346, 273}

\bibitem[\protect\citeauthoryear{{Villar-Mart{\'{\i}}n}, {Humphrey}, {De
  Breuck}, {Fosbury}, {Binette}  \& {Vernet}}{{Villar-Mart{\'{\i}}n}
  et~al.}{2007}]{VM2007a}
{Villar-Mart{\'{\i}}n} M.,  {Humphrey} A.,  {De Breuck} C.,  {Fosbury} R.,
  {Binette} L.,   {Vernet} J.,  2007, \mn@doi [\mnras]
  {10.1111/j.1365-2966.2006.11371.x}, \href
  {http://adsabs.harvard.edu/abs/2007MNRAS.375.1299V} {375, 1299}

\bibitem[\protect\citeauthoryear{{Willott}, {Rawlings}  \&
  {Blundell}}{{Willott} et~al.}{2001}]{Willott2001}
{Willott} C.~J.,  {Rawlings} S.,   {Blundell} K.~M.,  2001, \mn@doi [\mnras]
  {10.1046/j.1365-8711.2001.04209.x}, \href
  {http://adsabs.harvard.edu/abs/2001MNRAS.324....1W} {324, 1}

\bibitem[\protect\citeauthoryear{{Wylezalek} et~al.,}{{Wylezalek}
  et~al.}{2013}]{Wylezalek2013}
{Wylezalek} D.,  et~al., 2013, \mn@doi [\apj] {10.1088/0004-637X/769/1/79},
  \href {http://adsabs.harvard.edu/abs/2013ApJ...769...79W} {769, 79}

\bibitem[\protect\citeauthoryear{{van Ojik}}{{van Ojik}}{1995}]{Ojik1995}
{van Ojik} R.,  1995, PhD thesis, University of Leiden

\bibitem[\protect\citeauthoryear{{van Ojik}, {Rottgering}, {Miley}, {Bremer},
  {Macchetto}  \& {Chambers}}{{van Ojik} et~al.}{1994}]{Ojik1994}
{van Ojik} R.,  {Rottgering} H.~J.~A.,  {Miley} G.~K.,  {Bremer} M.~N.,
  {Macchetto} F.,   {Chambers} K.~C.,  1994, \aap, \href
  {http://adsabs.harvard.edu/abs/1994A%26A...289...54V} {289, 54}

\bibitem[\protect\citeauthoryear{{van Ojik}, {Roettgering}, {Miley}  \&
  {Hunstead}}{{van Ojik} et~al.}{1997}]{Ojik1997}
{van Ojik} R.,  {Roettgering} H.~J.~A.,  {Miley} G.~K.,   {Hunstead} R.~W.,
  1997, \aap, \href {http://adsabs.harvard.edu/abs/1997A%26A...317..358V} {317,
  358}

\makeatother
\end{thebibliography}

\bsp	
\label{lastpage}
\end{document}